\documentclass[aip,pop,reprint,amsmath,amssymb,floatfix]{revtex4-1}

\usepackage{epsfig}
\usepackage{color}
\usepackage{bm}

\newcommand{\be}{\begin{equation}}
\newcommand{\ee}{\end{equation}}
\newcommand{\lb}{\label}

\newcommand{\ba}{{\bf a}}

\newcommand{\bE}{{\bf e}}

\newcommand{\br}{{\bf r}}
\newcommand{\bu}{{\bf u}}

\newcommand{\bx}{{\bf x}}

\newcommand{\bz}{{\bf z}}

\newcommand{\bA}{{\bf A}}
\newcommand{\bB}{{\bf B}}

\newcommand{\bI}{{\bf I}}
\newcommand{\bJ}{{\bf J}}

\newcommand{\bX}{{\bf X}}
\newcommand{\bW}{{\bf W}}

\newcommand{\wt}{\widetilde}

\newcommand{\boJ}{{\mbox{\boldmath $\mathcal{J}$}}}

\newcommand{\boalpha}{{\mbox{\boldmath $\alpha$}}}
\newcommand{\boepsilon}{{\mbox{\boldmath $\varepsilon$}}}
\newcommand{\brho}{{\mbox{\boldmath $\rho$}}}
\newcommand{\boeta}{{\mbox{\boldmath $\eta$}}}
\newcommand{\boxi}{{\mbox{\boldmath $\xi$}}}

\newcommand{\bSigma}{{\mbox{\boldmath $\Sigma$}}}

\newcommand{\bdot}{{\mbox{\boldmath $\cdot$}}}
\newcommand{\btimes}{{\mbox{\boldmath $\times$}}}
\newcommand{\bcirc}{{\mbox{\boldmath $\circ$}}}
\newcommand{\bzed}{{\mbox{\boldmath $0$}}}
\newcommand{\grad}{{\mbox{\boldmath $\nabla$}}}
\newcommand{\bell}{{\mbox{\boldmath $\ell$}}}

\begin{document}


\title[Stochastic Flux-Freezing]{Stochastic Flux-Freezing and Magnetic Dynamo}



\author{Gregory L. Eyink}
\affiliation{Department of Applied Mathematics \& Statistics\\
and Department of Physics \& Astronomy \\
The Johns Hopkins University, USA}


\date{\today}

\begin{abstract}
We argue that magnetic flux-conservation in turbulent plasmas at high magnetic
Reynolds
numbers neither holds in the conventional sense nor is entirely broken, but
instead is valid
in a novel statistical sense associated to the ``spontaneous stochasticity''
of  Lagrangian
particle trajectories. The latter phenomenon is due to the explosive separation
of particles
undergoing turbulent Richardson diffusion, which leads to a breakdown of
Laplacian
determinism for classical dynamics. We discuss empirical evidence for
spontaneous stochasticity,
including our own new numerical results. We then use a Lagrangian path-integral
approach
to establish stochastic flux-freezing for resistive hydromagnetic equations and
to argue, based
on the properties of Richardson diffusion, that flux-conservation must remain
stochastic
at infinite magnetic Reynolds number. As an important application of these
results we consider
the kinematic, fluctuation dynamo in non-helical, incompressible turbulence at
unit magnetic Prandtl
number. We present results on the Lagrangian dynamo mechanisms by a stochastic
particle
method which demonstrate a strong similarity between the $Pr_m=1$ and $Pr_m=0$
dynamos. Stochasticity of field-line motion is an essential ingredient of both.
We finally
consider briefly some consequences for nonlinear MHD turbulence, dynamo and
reconnection.
\end{abstract}


\maketitle 


\section{Introduction}\lb{intro}

Hannes Alfv\'en in a seminal paper in 1942 introduced the notion of
flux-freezing in
magnetohydrodynamic plasmas at infinite conductivity, noting that ``every
motion
(perpendicular to the field) of the liquid in relation to the lines of force is
forbidden
because it would give infinite eddy currents'' \cite{Alfven42}. In the years
since, the
property of  flux-conservation has become a powerful tool in the analysis of
many
near-ideal plasma phenomena. For example, in his excellent monograph
\cite{Kulsrud05},
Kulsrud states that ``The most important property of an ideal plasma is
flux-freezing'',
before proceeding to illustrate its many applications.
Of course, physical plasmas in the the laboratory and in astrophysics
are subject to various forms of non-ideality, including Spitzer resistivity,
ambipolar
diffusion, etc. The general assumption in the field of plasma physics, however,
is that
as long as such non-ideality is sufficiently ``small'', then flux-freezing will
hold in an
approximate sense. This idea dominates the discussion of turbulent magnetic
dynamo
 at high kinetic and magnetic Reynolds numbers.
For example, it is commonplace to find statements in the literature
such as the following:  ``The small-scale turbulent dynamo is caused by the
random
stretching of the (nearly) frozen-in field lines by the ambient random flow''
\cite{Schekochihinetal04}, or  ``It is well established both analytically and
numerically
that a weak magnetic field can be amplified by the random motions of a highly
conducting fluid [1-3]. This occurs because magnetic-field lines are
generically
stretched by the random motions of the fluid in which they are (almost)
`frozen.' "
\cite{Boldyrevetal05}. The physical idea which underlies these statements is
that a tiny resistivity should diffuse field-lines only a short distance
through the plasma.
For example, the usual quantitative estimate is well expressed in this quote
from
Kulsrud's monograph, Ch.13, on magnetic reconnection:
\begin{quote}
``Flux freezing is a very strong constraint on the behavior of magnetic fields
in astrophysics.
As we show in chapter 3, this implies that lines do not break and their
topology is preserved.
The condition for flux freezing can be formulated as follows: In a time $t$, a
line of force can
slip through the plasma a distance
$$  \,\,\,\,\,\,\,\,\,\,\,\,  \,\,\,\,\,\,\,\,\,\,\,\,
\,\,\,\,\,\,\,\,\,\,\,\,\ell=\sqrt{\frac{\eta ct}{4\pi}}.
\,\,\,\,\,\,\,\,\,\,\,\,
     \,\,\,\,\,\,\, \,\,\,\,\,\,\,\,\,\,\,\, (1)$$
If this distance $\ell$ is small compared to $\delta$, the scale of interest,
then flux freezing
holds to a good degree of approximation.''
\end{quote}

We shall argue that these commonplace ideas on flux-freezing are wrong. They
contain an implicit assumption that the plasma fluid remains smooth and laminar
for
very small non-ideality. The quantitative estimate that field-lines slip
through a resistive
plasma only a diffusive distance $\propto \sqrt{\eta t}$ in time $t$ is
incorrect---by
many orders of magnitude---in a turbulent plasma. Since laminar flow at very
high kinetic
and magnetic Reynolds is unstable to development of turbulence,
flux-conservation
in the conventional sense must be the exception rather than the rule in
astrophysical
plasmas. Indeed, we shall show that the standard views on flux-freezing {\it
must} be
incorrect, because the very notion of a Lagrangian fluid particle trajectory
breaks
down in turbulent flow with a spatially ``rough'' velocity field (i.e. with a
power-law
kinetic energy spectrum similar to that of Kolmogorov). Recent research has
discovered
a novel phenomenon of ``spontaneous stochasticity'', according to which fluid
particle
trajectories are intrinsically random in high-Reynolds-number turbulence
\cite{Bernardetal98,GawedzkiVergassola00,Chavesetal03,
Kupiainen03,EvandenEijnden00,
EvandenEijnden01,LeJanRaimond02,LeJanRaimond04}. This surprising phenomenon
is a long overlooked consequence of the fluid-dynamical effect of Richardson
two-particle
turbulent dispersion \cite{Richardson26}. Because of spontaneous stochasticity,
it makes
no sense to assume that a field line follows ``the''  plasma fluid element,
because  there are
infinitely many distinct fluid trajectories starting from the same point! But
it is also not true
that flux-freezing is completely broken. We shall argue below that magnetic
flux-conservation
remains valid in the ideal limit of high Reynolds numbers, but in a novel
stochastic  sense
associated with the intrinsic stochasticity of the Lagrangian particle
trajectories.

A correct formulation of flux-freezing is fundamental to understand a number of
important astrophysical processes, such as turbulent dynamo and reconnection.
In previous work \cite{EyinkNeto10,Eyink10} we have shown how stochastic
flux-freezing is involved in the small-scale ``fluctuation dynamo'' for a
soluble
model problem: magnetic fields advected by the Kazantsev-Kraichnan ensemble
of velocity fields that are spatially rough and white-noise in time
\cite{Kazantsev68,
KraichnanNagarajan67,Kraichnan68}. It is worth noting, by the way, that
``spontaneous
stochasticity'' is a rigorously established phenomenon for this model
\cite{Bernardetal98,
GawedzkiVergassola00,Chavesetal03, Kupiainen03,EvandenEijnden00,
EvandenEijnden01,LeJanRaimond02,LeJanRaimond04}. It was shown
that presence of fluctuation dynamo effect at zero magnetic Prandtl number
depends crucially
on the degree of angular correlation between the infinite number of magnetic
field vectors
that are simultaneously advected by turbulence to the same spatial point
\cite{EyinkNeto10,Eyink10}.

Here we shall make a similar study for kinematic dynamo in non-helical,
hydrodynamic turbulence, by a Lagrangian numerical method
that employs data from a high Reynolds-number turbulent flow archived online
\cite{Lietal08,Perlmanetal07}. We present results for unit magnetic Prandtl
number
which demonstrate---and quantify---the effect of Richardson diffusion and
stochastic
flux-freezing on the small-scale turbulent dynamo.  We find, in fact,
remarkable
similarities between the Lagrangian mechanisms of small-scale dynamo in
hydrodynamic turbulence at unit Prandtl number and in the Kazantsev model
at zero Prandtl number. Previous numerical studies
\cite{Haugenetal04,Schekochihinetal04}
have instead found a close similarity of the unit Prandtl-number fluctuation
dynamo
with the solution of the Kazantsev model at {\it infinite} Prandtl-number,
observing,
in particular, the ``Kazantsev spectrum'' $k^{3/2}$ of magnetic energy at high
wavenumbers.  One study \cite{Schekochihinetal04} went so far as to claim that
`` It is clear that the kinematic dynamo in such [$Pr_m=1$] runs is of the
large-$Pr_m$
kind.'' This interpretation is ruled out by our new results at much higher
Reynolds
numbers, which show that the inertial-range phenomenon of Richardson diffusion
strongly affects the exponential growth rate of magnetic energy in the
kinematic regime.
However, stochastic flux-freezing is not a property of kinematic dynamo only
but will
hold also for fully nonlinear MHD turbulence and have important implications
there
for magnetic dynamo and reconnection.

The detailed contents of this paper are as follows: In the following section II
we shall
briefly review the phenomenon of ``spontaneous stochasticity'', both its
theoretical
bases and its present confirmation from simulations and experiments. We aso
present
new numerical results of our own which support the essential predictions. In
Section III
we discuss stochastic flux-freezing. We begin with a demonstration of the
stochastic
flux-conservation properties of resistive MHD, which are novel results
themselves.
We employ Lagrangian path-integral methods that provide good physical insight.
We then discuss the ideal case, via zero-resistivity and other limits.  In
section IV
we employ the new results to discuss the turbulent kinematic dynamo. After
reviewing
the Lagrangian theory of dynamo, we present our numerical results and their
comparison
with analytical results for the Kazantsev model at zero Prandtl number and with
previous
numerical studies. In section V we briefly discuss some implications and open
problems
for nonlinear MHD turbulence and section VI contains our final discussion. An
Appendix
sketches the derivation of the path-integral formulas used in the main text.

\section{Richardson Diffusion and Spontaneous Stochasticity}

\subsection{Richardson 2-Particle Dispersion}

We briefly review Richardson's theory \cite{Richardson26} of 2-particle or
relative turbulent dispersion, emphasizing perspectives of recent research.
See also more complete reviews
\cite{Sawford01,SalazarCollins09,ToschiBodenschatz09}.

The object of Richardson's study was the separation $\Delta
\bx(t)=\bx_1(t)-\bx_2(t)$
between a pair of passive Lagrangian tracer particles in a turbulent flow, such
as
ash particles in a volcanic plume. Richardson's approach was semi-empirical.
By estimating the ``effective diffusivity'' $K=\langle |\Delta\bx|^2\rangle/t$
as a
function of rms separation $\ell=\sqrt{\langle |\Delta\bx|^2\rangle},$ he
inferred from
data that there is a scale-dependent diffusivity coefficient
\be  K(\ell) \sim K_0 \ell^{4/3}.  \lb{eddy-diff} \ee
This is essentially a ``running coupling constant" in the modern sense of
renormalization
group theory.  Richardson proposed further that the probability density
function of the
separation vector $\bell=\bx_1-\bx_2$ would satisfy a diffusion equation
\be \partial_t P(\bell,t)=\frac{\partial}{\partial\ell_i} \left(K(\ell)
            \frac{\partial P}{\partial\ell_i}(\bell,t)\right). \lb{Rich-eq} \ee
Richardson observed that there is an exact similarity solution of his equation
given by a stretched-exponential PDF, which he wrote explicitly in one space
dimenson.
Here we note its form
\be P_*(\bell,t) = \frac{A}{(K_0 t)^{9/2}}\exp\left(-\frac{9\ell^{2/3}}{4K_0
t}\right) \lb{Rich-sol} \ee
in the more physically relevant case of three space dimensions.  All solutions
of
(\ref{Rich-eq}) approach this self-similar form asymptotically at long times
\cite{EyinkXin00}.
Averaging $\ell^2$ with respect to the self-similar density (\ref{Rich-sol})
yields
\be \langle\ell^2(t)\rangle = \gamma_0 t^3 \lb{Rich-law} \ee
with $\gamma_0=\frac{1144}{81}K_0^3.$ This is the famous Richardson $t^3$-law.

Richardson's work preceded the Kolmogorov 1941 (K41) theory of turbulence, but
it
was shown by Obukhov \cite{Obukhov41} to be fully consistent with that theory.
This can be seen by a toy calculation in one space dimension. Assume that
$\ell(t)$ satisfies the initial-value problem
$$ \frac{d}{dt} \ell(t) = \delta u(\ell)= \frac{3}{2}(g_0\varepsilon
\ell)^{1/3}, \,\,\,\,\ell(0)=\ell_0, $$
with velocity increment $\delta u(\ell)$ scaling as in K41 theory, where
$\varepsilon$
is the mean energy dissipation per unit mass.  Separation of variables gives
the exact solution
\be \ell(t)=\left[ \ell_0^{2/3} + (g_0\varepsilon)^{1/3}t \right]^{3/2}.
\lb{toy-sol} \ee
If one defines a time $t_0\equiv \ell_0^{2/3}/(g_0\varepsilon)^{1/3}$ which
characterizes the initial separation then, for $t\gg t_0,$
\be \ell^2(t) \sim g_0\varepsilon t^3. \lb{toy-law} \ee
For sufficiently long times,  the particles ``forget'' their initial separation
and the Richardson
law is obtained. The dimensionless parameter $g_0$ which appears in this form
of the $t^3$-law is usually called the Richardson-Obukhov constant. The
physical
mechanism of the explosive separation of particles, even faster than ballistic,
is the relative advection of the pairs by larger, more energetic eddies as
their
separation distance increases.

This physics seems relatively simple and benign, but it has extraordinary
consequences.
As first pointed out in a seminal paper of Bernard, Gaw\c{e}dzki, and Kupiainen
\cite{Bernardetal98}, Richardson's theory implies a breakdown in the usual
notion of
Laplacian determinism for classical dynamics! This may already be seen in
our toy calculation above. If we set $\ell_0=0,$ then the solution
(\ref{toy-sol})
becomes the Richardson law $\ell^2(t)= g_0\varepsilon t^3> 0$ for all positive
times $t.$ Thus, two particles started at the {\it same} point at time $0$
separate to
a finite distance at any time $t>0.$ The same oddity may be seen in
Richardson's
similarity solution (\ref{Rich-sol}), which satisfies at initial time $t=0$
$$ P_*(\bell,0) = \delta^3(\bell). $$
All particles start with separation $\ell(0)=0.$ However, $P_*(\bell,t)$
is a smooth density for $t>0,$ so that $\ell(t)>0$ with probability one at
later times.
Richardson's theory thus implies that two particles advected by the fluid
velocity
$\bu(\bx,t)$ which start at the {\it same} initial point $\bx_0$
$$ \frac{d}{dt}\bx(t)=\bu(\bx(t),t), \,\,\,\,\bx(0)=\bx_0 $$
can follow different trajectories. This seems to violate the theorem on
uniqueness of
solutions of initial-value problems for ODE's. However, such theorems assume
that
the advecting velocity $\bu(\bx,t)$ is H\"{o}lder-Lipschitz continuous in the
space
variable $\bx$
\be  |\bu(\bx_1,t)-\bu(\bx_2,t)|\leq C|\bx_1-\bx_2|^h \lb{hoelder} \ee
with exponent $h\geq 1.$ A turbulent velocity field in a Kolmogorov inertial
range
has instead H\"older exponent $h\doteq 1/3$ and the uniqueness theorem need not
apply. Our toy calculation earlier is just the standard textbook example for
failure of
uniqueness (see Hartman\cite{Hartman02}, p.2) . In that example, for any
non-negative ``waiting
time'' $\tau\geq 0$ $$  \ell(t)=(g_0\varepsilon)^{1/2}(t-\tau)_+^{3/2}$$
is a solution of the  initial-value problem with $\ell_0=0$. (Here $(x)_+=x$
for $x>0$ and $=0$ otherwise.)

The above considerations may seem fairly technical and mathematical. It has
been
shown, however, that this breakdown in uniqueness of trajectories can appear in
various physical limits for turbulent advection. Even more remarkably, the
solutions
of the deterministic classical dynamics become intrinsically stochastic! See
the important
series of papers
\cite{Bernardetal98,GawedzkiVergassola00,Chavesetal03,EvandenEijnden00,
EvandenEijnden01}. Advanced probablistic techniques have obtained the most
refined results \cite{LeJanRaimond02,LeJanRaimond04}.  A very clear and concise
review of the subject
\cite{Kupiainen03} is available, which the reader may consult for further
details. Below we
briefly describe the most basic theory and results.

\subsection{High-Reynolds-Number Limit and Spontaneous Stochasticity}

The easiest way to understand the phenomenon is via the problem of stochastic
particle advection,
\be  \frac{d}{dt}\wt{\bx}(t)=\bu^\nu(\wt{\bx}(t),t)+
     \sqrt{2\kappa}\,\wt{\boeta}(t), \,\,\,\,\bx(t_0)=\bx_0 \lb{SDE} \ee
with advecting velocity perturbed by a Gaussian white-noise $\wt{\boeta}(t)$
multiplied
by $\sqrt{2\kappa}$ and with velocity assumed spatially smooth at subviscous
length-scales
$\ell<\ell_\nu$ ,
for a finite viscosity $\nu.$ The transition probability for a single particle
in a fixed (non-random)
velocity realization $\bu^\nu$ can be written using a ``sum-over-histories''
approach as a
path-integral \cite{Drummond82,ShraimanSiggia94,Bernardetal98}:
\begin{eqnarray}
&& G^{\nu,\kappa}_\bu(\bx_f,t_f|\bx_0,t_0)
     = \int_{\bx(t_0)=\bx_0}\mathcal{D}\bx \,\, \delta^3(\bx_f-\bx(t_f)) \cr
&& \,\,\,\,\,\,\,\,\,\,\times
\exp\left(-\frac{1}{4\kappa}\int_{t_0}^t d\tau\,
|\dot{\bx}(\tau)-\bu^\nu(\bx(\tau),\tau)|^2\right).
\lb{P-pathint} \end{eqnarray}
Since this formula plays an important role in our analysis, we provide a
self-contained derivation
in the Appendix. A physical motivation to study such random advection is the
problem
of the evolution of a passive scalar, such as a temperature field or
dye concentration. These fields solve the scalar advection-diffusion equation
\be \partial_t\theta + (\bu^\nu\bdot\grad)\theta = \kappa\triangle\theta,
\lb{scalar-eq} \ee
with $\kappa$ the molecular diffusivity. The exact solution of
(\ref{scalar-eq}) is given
by the Feynman-Kac formula
\cite{ShraimanSiggia94,Bernardetal98,Drummond82,Sawford01}:
\begin{eqnarray}
&& \theta(\bx,t) = \int d^3x_0\, \theta(\bx_0,t_0)
G^{\nu,\kappa}_\bu(\bx_0,t_0|\bx,t) \cr
&& \,\,\,\,\, \,\,\,\,\, \,\,\,\,\,  \,\,\,= \int_{\ba(t)=\bx} \mathcal{D}\ba
\,\, \theta(\ba(t_0),t_0) \cr
&&   \,\,\,\,\, \,\,\,\,\, \,\,\,\,\,  \,\,\,\,\,
\times \exp\left(-\frac{1}{4\kappa}\int_{t_0}^t d\tau\,
|\dot{\ba}(\tau)-\bu^\nu(\ba(\tau),\tau)|^2\right)
\,\,\,\,\,\,\,
\lb{theta-FK} \end{eqnarray}
for $t_0<t.$ This corresponds to solving backward in time the stochastic
equation
$$ \frac{d}{d\tau}\wt{\ba}(\tau)
=\bu^\nu(\wt{\ba}(\tau),\tau)+\sqrt{2\kappa}\,\wt{\boeta}(\tau) $$
from $\tau=t$ to $\tau=t_0,$ with the condition $\wt{\ba}(t)=\bx.$ The present
value of the
scalar field is thus the average, along stochastic Lagrangian paths, of its
earlier values.

It naively appears by an application of the Laplace asymptotic method to
(\ref{P-pathint})
that the transition probability collapses to a delta-function
\be G^{\nu,\kappa}_\bu(\bx_f,t_f|\bx_0,t_0)\rightarrow \delta^3(\bx_f-\bx(t_f))
\lb{G-delta} \ee
as $\kappa\rightarrow 0,$ with $\bx(t)$ the solution of the ODE
$\dot{\bx}=\bu(\bx,t)$ for initial condition $\bx(t_0)=\bx_0.$
Only for such time-histories is the action vanishing in the exponent
of the path-integral.  However, it was shown
\cite{Bernardetal98,GawedzkiVergassola00,
Chavesetal03, Kupiainen03,EvandenEijnden00,EvandenEijnden01,
LeJanRaimond02,LeJanRaimond04} that (\ref{G-delta}) may not hold if
simultaneously
$\nu\rightarrow 0$ (or the Reynolds-number $Re=u_{rms}L/\nu\rightarrow \infty$)
and if in that limit the velocity field $\bu^\nu\rightarrow \bu,$ for a rough
(non-smooth,
singular) $\bu$.  In that case, as $\kappa,\nu\rightarrow 0,$
\be G^{\nu,\kappa}_\bu(\bx_f,t_f|\bx_0,t_0)\rightarrow
G_\bu(\bx_f,t_f|\bx_0,t_0) \lb{G-limit} \ee
for a nontrivial probability density $G_\bu.$ The Lagrangian trajectories can
remain random
as $\kappa,\nu\rightarrow 0$! This phenomenon has been called ``spontaneous
stochasticity''
\cite{Chavesetal03}, because of the analogy with spontaneous symmetry-breaking
in condensed
matter physics and quantum field-theory, where, for example, a ferromagnet may
retain a
non-vanishing magnetization even in the limit of vanishing external magnetic
field.
It is important to appreciate, however, that ``spontaneous stochasticity'' is a
very different
type of randomness than is usual in turbulence theory, associated to a random
ensemble
of velocity fields. Instead, the randomness in (\ref{G-limit}) is for a fixed
(non-random)
ensemble member $\bu$. The limiting distribution consists of time-histories
that are all
solutions of the same deterministic initial-value problem
\be \dot{\bx}=\bu(\bx,t),\,\,\,\,\bx(t_0)=\bx_0. \lb{iv-prob} \ee
As is clear from (\ref{P-pathint}), the limiting probability measure is in a
certain sense
the uniform or equal-weight distribution over all such solutions. We note in
passing that
Kneser's theorem in the mathematical theory of ODE's implies that, whenever
there is
more than one such solution, then there is in fact a continuous infinity of
solutions
(see Hartman \cite{Hartman02}, section II.4).

The above results have been rigorously established \cite{Bernardetal98,
GawedzkiVergassola00,Chavesetal03, Kupiainen03,EvandenEijnden00,
EvandenEijnden01,LeJanRaimond02,LeJanRaimond04}
for some model problems of turbulent advection, primarily the {\it Kraichnan
model}
of advection by a Gaussian random velocity field with zero mean and covariance
$$ \langle u_i^\nu(\bx,t) u_j^\nu(\bx',t')
\rangle=[D_0\delta_{ij}-S_{ij}^\nu(\bx-\bx')]\delta(t-t'). $$
The velocity fields are temporal white-noise and spatially rough for
$\ell_\nu<r<L_u,$
with a H\"older exponent $0<h<1,$ but smooth for $r<\ell_\nu.$ We discuss the
model
only for incompressible flow, in which case
\be S_{ij}^\nu(\br)=\left\{\begin{array}{ll}
        D_1 [(1+h)\delta_{ij}-h \hat{r}_i\hat{r}_j]r^{2h}, & \ell_\nu\ll r\ll
L_u, \cr
        D_1 \ell_\nu^{2h-2} [2\delta_{ij}-\hat{r}_i\hat{r}_j]r^{2}, & r\ll
\ell_\nu. \cr
        \end{array}\right. \lb{KK-space-corr} \ee
The velocity realizations $\bu^\nu(\bx,t)$ are divergence-free and H\"{o}lder
continuous
with exponent $h$ for $\nu \equiv D_1 \ell_\nu^{2h} \rightarrow 0.$ The kinetic
energy
spectra are power-laws $k^{-n}$ for the ``inertial range''  $1/L_u\ll k\ll
1/\ell_\nu,$ with
$n=1+2h$ and thus $1<n<3.$ The key feature which makes this model analytically
tractable is the Markovian property in time, which leads to the exact validity
of Richardson's
2-particle diffusion equation in the form
\be  \partial_t P(\br,t) =
\frac{\partial}{\partial r_i}\left(S_{ij}^\nu(\br)\frac{\partial P}{\partial
r_j}(\br,t)\right)
    +2\kappa \triangle_r P(\br,t), \lb{Kr-Rich-eq} \ee
with $\br=\bx_1-\bx_2.$  Note that Richardson's original equation
is obtained for $h=2/3$ rather than the Kolmogorov value $h=1/3,$  a
peculiarity of the
white-noise approximation. The main features of the solutions of this equation
can be
inferred from a study of the dispersion. Multiplying (\ref{Kr-Rich-eq}) by
$r_kr_\ell$ and
integrating over $\br$ leads to
\be \frac{d}{dt}\langle r_k(t)r_\ell(t)\rangle = 2\langle
S_{k\ell}^\nu(\br(t))\rangle + 4\kappa \delta_{k\ell}
\lb{disperse-eq} \ee
An analysis of (\ref{Kr-Rich-eq}) and (\ref{disperse-eq}) leads to the
following key results
for the ensemble of particles which are all started at the same point ($r_0=0$)
{}.

For $Pr\equiv \frac{\nu}{\kappa}<1,$ the dispersion is that of two independent
Brownian motions in three space dimensions
\be \langle r^2(t)\rangle \sim 12\kappa t \lb{rsq-kappa} \ee
for short times $t\ll t_\kappa=\frac{\ell_\kappa^2}{\kappa}$ with
$\ell_\kappa^2=
\left(\frac{\kappa}{D_1}\right)^{1/h}=\ell_\nu^2/(Pr)^{1/h},$ but of
Richardson-type
\be \langle r^2(t)\rangle \sim g_h(D_1 t)^{1/(1-h)} \lb{rsq-Rich} \ee
for longer times $t_\kappa\ll t\ll t_L=L^{2(1-h)}/D_1,$ with $g_h$ a constant
independent of $\kappa.$ For $Pr>1,$ the behavior is a bit more complex. The
result
(\ref{rsq-kappa}) still holds at very short times $t\ll t_\nu\equiv
\ell_\nu^2/\nu,$ but
at large Prandtl numbers there is an intermediate range of exponential growth
\be \langle r^2(t)\rangle\propto \kappa t_\nu e^{2\lambda_\nu t}
\lb{rsq-kap-nu} \ee
for  $t_\nu \ll t\ll \ln(Pr) t_\nu.$ Here $\lambda_\nu\propto t_\nu^{-1}$ is
the
leading Lyapunov exponent of the smooth advecting velocity field.
The dispersion at longer times again follows the Richardson law
(\ref{rsq-Rich}).
In either case, the Richardson law is valid once $\langle r^2(t)\rangle\gtrsim
\max\{\ell_\kappa^2,\ell_\nu^2\}$ and thus holds for arbitrarily small times
$t>0$
in the limit as $\nu,\kappa\rightarrow 0.$

A non-vanishing dispersion implies that the Lagrangian trajectories must stay
random
in the limit. Although the diffusion equation (\ref{Kr-Rich-eq}) has been
averaged over velocity
realizations $\bu,$ it must be the case that $P_\bu(\br,t|\bzed,0)\neq
\delta^3(\br)$ for $t>0$
and for a set of $\bu$ with nonzero probability, or otherwise the average over
$\bu$ would
also be a delta-function! The physical mechanism of spontaneous stochasticity
is clearly
the  ``forgetting'' of the length-scales $\ell_\kappa,\ell_\nu$ by Richardson
diffusion
for sufficiently long times, $t\gg t_\kappa,t_\nu,$ with those times also
vanishing in the
limit $\nu,\kappa\rightarrow 0.$
In the case of incompressible flow, the limiting distribution is completely
independent
of how the limit is taken. We note in passing that this is {\it not} true in
general for
compressible flows and that the possibility for Lagrangian particles to stick
as well
as to stochastically split allows there to be different limits, depending upon
the Prandtl
number $Pr,$ in the limit as $\nu,\kappa\rightarrow 0$
\cite{GawedzkiVergassola00,
EvandenEijnden01,LeJanRaimond04}. However, for incompressible flow the limiting
distribution is very universal and robust.

The same limit is obtained for incompressible flow even with $\kappa=0$
or $Pr=\infty,$ if the randomness is introduced through the initial conditions
rather than stochastic noise. Consider the solution of the initial-value
problem
for the Kraichnan ensemble of velocities
$$ \frac{d}{dt}\wt{\bx}(t)=\bu^\nu(\wt{\bx}(t),t),\,\,\,\, \wt{\bx}(t_0)=\bx_0
+
    \epsilon\wt{\brho}$$
where $\wt{\brho}$ is a zero-mean, unit-variance random vector with probability
density
$Q(\brho).$ One may interpret $\epsilon$ as the size of error in measuring the
initial
position of the particle. This corresponds to solving the Richardson diffusion
equation
(\ref{Kr-Rich-eq})  with $\kappa=0$ and initial condition
$P^\epsilon(\br,t_0)=\epsilon^{-3}
(Q*Q)(\br/\epsilon)$ so that $P^\epsilon(\br,t_0)\rightarrow \delta^3(\br)$ as
$\epsilon\rightarrow 0.$
However, if the limits are taken $\nu\rightarrow 0$ first and
$\epsilon\rightarrow 0$
subsequently, then the solution of the Richardson equation does not degenerate
to
a delta-function for $t>0.$ This may be seen by solving for the dispersion from
eq. (\ref{disperse-eq}) with $\kappa=0$ and $\langle r^2(0)\rangle=2\epsilon^2.
$
When $\epsilon<\ell_\nu$
\be \langle r^2(t)\rangle\sim 2\epsilon^2 e^{2\lambda_\nu t} \lb{rsq-nu} \ee
for times $t\ll t_\nu\ln(\ell_\nu/\epsilon)$  but for longer times follows
the Richardson law (\ref{rsq-Rich}). When $\ell_\nu<\epsilon$ instead, then the
short-time behavior is diffusive
\be  \langle r^2(t)\rangle\sim 2\epsilon^2 + (const.) D_1 \epsilon^{2h} t,
      \,\,\,\,\,\, \lb{rsq-vel} \ee
 for times $t\ll \frac{\epsilon^{2(1-h)}}{D_1}.$ Dispersion for such a ``random
cloud''
 of initial positions was first considered by Batchelor
\cite{Batchelor50,Batchelor52},
who obtained instead ballistic growth $\propto t^2$ for hydrodynamic
turbulence.
The diffusive result above is an artefact of the white-in-time velocity. At
times
$t\gg \frac{\epsilon^{2(1-h)}}{D_1}$ the Richardson law (\ref{rsq-Rich}) holds.
As in the previous cases, the Richardson law holds for any time $t>0$ if
$\nu,\epsilon
\rightarrow 0$ with $\epsilon$ vanishing slower than $\ell_\nu
e^{-O(D_1/\ell_\nu^{2(1-h)})}.$

As this last discussion should make clear, the phenomenon of ``spontaneous
stochasticity'' is not especially connected with the random perturbation of the
motion
equations in (\ref{SDE}). Instead it is the advection of the particle by a
rough velocity
field and the ``forgetting'' of the initial separations which makes the
Lagrangian
particle motions intrinsically stochastic.  Spontaneous stochasticity should
not be
confused with ``chaos'', as that term is used in dynamical systems theory
\cite{Chavesetal03,Kupiainen03}.  For chaotic dynamical systems with a
smooth velocity field, one sees only exponential growth of deviations as in
eq.(\ref{rsq-nu}). Because this result is proportional to $\epsilon^2,$ the
initial separation is never ``forgotten'' and, for all times, $\langle
r^2(t)\rangle
\rightarrow 0$ as $\epsilon\rightarrow 0.$  For chaotic dynamics, any
imprecision in the initial data is exponentially magnified, leading to loss of
predictability at long enough times.  Spontaneous stochasticity corresponds
instead
to $\lambda_\nu=+\infty$. The solution is unpredictable for {\it all} future
times, even
with infinitely precise knowledge of the initial conditions!

We remark finally that spontaneous stochasticity has very important
implications
for turbulent advection of a passive scalar \cite{Bernardetal98}. Note that
$$ \frac{d}{dt} \int d^3x\,\,\theta^2(\bx,t)= -2\kappa\int
d^3x\,\,|\grad\theta(\bx,t)|^2, $$
so that, naively, the integral is conserved for $\kappa=0.$
However, in the infinite Reynolds-number, fixed Prandtl-number limit
($\nu,\kappa\rightarrow 0$
with $Pr$ constant)
$$ \theta(\bx,t) = \int d^3x_0\, \theta(\bx_0,t_0) G_\bu(\bx_0,t_0|\bx,t), $$
using the results (\ref{theta-FK}) and (\ref{G-limit}).  The scalar at the
present time
is a nontrivial average along stochastic Lagrangian trajectories of its values
at an
earlier time, with molecular mixing replaced by turbulent mixing. In
particular,
the scalar intensity---for every velocity realization, without ensemble
averaging---
is still dissipated
$$ \int d^3x\,\,\theta^2(\bx,t)<\int d^3x\,\,
\theta^2(\bx,t_0),\,\,\,\,\,t>t_0$$
even as $\nu,\kappa\rightarrow 0$!
This is the scalar analogue of the dissipative anomaly of Onsager for fluid
turbulence
\cite{Onsager49,EyinkSreenivasan06,EyinkPhysD08}.
The Lagrangian mechanism is spontaneous stochasticity.

\subsection{Experimental and Numerical Results}

Our quantitative discussion above has been based upon the original Richardson
theory
and, in particular, his diffusion equation (\ref{Kr-Rich-eq}). This equation is
exact
for the Kraichnan white-in-time velocity ensemble but is only an approximation
for
hydrodynamic turbulence, where several of its quantitative predictions are
known to be
incorrect. We have already mentioned the diffusive short-time growth in
dispersion
for a particle cloud, eq.(\ref{rsq-vel}), which is ballistic for fluid
turbulence. (Note that
the ballistic regime is correctly predicted by Richardson's diffusion equation
for a
suitably time-dependent eddy diffusion tensor \cite{Batchelor52,Kraichnan66}.)
The diffusion equation (\ref{Kr-Rich-eq}) holds in the Kraichnan model also for
backward-in-time dispersion. However, actual dispersion rates are different
forward
and backward in time because of the negative skewness of turbulent velocity
increments
\cite{Kraichnan66,Sawfordetal05}. There is presently no quantitative theory of
turbulent dispersion which successfully accounts for all aspects of the
phenomenon.
It is necessary to stress that the prediction of ``spontaneous stochasticity''
has more general grounds in the mathematical theory of ODE's and is not
dependent
upon the diffusion approximation (\ref{Kr-Rich-eq}). Nevertheless, in the
absence of
any fully successful, quantitative theory, it is important to develop
understanding from
numerical simulations and laboratory experiments. We shall here briefly review
the
empirical studies of turbulent dispersion and the status of Richardson's
theory. In particular,
we shall present some new numerical results of our own on stochastic particle
advection
according to eq.(\ref{SDE}) for a turbulent velocity field.

We confine our discussion to just some of the latest studies by experiments
\cite{OttMann00,Bourgoinetal06,Ouelletteetal06,Xuetal08} and simulations
\cite{IshiharaKaneda02,
Biferaleetal05,Sawfordetal08} at the highest Reynolds numbers. For more
complete surveys
of the literature, see those papers and also recent review articles
\cite{SalazarCollins09,
ToschiBodenschatz09}. Athough the $t^3$-law (\ref{Rich-law}) and the
stretched-exponential
PDF (\ref{Rich-sol}) are probably the most famous predictions of Richardson's
theory, even
more important for our discussion is the ``forgetting'' of initial separations.
If $r_0$ is the
initial particle separation distance and $\varepsilon$ is energy dissipation
per mass, then
for times much greater than $t_0\equiv (r_0^2/\varepsilon)^{1/3}$ both $\langle
r^2(t)\rangle$
and $P(r,t)$ should become independent of $r_0.$ As we have seen, this is the
crucial
physical mechanism underlying spontaneous stochasticity. In general, it has
proved rather
difficult to observe in a completely consistent and convincing way all of these
predictions
of Richardson's theory.

Experiments of Ott and Mann at maximum Taylor-scale Reynolds number $Re_T=107$
observed
both a $t^3$ law and the Richardson PDF,  but varied $r_0$ only by a factor of
$1.5$
around the value $r_0=10\ell_\nu$ (for $\ell_\nu=(\nu^3/\varepsilon)^{1/4}$ the
Kolmogorov
dissipation length). Thus, they provide no information on collapse independent
of $r_0.$
A series of experiments by Bodenschatz and collaborators \cite{Bourgoinetal06,
Ouelletteetal06,Xuetal08} at substantially higher Reynolds numbers up to
$Re_T=815$
fail to see a $t^3$ law and instead produce results consistent with Batchelor's
ballistic $t^2$
range.  However, their smallest achievable value of the initial separation was
only about
$r_0=30\ell_\nu,$ so that it is arguable that they need longer times and still
higher Reynolds
numbers.  At their smallest value of $r_0$ they did observe Richardson's PDF
(\ref{Rich-sol})
and for $r_0=20$--$150\ell_\nu$ they see results roughly consistent with
Richardson's
predictions for the quantity $\langle r^{2/3}(t)\rangle-r_0^{2/3}$ and also
some tendency
to collapse independent of $r_0$ (see Xu et al. \cite{Xuetal08}, Fig.~1).

Numerical simulations of  Ishihara and Kaneda \cite{IshiharaKaneda02} at
$Re_T=283$
claimed to observe an inertial-range  $t^3$ law for $r_0$ in a range between
$5$--$45\ell_\nu,$
but no tendency whatsoever for collapse at long times independent of $r_0.$
However,
Biferale et al. \cite{Biferaleetal05} in a simulation at nearly identical
$Re_T=284$ and
with $r_0$ in a range of $1.2$--$19.6\ell_\nu$ see exactly the opposite: only a
slight indication
of a $t^3$-law for their smallest separation but a strong tendency to collapse
at long times.
They also observe Richardson's PDF (\ref{Rich-sol}) for $r_0=1.2\ell_\nu.$ More
recently,
Sawford et al. \cite{Sawfordetal08} have performed simulations with maximum
$Re_T=650.$
They find the best evidence yet of Richardson's predictions for the dispersion
(see their Fig.4), with
a reasonable $t^3$ range for $r_0=4\ell_\nu.$ Other values of $r_0$ in the
range of $0.25$--$256
\ell_\nu$ do not give a convincing $t^3$-law but do verify a tendency for
collapse at long times.

In view of the incomplete verification of Richardson's predictions, we decided
to undertake
our own numerical investigation. Unlike previous works, however, which have
nearly all
studied deterministic fluid particles with variable initial separations $r_0,$
we have instead
studied the problem of stochastic particle advection according to the
eq.(\ref{SDE}). The
velocity field is obtained from a $1024^3$ pseudospectral numerical simulation
of forced,
statistically stationary turbulence at  $Re_T=433.$ The flow data is available
online
at {\tt http://turbulence.pha.jhu.edu} and fully documented there and in papers
\cite{Lietal08,
Perlmanetal07}. The entire flow history for about one large-eddy turnover time
$L_u/u'$
is archived at a time resolution suitable for particle-tracking experiments,
with spatial and
temporal interpolation implemented within the database. This is very convenient
for our purposes, since it permits us to study particle dispersion backward in
time as well as
forward. As we have discussed above, it is backward dispersion that is most
relevant
for turbulent mixing \cite{Sawford01}.

We have studied two values of the Prandtl number $Pr=\nu/\kappa=1$ and $0.1$.
We solved (\ref{SDE}) using the simplest Euler-Maruyama scheme and also, for
convergence analysis,  an explicit, 1.5th-order strong scheme (Kloeden and
Platen
\cite{KloedenPlaten92}, Section 11.2). We took time discretization
$dt=10^{-3},$
which guaranteed that particles moved a fraction of $\ell_\nu$ under both
turbulent advection and Brownian diffusion at each time-step. The velocity
field between stored data-points was interpolated by 6th-order Lagrange
polynomials in space and piecewise-cubic Hermite polynomials in time. The
results were
verified to be converged in $dt$ both by comparison with the higher-order
method and
with the Euler scheme at a halved step size. For each initial location $\bx_0$
we evolved
$N=1024$ independent particle realizations, giving 523,776 particle pairs, over
the whole
time range of the database. For forward tracking we started particles at
$t_0=0$ and for
backward tracking at $t_f=2.048$ (the final time in the database). We then
averaged all
results over 256 initial locations $\bx_0$ obtained by choosing 4 independent,
uniformly
distributed points from each of $4^3=64$ subcubes of the whole flow domain.

\begin{figure}
\begin{center}
\includegraphics[width=250pt,height=450pt]{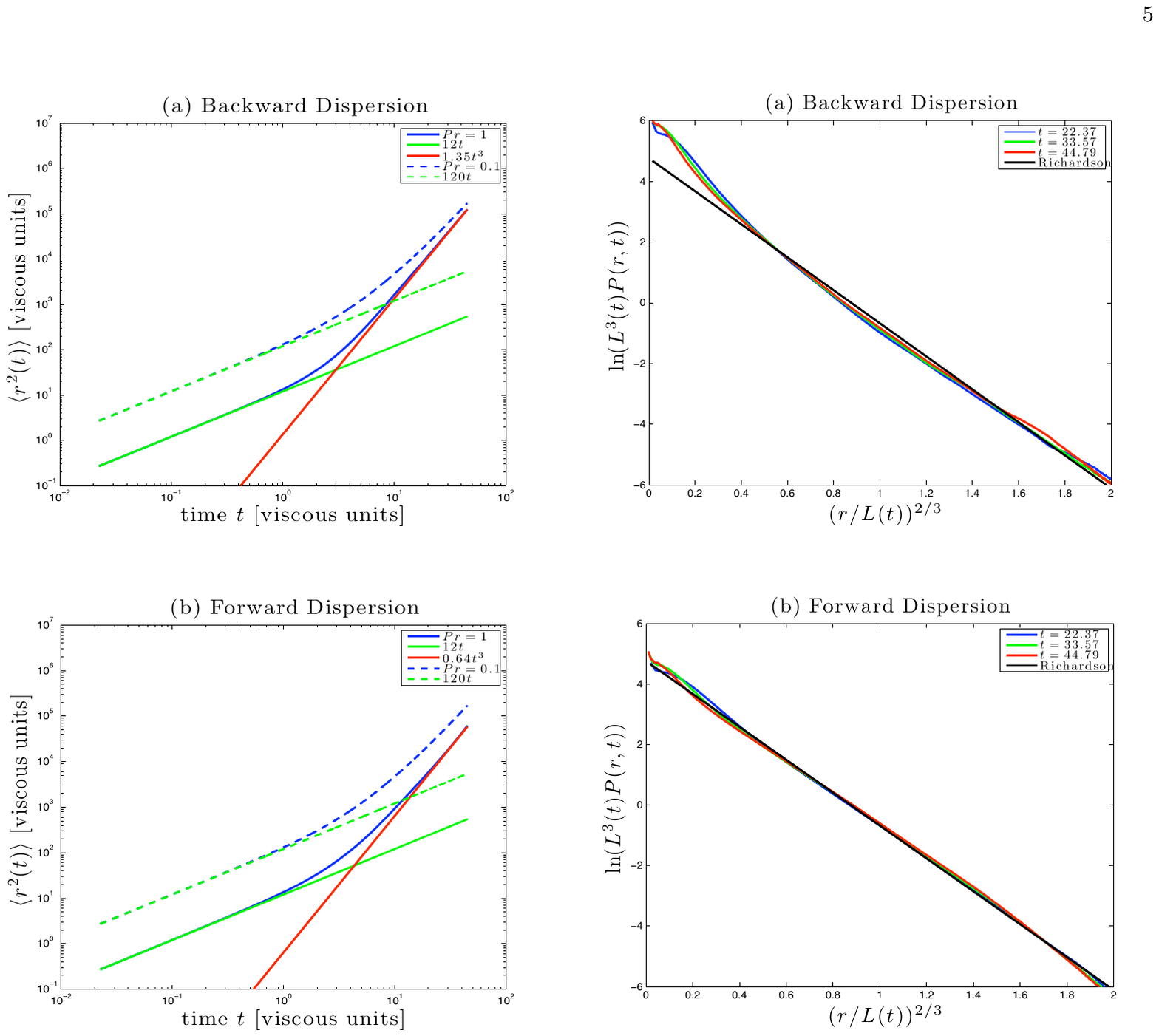}\\
\end{center}
\caption{Mean dispersion of particle pairs: (a) backward dispersion, (b)
forward dispersion.
The backward in time results are plotted against $t'=t_f-t$ and all quantities
are non-dimensionalized with viscous units (see text). The $Pr=1$
results are plotted with solid lines ({\bf \textemdash\textemdash}), $Pr=0.1$
results
with dashed lines ({\boldmath $-\,-\,-$}). We color code the lines with
\textcolor{blue}{blue}
for raw data, \textcolor{green}{green} for short-time molecular diffusion, and
\textcolor{red}{red} for long-time Richardson diffusion.}
\end{figure}

Our results for particle dispersion are given in Fig.1. We present there a
log-log
plot of $\langle r^2(t)\rangle$ (normalized by $\ell_\nu^2)$ versus time $t$
(normalized
by $t_\nu=\ell_\nu^2/\nu$). For backward dispersion we take $t\rightarrow
t'=t_f-t,$
to facilitate comparison with the forward-in-time results.  In the viscous
units that we
employ, the early-time diffusive separation (\ref{rsq-kappa}) becomes $\langle
r^2(t)
\rangle\sim 12t/Pr.$ This regime is clearly seen for both backward and forward
cases
and for both $Pr=1$ and $0.1.$ Furthermore, for the unit Prandtl number cases,
we
see a convincing transition to a $t^3$-law for $t\gtrsim 1.$ This occurs
slightly earlier
for backward dispersion than for forward. Also, we find that the asymptotic
Richardson-Obukhov constant is greater for backward dispersion than for
forward,
in agreement with earlier results \cite{Sawfordetal05}. An average of the local
constants $g(t)\equiv d\langle r^2(t)\rangle/d(t^3)$ in the $t^3$-scaling range
gives
values of $g_0=1.35$ for backward dispersion and $g_0=0.64$ for forward
dispersion.
The latter agrees perfectly with a recent theoretical prediction
\cite{FranzeseCassiani07}
and both are generally consistent with previous values \cite{Sawfordetal08}.
Pure
cube scaling laws with these coefficients are plotted in Fig.1 for comparison
with
the numerically obtained mean dispersions. The agreement is obviously quite
good at
long times, especially for the backward case. Even more importantly, the
$Pr=0.1$
dispersions show a very clear trend to approach the same cubic laws at
sufficiently
large times $t\gg t_\kappa=(\kappa^3/\varepsilon)^{1/4}$ or, in viscous units,
$t\gg (Pr)^{-3/4}.$ Our Fig.1 thus provides strong evidence of the
``forgetting'' of the
molecular diffusion time-scale $t_\kappa$ by turbulent Richardson diffusion at
long
times, which is the essential ingredient of spontaneous stochasticity.

To be completely conclusive, we would need to see collapse of the dispersion
curves for
different $Pr$ in a range where both show $t^3$-scaling. We see no clear
Richardson $t^3$
for the $Pr=0.1$ cases in the time ranges plotted. We cannot continue the
time-integration
further for two reasons. First, the velocity field from the turbulence archive
contains no data
for longer times. Second, the rms dispersion distance $L(t)=\sqrt{\langle
r^2(t)\rangle}$ at
the final time has reached a value $L(t_f)\doteq 1$, just slightly smaller than
the velocity
integral length-scale $L_u=1.376$ for the flow. (This is expected, since $t_f$
is about
one large-eddy turnover time.) To integrate further to see a conclusive
collapse, we
would need a numerical simulation at higher Reynolds numbers and integrated to
longer
times. However, our confirmation of Richardson diffusion at $Re_T=433$ using
stochastic Lagrangian trajectories is comparable to, or even better than, the
results of
Sawford et al. \cite{Sawfordetal08} at $Re_T=650$ using deterministic
Lagrangian
trajectories. (See Fig.4 in that paper). There are two plausible arguments why
this should
be so. In the first place, using stochastic trajectories, all the particles
start at the same point.
At the time $t\approx t_\nu$ when $\langle r^2(t)\approx \ell_\nu^2$ (for
$Pr=1$), the particles
are not randomly placed in the flow, but have already been experiencing
relative advection
by different-sized eddies at the onset of the inertial range. Thus, they begin
to experience
Richardson diffusion at that time. However, using the usual technique of
seeding the flow
with particles at initial separations $r_0\doteq \ell_\nu,$ one would still
need to wait some
additional time for the initial configuration to be ``forgotten''. A second
reason is that backward
dispersion is faster than forward dispersion, so that the range of $t^3$
scaling occurs even
earlier in that case. The technique of stochastic Lagrangian trajectories
appears to be
promising in the numerical study of Richardson diffusion.

%
%
%
%
%
%

In order to make a completely convincing case that we are observing Richardson
diffusion,
we have also numerically calculated the PDF $P(r,t)$ of the
particle-separations. Our
results for $Pr=1$ are presented in Fig.2, with the normalization
$\int_0^\infty dr\,\,r^2P(r,t)=1.$
As has been previously observed
\cite{OttMann00,Biferaleetal05,SalazarCollins09},
Richardson's analytical formula for the long-time PDF of separation distances
implies that
all the PDF's at different times will collapse when scaled with
$L(t)=\sqrt{\langle r^2(t)\rangle}.$
In fact, equation (\ref{Rich-sol}) is equivalent to
$$ L^3(t) P(r,t) =\exp\left[-\alpha
\left(\frac{r}{L(t)}\right)^{2/3}+\beta\right] $$
with numerical values
$$\alpha=(1287/8)^{1/3}\doteq 5.4387$$
and
$$\beta=\ln\left(\frac{3}{35}(143)^{3/2}\sqrt\frac{2}{\pi}\right)\doteq
4.7617.$$
Thus, Richardson's theory makes a parameter-free prediction that a log-linear
plot of
$L^3(t) P(r,t)$ versus $(r/L(t))^{2/3}$ should give a straight line with
slope $-\alpha$ and $y$-intercept $\beta.$ In Fig.2, therefore, we have plotted
our
PDF's in this way, at three times $t=22.37,\,33.57,\,44.79$ all lying in the
range
of $t^3$ scaling. We have also plotted the straight line predicted by
Richardson's
theory. We see that the PDF's scaled in this way collapse very nicely.
Furthermore,
except for some deviation at small $r$ in the backward dispersion case, they
very
closely agree with the predictions of Richardson's theory.

\begin{figure}
\begin{center}
\includegraphics[width=250pt,height=450pt]{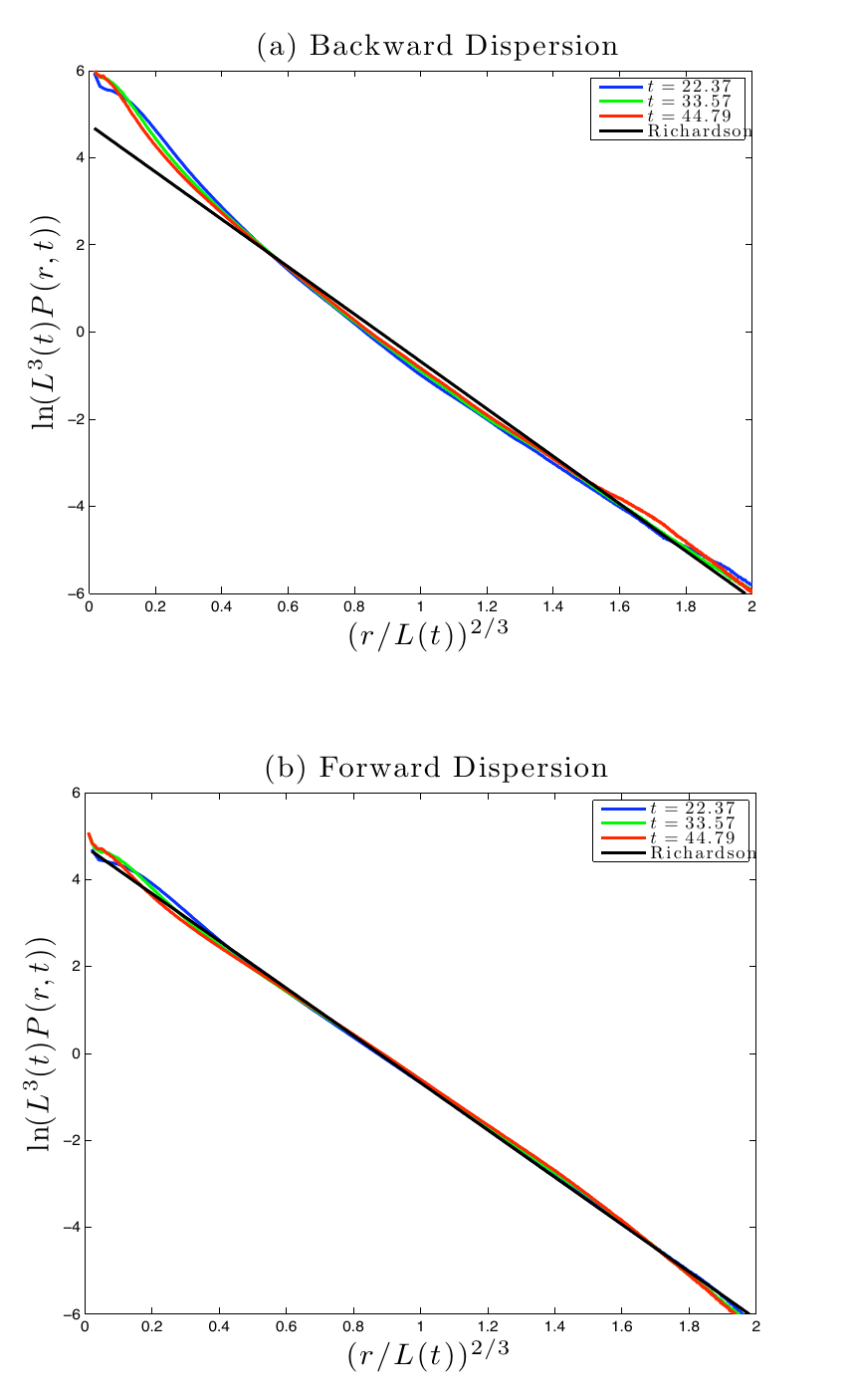}\\
\end{center}
\caption{Probability densities of pair-separation distances:
(a) backward dispersion, (b) forward dispersion. The quantities at different
times are normalized by $L(t)=\sqrt{\langle r^2(t)\rangle}$ as shown. We plot
densities for
three different times in the $t^3$ scaling range, with  \textcolor{blue}{blue}
for $t=22.37$,
\textcolor{green}{green} for $t=33.57$, and \textcolor{red}{red} for $t=44.79$.
In the backward case these times correspond to $t'=t_f-t.$ The solid {\bf
black} line
gives Richardson's analytical prediction for the density (see text).}
\end{figure}

\section{Stochastic Flux-Freezing}

The standard views on flux-freezing in high-conductivity plasmas are
inconsistent with the
phenomenon of spontaneous stochasticity.  It is nearly ubiquitously argued that
flux-freezing
should hold better as magnetic diffusivity $\lambda\rightarrow 0.$ However,
high magnetic
Reynolds numbers are usually associated also with high kinetic Reynolds
numbers.  If kinematic
viscosity $\nu\rightarrow 0$ simultaneously with the resistivity and the plasma
becomes turbulent,
then Lagrangian trajectories will no longer be unique. Which fluid trajectory
shall a magnetic field
line follow if  there are infinitely many such trajectories? This is the
paradox of flux-freezing.

As we shall argue below, a form of flux-freezing {\it does} survive at small
resistivities and
viscosities, but in a novel stochastic sense. Before we make this argument,
however, we shall
first discuss the related subject of flux-freezing properties of resistive
hydromagnetics.

\subsection{Resistive Hydromagnetics}

In this subsection we shall discuss magnetic fields that satisfy the resistive
induction equation
\be \partial_t\bB =\grad\btimes(\bu\btimes\bB -\lambda\grad\btimes\bB),
\lb{induct-res} \ee
with $\lambda=\eta c/4\pi$ the magnetic diffusivity. It is important to stress
that our analysis
applies here applies to a very general velocity field $\bu.$  It may be
incompressible or
compressible. It may be externally prescribed or it may satisfy a dynamical
equation that contains
$\bB$ itself. For example, $\bu$ may be the plasma velocity that obeys the
standard
magnetohydrodynamic momentum equation or it may be taken to be $\bu_e=\bu-
\frac{c}{4\pi e n}\grad\btimes\bB,$ the electron fluid velocity in Hall
magnetohydrodynamics
\footnote{We take this opportunity to correct a typo in our previous paper
\cite{Eyink09} where
it was incorrectly written that  $\bu_e=\bu-\frac{m_e c}{4\pi e
\rho}\grad\btimes\bB,$
with $\rho$ the mass density and $m_e$ the electron mass. Instead the latter
should
have been $m_i$ the ion mass.}. Our only assumption in this section shall be
that $\bu$
is a smooth vector field.

{\it A priori} there is no obvious way how to describe magnetic-line motion for
non-ideal
plasmas. One approach that has been widely employed in discussions of magnetic
reconnection \cite{Boozer90} (or Kuslrud \cite{Kulsrud05}, Section 3.4) is to
introduce
a  ``slip velocity'' $\Delta\bu = \bB\btimes (\lambda\grad\btimes\bB)/B^2$. In
that case, one may
attempt to introduce an ``effective velocity'' $\bu_*=\bu+\Delta\bu$ of the
field lines. Unfortunately,
this approach is not generally successful because $(\Delta
\bu)\btimes\bB=-\lambda\grad\btimes\bB$
if and only if $\bB\bdot{\bf E}=0$ (or $\bB\bdot\bJ=0$ for Ohmic non-ideality).
As has been emphasized
\cite{Priestetal03,HornigPriest03}, no effective-velocity approach is
satisfactory for discussions
of three-dimensional magnetic reconnection. In fact, those authors show that,
even if the non-ideality
is spatially localized, there generally exists {\it no} smooth velocity field
$\bu_*$ whatsoever such that  $\partial_t\bB=\grad\btimes(\bu_*\btimes\bB)$ 
for a non-ideal plasma.

For magnetic fields that obey (\ref{induct-res}), however, there is a natural
and consistent way
to describe line-motion as a process of {\it stochastic advection}. Such
approaches have already
been employed for  some time in discussion of kinematic magnetic dynamos, at
least
for incompressible velocity fields \cite{Molchanovetal84,DrummondHorgan86}.
Recently,
we gave a rigorous proof of stochastic flux-conservation properties for
nonlinear
hydromagnetic models using mathematical methods of stochastic analysis
\cite{Eyink09}.
We shall present here a more physical demonstration of these results using
path-integral methods
which, also, extends their validity to compressible fluid models.

To begin, we note that the induction equation (\ref{induct-res}) may be
rewritten as
\be \partial_t\bB + (\bu\bdot\grad)\bB = (\bB\bdot\grad)\bu
      -\bB(\grad\bdot\bu)+\lambda\triangle\bB. \lb{magnetic-eq} \ee
In this form it is the same as the scalar advection equation (\ref{scalar-eq}),
except for
the additional two terms on the righthand side. The path-integral formula
(\ref{theta-FK})
for the scalar solution may thus be easily adapted to this situation.  The
solution of
(\ref{magnetic-eq}) with initial condition $\bB(t_0)=\bB_0$ is given by the
``sum-over-histories''
formula
\begin{eqnarray}
\bB(\bx,t) &=&
 \int_{\ba(t)=\bx} \mathcal{D}\ba \,\, \bB_0(\ba(t_0))\bdot \boJ(\ba,t) \cr
&& \,\,\,\,\times
\exp\left(-\frac{1}{4\lambda}\int_{t_0}^t d\tau\,
|\dot{\ba}(\tau)-\bu^\nu(\ba(\tau),\tau)|^2\right)
\,\,\,\,\,\,\,\,\,\,\,\,\,\,
\lb{B-FK} \end{eqnarray}
where $\boJ(\ba,\tau)$ where is a $3\times 3$ matrix and $\bB$ is interpreted
as a
3-dimensional row vector. $\boJ$ satisfies the following ODE along the
trajectory $\ba(\tau)$:
\begin{eqnarray}
\frac{d}{d\tau}\boJ(\ba,\tau)&=&\boJ(\ba,\tau)\grad_x\bu(\ba(\tau),\tau) \cr
   && \,\,\,\,\,\,\,\,\,\, -\boJ(\ba,\tau)(\grad_x\bdot\bu)(\ba(\tau),\tau),
\lb{J-eq}
\end{eqnarray}
with initial condition $\boJ(\ba,t_0)={\bf I}.$ It is easy to check by taking
the time-derivative
of (\ref{B-FK}) and using (\ref{J-eq}) to show that the induction
eq.(\ref{magnetic-eq}) is
satisfied. Just as for the scalar problem, the condition $\ba(t)=\bx$ on the
path-integral
trajectories implies that they correspond to solutions of the stochastic
equation
\be \frac{d}{d\tau}\wt{\ba}(\tau)
=\bu(\wt{\ba},\tau)+\sqrt{2\lambda}\,\wt{\boeta}(\tau),
 \,\,\,\, \wt{\ba}(t)=\bx \lb{a-stoch-eq} \ee
integrated backward in time from $\tau=t$ to $\tau=t_0.$

However, the stochastic equation (\ref{a-stoch-eq}) may also
be integrated forward in time from $\tau=t_0$
to $\tau=t$. In that case, the same ensemble of trajectories may be obtained by
considering
only those particles with initial locations carefully selected to arrive at
$\bx$ at time $t$,
for a given realization of the white-noise $\wt{\boeta}(t).$ With a slight
change of notation,
we may characterize this ensemble of time-histories as those $\wt{\bx}(\tau)$
which solve
\be \left\{\begin{array}{l}
\frac{d}{d\tau}\wt{\bx}(\ba,\tau)
=\bu(\wt{\bx}(\ba,\tau),\tau)+\sqrt{2\lambda}\,\wt{\boeta}(\tau),
  \,\,\tau>t_0 \cr
 \,\,\,\,\,\,\,\wt{\bx}(\ba,t_0)=\ba \cr
\end{array}\right.  \lb{stoch-lag} \ee
such that the inverse map $\wt{\ba}(\bx,\tau)$ to $\wt{\bx}(\ba,\tau)$
specifies the starting
point by $\ba=\wt{\ba}(\bx,t).$ Notice that (\ref{stoch-lag}) is a stochastic
generalization of the
usual equation for a Lagrangian flow map $\wt{\bx}(\ba,t)$ of a particle with
initial ``label''
$\ba$ and that  $\wt{\ba}(\bx,\tau)$ is the ``back-to-labels'' map. It is easy
to show furthermore
by applying $\grad_a$ to (\ref{stoch-lag}) that
\be \wt{\boJ}(\ba,t)\equiv \frac{1}{{\rm det}\,(\grad_a\wt{\bx}(\ba,t))}
     \grad_a\wt{\bx}(\ba,t). \lb{J-sol} \ee
solves equation (\ref{J-eq}) with initial condition $\wt{\boJ}(\ba,t_0)={\bf
I}.$ It is
therefore possible to re-express the path-integral formula (\ref{B-FK}) as
\be \bB(\bx,t) =\overline{\,\,\,\frac{1}{{\rm det}\,(\grad_a\wt{\bx}(\ba,t))}
     \bB_0(\ba)\bdot\grad_a\wt{\bx}(\ba,t)|_{\wt{\ba}(\bx,t)}\,\,\,}.
\lb{stoch-Lundquist} \ee
The overbar $\overline{\,\,\,\cdot\,\,\,}$ represents the average over
realizations of the
random white noise process $\wt{\boeta}(t)$ in (\ref{stoch-lag}).

 We shall call the above result the {\it stochastic Lundquist formula}, since
it is the
stochastic generalization of the standard Lundquist formula \cite{Lundquist51}
(or Kulsrud \cite{Kulsrud05}, section 4.8). It may be cast into a more familiar
form by noting
that the determinant that appears there can be interpreted as the ratio of
initial and
final mass densities
\footnote{This interpretation requires some caution. The quantity
$$\tilde{\rho}(\bx,t)\equiv \left.
     \frac{\rho_0(\ba)}{{\rm
det}\,(\grad_a\tilde{\bx}(\ba,t))}\right|_{\tilde{\ba}(\bx,t)} $$
satisfies the stochastic (Stratonovich) equation
$$\partial_t\tilde{\rho}(\bx,t)+\grad\bdot\big[ \big(\bu(\bx,t)
+\sqrt{2\lambda}\tilde{\boeta}(t)\big)\circ\tilde{\rho}(\bx,t)\big]=0$$
with initial condition
$\tilde{\rho}(\bx,t_0)=\rho_0(\bx).$
It thus represents the mass density in
the random flow with noise realization
$\tilde{\eta}(t).$
However, the noise-average
$\bar{\rho}(\bx,t)\equiv \overline{\,\tilde{\rho}(\bx,t)\,}$
is {\it not} the physical mass density!
In fact, by converting the previous stochastic equation to Ito form and taking
the average,
it follows that
$$\partial_t\bar{\rho}(\bx,t)+\grad\bdot(\bu(\bx,t)\bar{\rho}(\bx,t))=
      \lambda\triangle\bar{\rho}(\bx,t), $$
which is not the correct continuity equation.}:
$$ {{\rm det}\,(\grad_a\wt{\bx}(\ba,t))} =
\frac{\rho_0(\ba)}{\wt{\rho}(\wt{\bx}(\ba,t),t)}$$
It follows that the vector field $\wt{\bB}/\wt{\rho}$ is stochastically
``frozen-in'' and advected
along stochastic Lagrangian trajectories, where $\wt{\bB}$ is defined to be the
quantity
under the overbar in (\ref{stoch-Lundquist}). Notice, therefore, that the
average in
(\ref{stoch-Lundquist}) is not over the frozen-in field $\wt{\bB}/\wt{\rho},$
but rather over the
magnetic field $\wt{\bB}$ itself. This is necessary in order to reproduce the
Laplacian term
in (\ref{magnetic-eq}), which has the form $\lambda\triangle\bB$ and not
$\lambda\triangle(\bB/\rho).$

\begin{figure}
\begin{center}
\includegraphics[width=160pt,height=360pt]{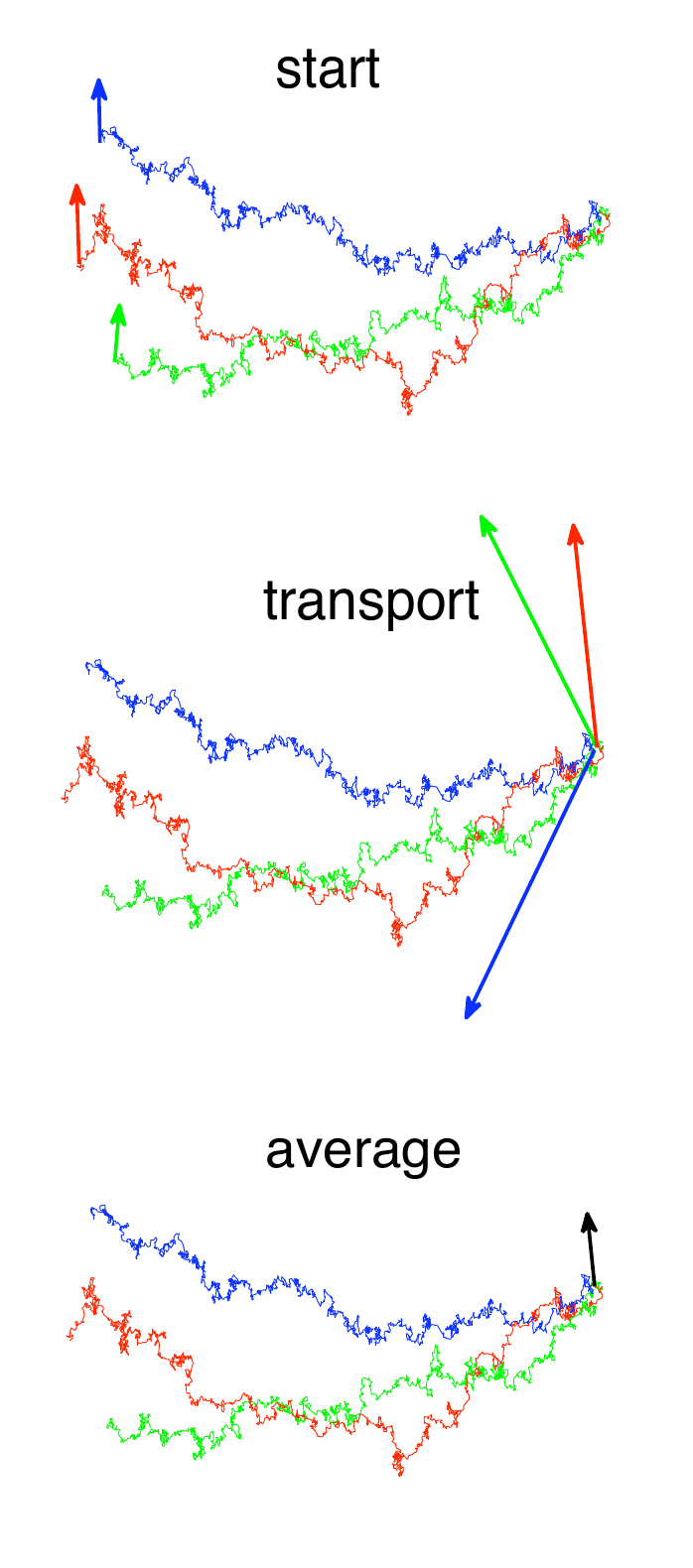}\\
\end{center}
\caption{Illustration of the stochastic Lundquist formula. Three stochastic
Lagrangian trajectories running backward in time from a common
point are shown in \textcolor{red}{red},  \textcolor{green}{green},
\textcolor{blue}{blue}.
Starting field vectors, represented by correspondingly colored arrows, are
transported
along the trajectories, stretched and rotated, to the common final point. These
are then
averaged to give the resultant magnetic field at that point, indicated by the
{\bf black} arrow.}
\end{figure}

The use of the stochastic Lunquist formula is illustrated in Fig.~3. The aim is
to calculate
the magnetic field $\bB$ at spacetime point $(\bx,t).$ The first step is to
generate
an ensemble of stochastic Lagrangian trajectories solving (\ref{a-stoch-eq})
backward
in time from $\bx$ at time $t$ to random locations $\wt{\ba}(t_0)$ at the
initial time $t_0.$
The path-integral formula (\ref{B-FK}) sums over all such random
time-histories.
We show in  Fig.~3 (top) three stochastic trajectories generated numerically
from the
turbulence database together with the starting magnetic field vectors $\bB_0,$
indicated by arrows at the starting locations $\wt{\ba}(t_0).$ The next step is
to transport
each of the field vectors in the usual ``frozen-in'' fashion along the
stochastic
Lagrangian trajectories to the final spacetime point $(\bx,t).$ The result is
an ensemble
of field vectors $\wt{\bB}$ at that point, stretched and rotated by the flow.
These are
illustrated in Fig.~3 (middle) by the collection of three arrows at $(\bx,t),$
obtained
by transporting the three initial vectors. In the usual deterministic Lundquist
formula
there would be just one trajectory and one vector $\wt{\bB}$ at the final
point, which
would give the desired magnetic field. Now however as the final step one must
average
over the ensemble of random vectors $\wt{\bB}$ in order to obtain the resultant
magnetic
field $\bB(\bx,t).$ This is illustrated by the black arrow in Fig.~3 (bottom).
In contrast to
the previous transport step which preserved line-topology (in each individual
realization),
the final averaging step resistively ``glues'' the transported lines together
and changes
the magnetic field-line topology.

%
%

There is an elegant reformulation of the stochastic Lundquist formula which
must
be mentioned here, both because of its conceptual simplicity and also because
of its
potentially greater generality (see next subsection).  Consider any smooth,
oriented surface
$S$ at final time $t.$ Then the formula (\ref{stoch-Lundquist}) may integrated
in $\bx$
over the surface $S,$ with respect to the vector area element
$d\bA(\bx)=d\bx\btimes d\bx,$
and the ensemble-average and surface-integration interchanged
on the righthand side. Because the expression under the overbar is the one that
appears
in the usual Lundqust formula, the standard multi-variable calculus
manipulations
convert this into a surface integral over $\wt{\ba}(S,t),$ the surface $S$
randomly advected
backward in time to the initial time $t_0.$ As before
$\wt{\ba}(\cdot,t)=\wt{\bx}^{-1}(\cdot,t)$
is the ``back-to-label map'' for the stochastic forward flow.  The result is
the following
{\it stochastic Alfv\'{e}n theorem}
\be \int_S \bB(\bx,t)\bdot d\bA(\bx)=
     \overline{\,\,\,\int_{\wt{\ba}(S,t)} \bB_0(\ba)\bdot d\bA(\ba)\,\,\,},
\,\,\,\, t>t_0.
     \lb{stoch-Alfven} \ee
This result generalizes  a previous theorem \cite{Eyink09} to compressible
plasmas.
Eq.(\ref{stoch-Alfven}) expresses the conservation of magnetic flux on average,
as illustrated
in Fig.~4. An initial loop $C,$ boundary of the surface $S,$ is shown there in
black. This is
stochastically advected backward in time to give an infinite ensemble of loops
at the initial
time $t_0.$ These are represented by the three colored loops. The
ensemble-average of
the magnetic flux through the collection of loops at the initial time $t_0$ is
equal to the magnetic
flux through the loop $C$ at the final time $t.$

 The stochastic Alfv\'en theorem is an example of what is called a ``martingale
property''
 in probability theory. The magnetic flux through each advected loop at the
earlier time
 $t_0$ is unequal to the magnetic flux through $C$ at time $t.$ Nevertheless,
the mean
 flux remains the same. Note that this result implies an irreversibility or an
``arrow of time''
 since it only holds for backward stochastic advection of loops.
Backward-in-time is the
 causal direction, since the magnetic flux at the present must be obtained as
an
 average of past values and not of future values. If we assumed a  ``forward
 martingale''  property then we would obtain instead the magnetic induction
 equation (\ref{magnetic-eq}) with a {\it negative} resistivity term
$-\lambda\triangle\bB.$
 Note, in fact, that the stochastic Alfv\'en theorem (backward in time) is
mathematically
 equivalent to the usual resistive induction equation (\ref{induct-res}) or
(\ref{magnetic-eq})
 \cite{Eyink09}.

\begin{figure}
\begin{center}
\includegraphics[width=230pt,height=150pt]{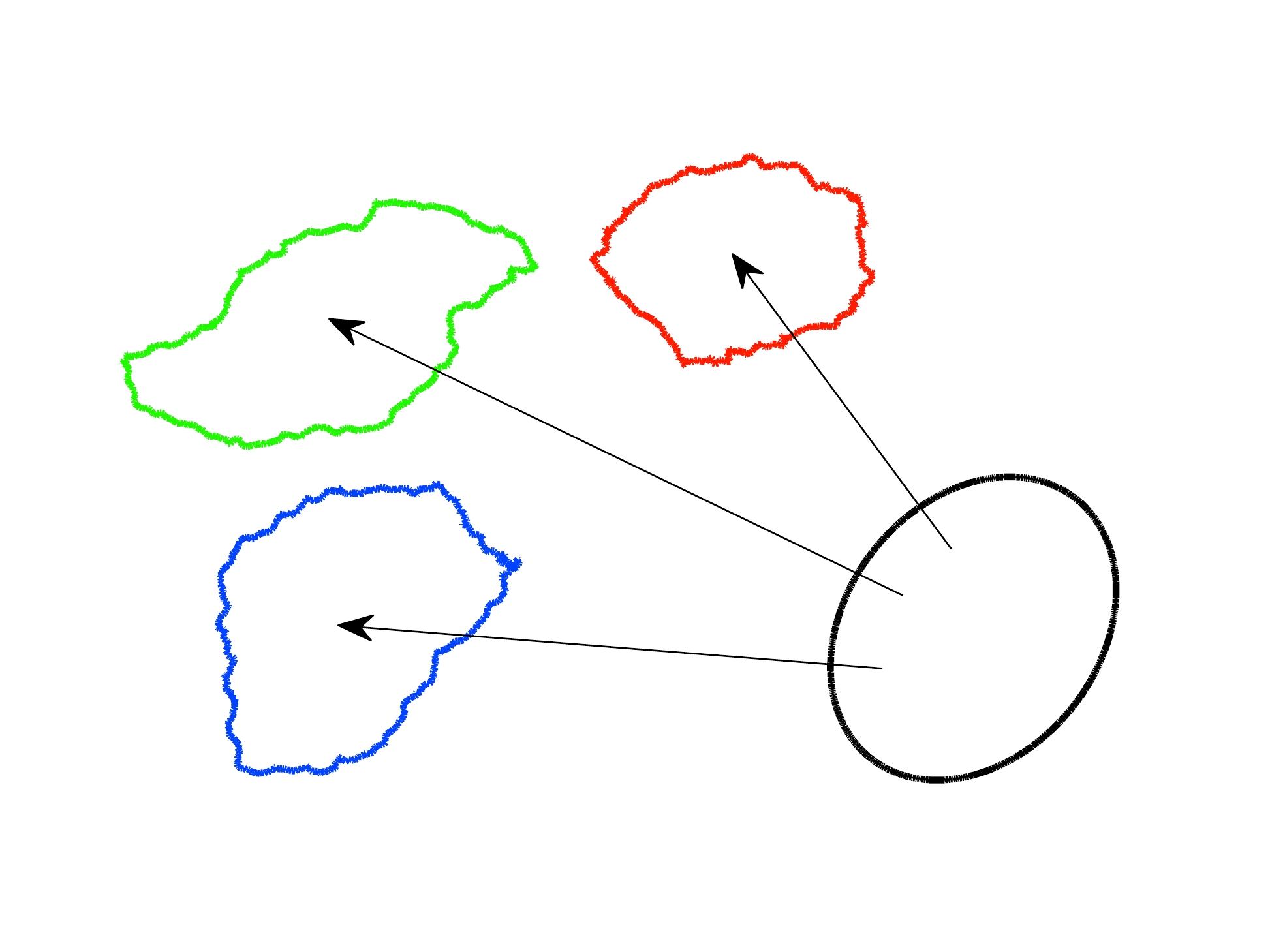}\\
\end{center}
\caption{Illustration of the stochastic Alfv\'en theorem. Shown are
three members (\textcolor{red}{red},  \textcolor{green}{green},
\textcolor{blue}{blue})
of  the infinite ensemble of loops obtained by stochastic advection of a loop
$C$ ({\bf black})
at time $t$ backward in time to $t_0.$ The average of the magnetic flux through
the
ensemble of loops is equal to the magnetic flux through $C.$}
\end{figure}

%
%
%

\subsection{High-Reynolds-Number Limit}

We now consider the limit of large kinematic and magnetic Reynolds numbers. For
simplicity
we shall assume that $Pr_m=\nu/\lambda$ remains fixed as
$\nu,\lambda\rightarrow 0.$

Consider the Feynman-Kac formula (\ref{B-FK}). By a naive application of the
Laplace
method,  one would assume that the path-integral collapses to a single
deterministic
trajectory as $\lambda\rightarrow 0,$ with rms fluctuations of order $(\lambda
t)^{1/2}$
for small but nonzero $\lambda.$ This is precisely the heuristic estimate of
line-slippage
made by Kulsrud \cite{Kulsrud05} which was quoted in the Introduction. This
estimate
is rigorously correct if the velocity and magnetic fields are assumed to remain
smooth
in the limit $\nu,\lambda\rightarrow 0.$ Thus, the heuristic estimate is
correct if the
plasma flow remains laminar, but this will be the exception rather than the
rule at high
Reynolds numbers. In a turbulent flow, the behavior will be quite different. As
we can
see from our Fig.~1 for incompressible hydrodynamic turbulence, the heuristic
estimate
is only valid for very short times smaller than the resistive time $t_\lambda=
(\lambda^3/\varepsilon)^{1/4}.$
At longer times,  the rms slip distance of the field lines instead follows the
Richardson
law $\sim (\varepsilon t^3)^{1/2},$ independent of $\nu$ and $\lambda.$ The
quantitative behavior will be different in plasmas with strong magnetic fields,
due to the effects of the Lorentz force, as discussed more in section V.
However, the
qualitative behavior must be the same
whenever the advecting velocity is turbulent and spatially rough.

The Feynman-Kac formula (\ref{B-FK}) is not very well-suited to analyzing
the limit of high Reynolds numbers, however, because the velocity-gradients
that appear in the definition of the matrix $\boJ$ diverge in that limit.
Likewise,
the gradients of the Lagrangian flow map that appear in the definition
(\ref{J-sol})
of $\wt{\boJ}$ are expected to diverge. The integrated form of
flux-conservation,
the stochastic Alfv\'{e}n theorem (\ref{stoch-Alfven}), is more likely to
remain meaningful
in the limit of infinite Reynolds number. The backward-advected loops
$\wt{\ba}^\lambda(C,t)$
at finite values of $\lambda,\nu$ are expected to approach well-defined curves
$\wt{\ba}(C,t)$ as $\lambda\rightarrow 0,$ which, however, are not rectifiable
but
fractal \cite{Mandelbrot76,SreenivasanMeneveau86,FungVassilicos91,
VillermauxGagne94,Kivotides03,NicolleauElmaihy04}. To make mathematical sense
of magnetic flux through such fractal loops, we may introduce the vector
potential
$\bA_0=({\rm curl})^{-1}\bB_0$ and rewrite the flux through surface
$\wt{\ba}(S,t)$
as a line-integral around its perimeter $\wt{\ba}(C,t).$ We may then further
transform
by change of variables to a line-integral around the original loop $C,$ as:
$$ \oint_{\wt{\ba}(C,t)} \bA_0(\ba)\bdot d\ba =
   \oint_C \bA_0(\wt{\ba}(\bx,t))\bdot d\wt{\ba}(\bx,t). $$
The integral on the right may be interpreted as a generalized Stieltjes
integral,
which is well-defined as long as the map $\wt{\ba}(\bx,t)$ is
suitably H\"{o}lder continuous \cite{Young36,Young38}.

It may seem from our arguments to this point that the validity  at very high
Reynolds numbers
of the stochastic flux-freezing result (\ref{stoch-Alfven}) is dependent upon
the particular
stochastic representation of resistive effects  employed in
eqs.(\ref{stoch-lag}) and
(\ref{stoch-Lundquist}). However, the same ``martingale property'' can be
obtained in
the limit $\nu,\lambda\rightarrow 0$ by a different argument that employs only
the standard
Lagrangian flow \cite{Eyink07}. Define the deterministic flow, as usual, by
\be \left\{\begin{array}{l}
 \frac{d}{d\tau}\bx^\nu(\ba,\tau)=\bu^\nu(\bx^\nu(\ba,\tau),\tau),
  \,\,\,\,\,\,\,\,\,\,\,\tau>t_0 \cr
 \,\,\,\,\,\,\,\bx^\nu(\ba,t_0)=\ba \cr
\end{array}\right.  \lb{det-lag} \ee
where the superscript $\nu$ is a reminder that the dynamical equation for the
advecting
velocity field contains a certain viscosity $\nu=(Pr)\lambda.$ Correspondingly
one defines
the inverse map $\ba^\nu(\cdot,t)=(\bx^\nu)^{-1}(\cdot,t).$ Stochasticity can
be introduced
by assuming small random perturbations of the loop, taking $C\rightarrow
C+\epsilon\,\wt{C},$
where $\wt{C}$ is a random loop from a well-behaved ensemble \footnote{Note
that
the addition of loops is pointwise. That is, if two loops $C,C'$ are given
parametrically
by periodic functions $C,C': \,s\in [0,1]\mapsto  C(s),C'(s)\in {\mathbb R}^3,$
then the sum-loop
$C+C'$ is parameterized by $(C+C')(s)=C(s)+C'(s).$}. Thus, $\epsilon$ can be
regarded as
the spatial-resolution in determining the precise form of the loop $C.$ We may
then
argue that, at least for incompressible flow, the ensemble of loops
$\wt{\ba}^\lambda(C,t)$
obtained from stochastic advection in the limit $\lambda\rightarrow 0$
coincides with the
ensemble of loops obtained from deterministic advection
$\ba^\nu(C+\epsilon\,\wt{C},t)$
taking the limits first $\nu,\lambda\rightarrow 0$ and then
$\epsilon\rightarrow 0.$ As
discussed in the previous section, this is rigorously known to be true for
point-particles
advected by velocities selected from the Kraichnan white-in-time ensemble
\cite{EvandenEijnden00,LeJanRaimond04}. The physical mechanism is just
turbulent Richardson diffusion. We therefore conjecture that the same result
holds
for loops. If this is so, then the double limit
$$ \lim_{\epsilon\rightarrow 0}\lim_{\nu,\lambda\rightarrow 0}
     \oint_{\ba^\nu(C+\epsilon\,\wt{C},t)} \bA_0(\ba)\bdot d\ba $$
gives precisely the same ensemble of fluxes with the same distribution as in
the previous
approach. In that case, (\ref{stoch-Alfven}) must again hold in the limit, or,
more precisely,
\be \oint_C \bA(\bx,t)\bdot d\bx=  \lim_{\epsilon\rightarrow
0}\lim_{\nu,\lambda\rightarrow 0}
    \overline{ \oint_{\ba^\nu(C+\epsilon\,\wt{C},t)} \bA_0(\ba)\bdot d\ba},
\lb{alfven-resist} \ee
where $\bA=\lim_{\nu,\lambda\rightarrow 0} \bA^{\lambda}$ and the overbar now
indicates
average over the ensemble of loop perturbations $\widetilde{C}.$  This is a
nontrivial
result because for an individual loop $C$
$$ \frac{d}{dt}\oint_{\bx^\nu(C,t)} \bA^\lambda(\bx,t)\bdot d\bx=
      -\lambda \oint_{\bx^\nu(C,t)} (\grad\btimes\bB^\lambda)(\bx,t)\bdot d\bx.
$$
Since $\grad\btimes\bB^\lambda$ diverges in the limit $\nu,\lambda\rightarrow
0,$
there is no reason to expect that the righthand size vanishes in that limit.
This is the
standard argument how flux-freezing can be violated in thin current-sheets. It
stands to
reason that magnetic flux through an individual Lagrangian loop will fluctuate
in time and
not be conserved.  Nevertheless, our arguments lead us to conclude that
flux-freezing
in turbulent flow is still preserved in the mean sense (\ref{alfven-resist}) at
infinite
Reynolds number.

A scalar resistivity of the form in (\ref{induct-res}) or (\ref{magnetic-eq}),
in fact, plays
no essential role in our arguments. Any microscopic plasma mechanism of ``line
slippage''
will be accelerated by turbulent advection as soon as the lines have separated
by a distance
of order $\ell_\nu,$ the viscous length. More realistic mechanisms of
line-slippage such
as anisotropic resistivity in the Braginski equations  \cite{Braginski65} (also
Kulsrud
\cite{Kulsrud05}, Ch.8) or the Hall effect often invoked in theories of fast
reconnection
\cite{ZweibelYamada09} may all serve the same role. After a very short time the
microscopic
plasma mechanism of line-slippage, whatever it may be, will be ``forgotten''
and replaced
by turbulent Richardson diffusion.

\section{Turbulent Magnetic Dynamo}

The stochasticity of flux-freezing plays an essential role in the operation of
the turbulent magnetic
dynamo. We have already made a detailed analysis of this in the
Kazantsev-Kraichnan model
of fluctuation kinematic dynamo for a non-helical,  incompressible velocity
field, with
$Re_m=\infty$ and $Pr_m=0$  \cite{EyinkNeto10,Eyink10}. Here we shall present a
more general
study. We first discuss how our path-integral approach relates to the standard
Lagrangian formulations
of magnetic dynamo
\cite{Moffatt74,Kraichnan76a,Kraichnan76b,Molchanovetal84,DrummondHorgan86,
VainshteinKichatinov86,Kichatinov89}, in a framework that encompasses helical
and/or compressible
flows and turbulent velocity fields with realistic time-correlations. We focus
on the kinematic dynamo
here but much of our discussion carries over also to the nonlinear dynamo
(which is considered more
specifically in section V). We then present numerical results for a particular
case, the kinematic
fluctuation dynamo in a non-helical,  incompressible turbulent velocity field
for $Pr_m=1,$ using the
same hydrodynamic turbulence database that was employed in our study of
Richardson diffusion
(section II.C). We compare our results with earlier numerical studies
\cite{Haugenetal04,
Schekochihinetal04,Choetal09,ChoRyu09} at lower Reynolds number and, also, with
our previous analytical
results in the Kazantsev-Kraichnan model at $Pr_m=0$ \cite{Eyink10}. As we
shall see, spontaneous
stochasticity and Richardson diffusion play a very similar role in the
fluctuation dynamo
for both $Pr_m=0$ and $Pr_m=1.$

\subsection{Lagrangian Description of Dynamo}

The Feynman-Kac formula (\ref{B-FK}) for the magnetic field may be rewritten as
\be B^i(\bx,t)=\int d^3a \, B_0^k(\ba) \hat{F}^i_k(\ba,t_0|\bx,t;\bu)
\lb{B-F-eq} \ee
with the definition
\begin{eqnarray}
&& \hat{F}^i_k(\ba,t_0|\bx,t;\bu) = \int_{\boalpha(t)=\bx} \mathcal{D}\boalpha
\,\,
\delta^3(\boalpha(t_0)-\ba) \mathcal{J}^i_k(\boalpha,t) \cr
&& \,\,\,\,\,\,\,\,\,\,\,\,\,\times
 \exp\left(-\frac{1}{4\lambda}\int_{t_0}^t d\tau\, |\dot{\boalpha}(\tau)
 -\bu^\nu(\boalpha(\tau),\tau)|^2\right).
 \lb{Fhat-def} \end{eqnarray}
This latter quantity is a generalization to a compressible flow of the
(Eulerian) magnetic Green's
function considered by Lerche \cite{Lerche73} and Kraichnan
\cite{Kraichnan76a}, expressed as
a Lagrangian path-integral. It completely encodes all the effects of the
advecting flow.

A description of the {\it mean-field dynamo} is obtained if one averages over
the ensemble
of velocity fields and the random initial conditions of the magnetic field.
Assuming that
these are statistically independent (which requires that the effects of the
Lorenz force
be negligible),
\be \langle B^i(\bx,t)\rangle=\int d^3a\,
     \langle B_0^k(\ba)\rangle  F^i_k(\ba,t_0|\bx,t) \lb{MF-dyn} \ee
with $F^i_k(\ba,t_0|\bx,t)\equiv \langle \hat{F}^i_k(\ba,t_0|\bx,t)\rangle$
the mean magnetic
Green's function.  The same result holds without the assumption of kinematic
dynamo if
the initial magnetic field is non-random and the mean Green's function is
defined by a conditional
average for fixed $\bB_0$  \cite{Kichatinov89}. Of course, in that case the
mean Green's function
becomes dependent upon the magnetic field. The mean Green's function involves
Taylor
1-particle diffusion, or absolute diffusion with respect to a starting point
$\bx$ at time $t,$
with the stochastic fluid particle moving backward to time $t_0.$ In the case
of homogeneous
velocity statistics,
$$ F^i_k(\brho;t_0,t)\equiv F^i_k(\brho,t_0|\bzed,t) $$
becomes a function of the single variable $\brho=\ba-\bx.$ The large-$\rho$
behavior
of the mean magnetic Green's function is well-known if the velocity statistics
are also
isotropic (but  reflection non-symmetic), in which case the usual $\alpha$ and
$\beta$
effects of mean-field electrodynamics determine the large-distance decay (see
Kraichnan
\cite{Kraichnan76a}, eq.(3.17)). In particular,  the mean Green's function is
non-negligible
only for $\rho=O\left((\beta |t-t_0|)^{1/2}\right),$ with $\beta$ the
eddy-diffusivity of the
mean magnetic field. If the magnetic field statistics are also homogeneous,
then
the mean field  $\langle{\bB}(\bx,t)\rangle=\langle \bB_0\rangle$ becomes
time-independent
(no mean-field dynamo), in which case (\ref{MF-dyn}) yields the sum-rule
\be \int d^3\rho \,F^i_k(\brho;t_0,t)=\delta^i_k. \lb{MF-sumrule} \ee
The exact Lagrangian formulas for the mean-field electrodynamics
coefficients, $\alpha,\beta,$ etc., that were derived by Moffatt
\cite{Moffatt74} and Kraichnan
\cite{Kraichnan76a,Kraichnan76b} at infinite conductivity, hold also for
positive resistivity
within the present stochastic framework \cite{Molchanovetal84,DrummondHorgan86,
Kichatinov89}.  These formulas involve the stochastic displacement field
$\wt{\boxi}(\bx,t)=\bx-\wt{\ba}(\bx,t)$ of 1-particle turbulent diffusion
\footnote{The important
papers of Vainshtein and Kichatinov \cite{VainshteinKichatinov86,Kichatinov89}
employ a representation in terms of 2-particle turbulent diffusion, but this is
a
technical device to avoid the direct appearance of space-gradients. We might
note
also that these these authors believed that for $\eta=0$ the transition
probability
$p_2(\bx,{\,\!}^1\bx|\bz,{\,\!}^1\bz,t)$ of two fluid particles should be
``going to zero when
$\bz={\,\!}^1\bz$ and $\bx\neq {\,\!}^1\bx$''. They thus miss the phenomenon of
spontaneous stochasticity. They furthermore wrote on turbulent dynamo at high
Reynolds numbers that: ``For the magnetic field, there also exists a region of
scales
in which $R_m\gg 1,$ i.e. the frozen-in condition is fulfilled''. Their paper
thus provides
another example in which flux-freezing is assumed to hold, incorrectly,
for Reynolds numbers $Re,Re_m\gg 1.$}.

As a side remark, we note that the results on the mean Green's function in
homogeneous,
isotropic turbulence which we reviewed above may have limited relevance to the
description
of astrophysical dynamos  and laboratory dynamo experiments. The separation of
scales
required for the applicability of mean-field electrodynamics often does not
occur in
practice and large-scale magnetic fields might not be understood without
reference
to object-specific features, global flow geometry and boundary-conditions. In
fact,
as a general rule, mean-field dynamo effect requires not just turbulent
diffusion
of field-lines but also a globally organized motion of magnetic fields. To show
this,
we present an argument based on Faraday's law for the mean field
$$ \partial_t\langle \bB\rangle+\grad\bdot\bSigma=0, $$
rewritten as a local conservation law for the magnetic field vector. Here
$$ \Sigma^{ij}=\langle u^i B^j -u^j B^i\rangle
    +\lambda\left(\frac{\partial \langle B^i\rangle}{\partial x_j}
    -\frac{\partial \langle B^j\rangle}{\partial x_i}\right) $$
represents the spatial flux of the $j$th component of mean magnetic field in
the $i$th
coordinate direction. The space-average of $\langle \bB(\bx,t)\rangle$ over a
volume
$V$ can only change in time by a transport of magnetic field lines through its
surface $\partial V$.
Furthermore, a simple calculation with Ampere's law $\langle
\bJ\rangle=\frac{1}{4\pi}\grad\btimes
\langle\bB\rangle$ gives
$$ \frac{d}{dt} \int \frac{1}{8\pi} |\langle \bB(\bx,t)\rangle|^2 \,d^3x =
     \int \frac{1}{2} \epsilon_{ijk}\langle J^i\rangle \Sigma^{jk} \, d^3x, $$
where the integral is over all of space. We see that  energy in the mean-field
$\langle\bB\rangle$
grows when the transport of magnetic flux across the closed lines of mean
electric current
reinforces the Amperian fields induced by those currents (using the righthand
rule).  We
thus see that mean-field dynamo action requires coherent motion of magnetic
field
lines, coordinated over large spatial scales. Note that very similar ideas are
widely used in
condensed matter physics to explain, for example, the decay of magnetic flux
through a
superconducting ring by ``phase-slippage'' of quantized field lines
\cite{Huggins70,Huggins94,Eyink08}.

Returning to the discussion of small-scale turbulence, we note that a formula
for
the magnetic correlation function analogous to (\ref{MF-dyn}) for the mean
field
can be derived, under precisely the same assumptions:
\begin{eqnarray}
\langle B^i(\bx,t)B^j(\bx',t) \rangle
& = & \int d^3a\int d^3a' \, \langle B_0^k(\ba)B_0^\ell(\ba')\rangle \cr
 && \,\,\,\,\,\,\,\,  \times F^{ij}_{k\ell}(\ba,\ba',t_0|\bx,\bx',t)
\lb{Bcorr} \end{eqnarray}
with\footnote{This quantity was previously \cite{EyinkNeto10}
denoted $\bar{F}^{ij}_{k\ell}(\ba,\ba',t_0|\bx,\bx',t).$}
\be F^{ij}_{k\ell}(\ba,\ba',t_0|\bx,\bx',t) =
     \langle \hat{F}^i_k(\ba,t_0|\bx,t) \hat{F}^j_\ell(\ba',t_0|\bx,t) \rangle.
\lb{2-GF} \ee
The behavior of this ``two-body'' Green's function is determined by the
properties of
turbulent 2-particle (Richardson) diffusion effects. E.g. setting  $\bx=\bx'$
in the above formula leads to an expression for mean magnetic energy density
$\langle B^2(\bx,t)\rangle,$ which is related to pairs of stochastic Lagrangian
trajectories
(with independent Brownian motions) that start at points $\ba$ and $\ba'$ at
time $t_0$ and
both end at $\bx$ at time $t.$

Combining (\ref{MF-dyn}) and (\ref{Bcorr}) gives a formula
for the correlation of the magnetic fluctuations:
\begin{eqnarray}
&&  \langle B^i(\bx,t)B^j(\bx',t)\rangle-\langle B^i(\bx,t)\rangle
     \langle B^j(\bx',t)\rangle \cr \cr
&& = \int d^3a\int d^3a' \, \left[\langle B_0^k(\ba)B_0^\ell(\ba')\rangle
    - \langle B_0^k(\ba)\rangle \langle B_0^\ell(\ba')\rangle \right] \cr
&&   \,\,\,\,\,\,\, \,\,\,\,\,\,\, \,\,\,\,\,\,\, \,\,\,\,\,\,\, \,\,\,\,\,\,\,
 \,\,\,\,\,\,\,
    \times   F^{ij}_{k\ell}(\ba,\ba',t_0|\bx,\bx',t) \cr \cr
&&  + \int d^3a\int d^3a' \, \langle B_0^k(\ba)\rangle
         \langle B_0^\ell(\ba')\rangle \cr
&& \times \left[F^{ij}_{k\ell}(\ba,\ba',t_0|\bx,\bx',t) -
      F^i_k(\ba,t_0|\bx,t) F^j_\ell(\ba',t_0|\bx,t)\right].
\,\,\,\,\,\,\,\,\,\,\,
\lb{Bfluc-corr} \end{eqnarray}
The first term on the righthand side represents {\it fluctuation dynamo} due to
growth of magnetic fluctuations, whereas the second term represents {\it
magnetic
induction}, or the generation of magnetic fluctuations from the mean field by
random advection.   Note that for $|\bx-\bx'|\gg L_u,$ the integral correlation
length of the velocity field,
$$ F^{ij}_{k\ell}(\ba,\ba',t_0|\bx,\bx',t) \simeq
      F^i_k(\ba,t_0|\bx,t) F^j_\ell(\ba',t_0|\bx,t), $$
because the two stochastic particle trajectories become statistically
independent.
As a consequence, the second magnetic induction term in (\ref{Bfluc-corr})
always
goes to zero for $|\bx-\bx'|\rightarrow \infty.$ The first term will also
vanish in
that limit if $\langle B_0^k(\ba)B_0^\ell(\ba')\rangle-\langle
B_0^k(\ba)\rangle
\langle B_0^\ell(\ba')\rangle\rightarrow 0$ for $|\ba-\ba'|\rightarrow \infty$
(statistical ``clustering'' of initial data).

The above formulas simplify in the special case of spatially homogeneous
statistics
for both the velocity and magnetic fields. In particular, (\ref{Bcorr}) becomes
\be \langle B^i(\br,t)B^j(\bzed,t) \rangle=
    \int d^3\rho\, \langle B^k_0(\brho)B^\ell_0(\bzed)\rangle \,
     F^{ij}_{k\ell}(\brho,t_0|\br,t). \lb{Bcorr-hom} \ee
with the homogeneous 2-body mean Green's function
\be F^{ij}_{k\ell}(\brho,t_0|\br,t)\equiv \int d^3a \,
     F^{ij}_{k\ell}(\ba,\ba+\brho,t_0|\bx,\bx+\br,t). \lb{2-GF-hom} \ee
For $|\br|\gg L_u,$
$$ F^{ij}_{k\ell}(\brho,t_0|\br,t)\simeq \int d^3a \,
     F^i_k(\ba+\brho;t_0,t) F^j_\ell(\ba+\br;t_0,t), $$
and the 2-body Green's function is non-negligible only for $|\brho-\br|=
O\left((\beta |t-t_0|)^{1/2}\right).$
Then (\ref{MF-sumrule}) implies that
\be \lim_{|\br|\rightarrow\infty} \int d^3\rho\,
      F^{ij}_{k\ell}(\brho,t_0|\br,t) = \delta^i_k\delta^j_\ell.
\lb{2-GF-sumrule} \ee
These properties will be used in our discussion of the numerical results below.

\subsection{Numerical Study of Kinematic Dynamo}

We now present a numerical study of small-scale turbulent kinematic dynamo at
$Pr_m=1.$
We employ the same database of non-helical, incompressible fluid turbulence
that was used
in our investigation of Richardson diffusion in section II.C. This might appear
to be a poor
choice at first sight, since the conventional view
\cite{Haugenetal04,Schekochihinetal04} is
that the kinematic fluctuation dynamo at $Pr_m=1$ is a phenomenon of
sub-viscous scales.
Schekochihin et al. conclude explicitly: ``This dynamo is driven by the
viscous-scale
eddies and whatever the inertial-range velocities might do is guaranteed to
happen at
a slower rate'' \cite{Schekochihinetal04}. For this reason, the previous
numerical studies
have taken special pains to resolve well the viscous range, at a sacrifice of
Reynolds number.
For example, the highest resolution $1024^3$ simulation of Haugen et al.
\cite{Haugenetal04}
had a Taylor-scale Reynolds number $Re_T=230$ which is nearly half that of the
database that we employ, for which $Re_T=433$ \cite{Lietal08,Perlmanetal07}.
The previous numerical results seemed to verify the idea that viscous scales
played the dominant
role;  for example, the magnetic energy spectrum in the kinematic regime was
found to be peaked
at wavenumbers a little higher than the viscous Kolmogorov wavenumber $k_\nu.$
See Haugen
et al. \cite{Haugenetal04}, Fig.~4 and Schekochihin et
al.\cite{Schekochihinetal04}, Fig.~22(a).
The viscous range is not so well-resolved in the database that we employ, with
the grid spacing
$\Delta x$ of the simulation being slightly greater than $2\ell_\nu.$
Nevertheless, our study was
designed to show the critical role of inertial-range advection to the
small-scale turbulent dynamo
at  high Reynolds numbers and, thus, was forced to sacrifice resolution of the
viscous range.
Our results will show that both ranges play a critical role at $Pr_m=1$ and, we
will argue,
even for $Pr_m$ much larger.

\subsubsection{Methods}

Our Lagrangian numerical approach is based upon the results in section III.A.
We construct
an ensemble of stochastic particles that solve eq.(\ref{a-stoch-eq}) backward
in time from
common starting point $\bx$. The Feynman-Kac formula (\ref{B-FK}) then yields
\begin{eqnarray*}
&& B^2(\bx,t_f) = \cr
 &&    \,\,\,\,\,\,\,\,\,
\overline{\overline{\,B_0^i(\wt{\ba}(t_0))B_0^j(\wt{\ba}'(t_0))
\left(\boJ(\wt{\ba},t_f,t_0)\boJ^\top(\wt{\ba}',t_f,t_0)\right)_{ij}\,}}',
     \,\,\,\,\,\,\,\,\,\,\,\,\,\,\,\,\,\,\,
\end{eqnarray*}
where the double-overline indicates an average over two ensembles of
trajectories
$\wt{\ba}(t),$ $\wt{\ba}'(t)$ with independent realizations of the Brownian
noise.   The
$\boJ$ matrix satisfies  equation (\ref{J-eq}) for $\grad_x\bdot\bu=0:$
$$ \frac{d}{dt_f}\boJ(\wt{\ba},t_f,t_0)=
\boJ(\wt{\ba},t_f,t_0)\grad_x\bu(\wt{\ba}(t_f),t_f),\,\,\,\,
     \boJ(\wt{\ba},t_0,t_0)=\bI. $$
The exact solution of this equation is an anti-time-ordered exponential of the
velocity-gradient
from $t_0$ to $t_f$ and we have here indicated explicitly the dependence of
$\boJ$ upon both
times. As a matter of fact, it is numerically easier to use the ODE in the
initial time $t_0$
\be \frac{d}{d\tau}\boJ(\wt{\ba},t_f,\tau)=-\grad_x\bu(\wt{\ba}(\tau),\tau)
     \boJ(\wt{\ba},t_f,\tau),\,\,\,\,\boJ(\wt{\ba},t_f,t_f)=\bI, \lb{J-eq-2}
\ee
which may be solved backward in time from $\tau=t_f$ to $\tau=t_0$ along with
the stochastic
equations (\ref{a-stoch-eq}). We then average over an ensemble of initial
conditions $\bB_0$ (the same for each $\tau$) to obtain the mean magnetic
energy
\begin{eqnarray*}
&& \langle B^2(\bx,t_f)\rangle_\tau= \cr
 &&    \,\,\,\,\,\,\,\,\,\overline{\overline{\,
      \langle B_0^i(\wt{\ba}(\tau))B_0^j(\wt{\ba}'(\tau))\rangle
\left(\boJ(\wt{\ba},t_f,\tau)\boJ^\top(\wt{\ba}',t_f,\tau)\right)_{ij}\,}}'.
     \,\,\,\,\,\,\,\,\,\,\,\,\,\,\,\,\,\,\,
\end{eqnarray*}
In effect we are solving for the growth of magnetic field by moving the time
$\tau$ of the initial
conditions backward rather than advancing $t_f$ forward. Thus, our results
below shall
be plotted with respect to the difference variable $t=t_f-\tau$. Assuming
ergodicity, an
average over space
$$  \langle B^2(t)\rangle\equiv \frac{1}{V}\int d^3x\,\,\langle
B^2(\bx,t_f)\rangle_\tau $$
is equivalent to an average over an ensemble of velocities. In a statistical
steady state,
this average should indeed be a function only of the difference variable
$t=t_f-\tau.$

To further simplify matters, we take as our initial seed field for the dynamo a
spatially uniform
magnetic  field $\bB_0$ which is still random, however, and statistically
isotropic. The
covariance choice $\langle B_0^i B_0^j\rangle =\frac{1}{3}\delta^{ij}$ implies
a magnetic
energy initially equal to one. This is not very small, but there is no
requirement of small
field strength in our kinematic problem. The formula for the mean magnetic
energy
then factorizes as
\begin{eqnarray}
\langle B^2(\bx,t_f)\rangle_\tau &=& \overline{\overline{
\frac{1}{3}{\rm
tr}\,\left(\boJ(\wt{\ba},t_f,\tau)\boJ^\top(\wt{\ba}',t_f,\tau)\right)\,\,}}'
\cr
&=&\frac{1}{3}{\rm
tr}\,\left(\bar{\boJ}(t_f,\tau)\bar{\boJ}^\top(t_f,\tau)\right),
\lb{Bsq-uni} \end{eqnarray}
where $\bar{\boJ}(t_f,\tau) \equiv \overline{\,\boJ(\wt{\ba},t_f,\tau)\,}.$
Note that ${\bf F}=
\bar{\boJ}\,\bar{\boJ}^\top$ is a positive-definite, symmetric matrix which
formally reduces
in the limit of vanishing noise to the usual (left) Cauchy-Green or Finger
deformation tensor
of continuum mechanics. The initially uniform magnetic field does not stay
uniform but
develops small-scale fluctuations by an induction effect.  There is, in fact,
no very precise
distinction between ``magnetic induction'' and ``fluctuation dynamo'',  as we
have discussed
elsewhere \cite{Eyink10}, and weak uniform seed fields have been used in many
previous
studies of turbulent magnetic dynamo \cite{ChoVishniac00,Choetal09,ChoRyu09}.
Thus, at fixed $\br,$ the magnetic correlation function with this initial seed
field is dominated
at long times $t$ by the leading dynamo eigenmode $\mathcal{E}$,
$$ \langle B^i(\br,t) B^j(\bzed,t)\rangle  \sim ({\rm const.}) e^{\gamma t}
\mathcal{E}^{ij}(\br),
     \,\,\,\,\,t\rightarrow\infty$$
with $\gamma$ the dynamo growth rate. In the opposite limit of large distances
for fixed $t$ it follows from (\ref{Bcorr-hom}),(\ref{2-GF-sumrule}) that
$$ \langle B^i(\br,t) B^j(\bzed,t)\rangle \sim \frac{1}{3}\delta^{ij},
     \,\,\,\,\, |\br|\rightarrow\infty$$
for our choice of initial seed field.

We implemented this scheme numerically by solving the SDE (\ref{a-stoch-eq})
backward in time for $N=1024$ samples $\wt{\ba}^n(\tau),$ $n=1,2,\dots N,$ all
started
from point $\bx$ at time $t_f$ with independent realizations of the noise. We
took
$t_f=1.5$ and $t_0=0.5,$ because the spatially-averaged energy dissipation
$\varepsilon(\tau)$ is very constant for the interval of time $t_0<\tau<t_f$ in
the database,
varying by $<1\%$ from its space-time mean value $\bar{\varepsilon}=0.0919$
over
that interval.  As in our study of Richardson diffusion, we solved
(\ref{a-stoch-eq})
using the Euler-Maruyma scheme with $dt=10^{-3}$ and solved the equation
(\ref{J-eq-2}) for $\boJ$ with the Euler method. Velocity-gradients are
calculated
by a 4th-order finite-difference scheme with 4th-order Lagrange interpolation
in space.
We checked convergence in $dt$ for several $\bx$ values both by taking smaller
$dt$
and by comparison with the 1.5th-order of Platen for (\ref{a-stoch-eq}) and
with a
consistent equation for the matrix $\boJ.$ We then approximated
\begin{eqnarray}
&& \langle B^2(\bx,t_f)\rangle_\tau \doteq \cr
&& \,\,\,\, \frac{2}{N(N-1)}\sum_{1\leq n<m\leq N}
       \frac{1}{3}{\rm tr}\,\left(\boJ(\wt{\ba}^n,t_f,\tau)
        \boJ^\top(\wt{\ba}^m,t_f,\tau)\right), \cr
&&
        \lb{Bsq-tau-approx}
        \end{eqnarray}
by a sum over the $N(N-1)/2=523,776$  number of pairs of samples. Note that
Hoeffding's
law of large numbers for $U$-statistics \cite{Hoeffding61,Serfling80} implies
that this
pair-average converges for $N\rightarrow\infty $ to the double-average in
(\ref{Bsq-uni})
over two independent realizations of the white-noise. We then furthermore
averaged in
space over $S=1600$ points $\bx_s,$ $s=1,2,\dots S, $ with 25 points chosen
randomly
from each of $64=4^3$ subcubes of the whole domain. We obtain
\be \langle B^2(t)\rangle \doteq \frac{1}{S}\sum_{s=1}^S \langle
B^2(\bx_s,t_f)\rangle_\tau
\lb{Bsq-approx} \ee
as our final approximation to the magnetic energy. More space-averaging was
required
for the kinematic dynamo than for Richardson diffusion because of intermittency
of the
velocity-gradients involved in line-stretching \footnote{There was one
additional complication
in the calculation. As a modest requirement for accuracy, we required that the
magnetic energy
at each of the 1600 points $\bx_s$ should be positive and with relative error
less than $50\%$
for times $t=t_f-\tau$ less than $10t_\eta.$ This criterion was easily
satisfied for most points,
with $N=1024$ particles. However, there were 7 points of the 1600 for which
this number
$N$ did not suffice. These were points of very large magnitude of
velocity-gradient, two or three times
the rms value, associated to turbulent intermittency. For such points, we had
to use more particles,
up to $N=65,536$ or more than 2 billion particle-pairs in the worst case.}.

\subsubsection{Results}

\begin{figure}
\begin{center}
(a)
\includegraphics[width=240pt,height=200pt]{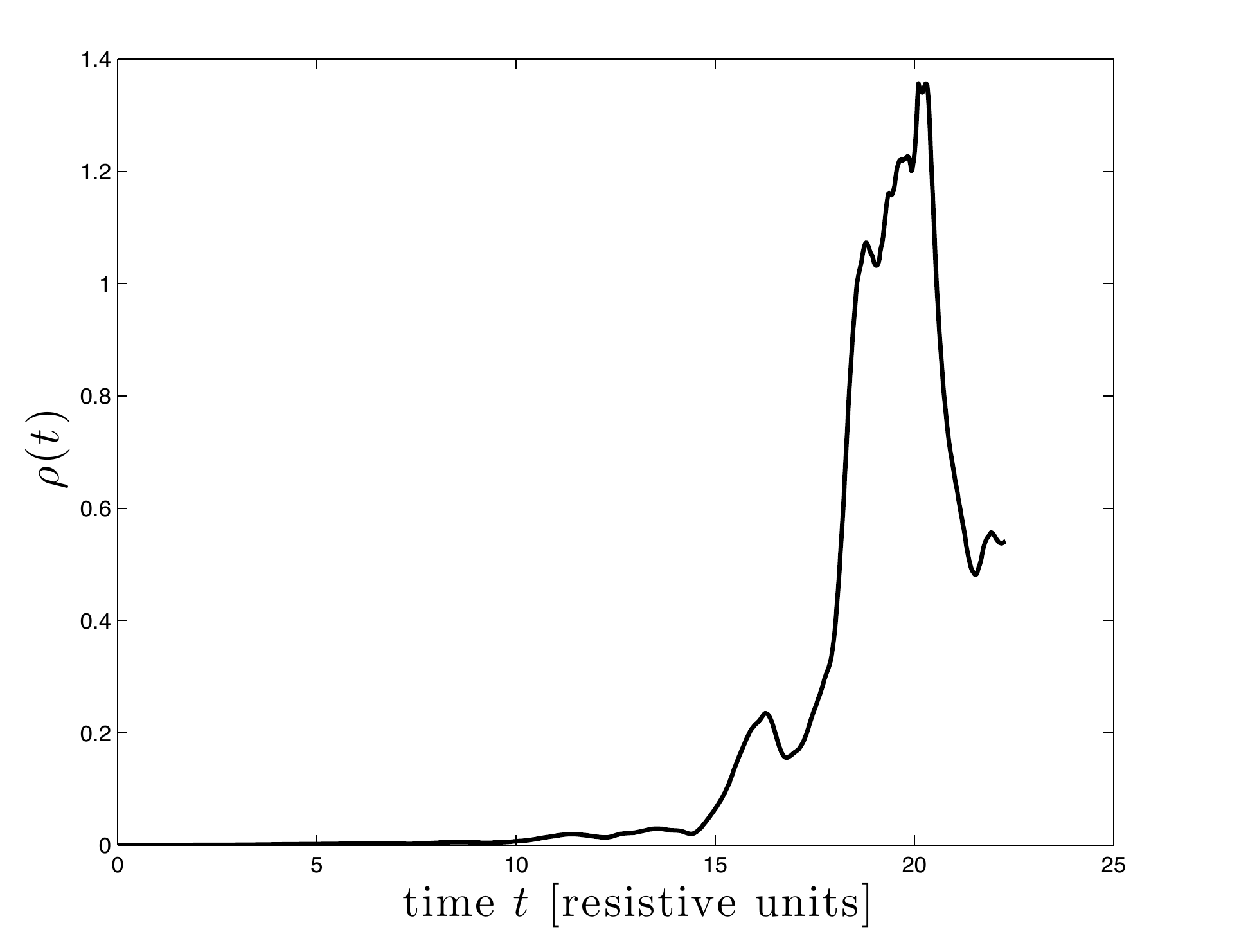}\\
(b)
\includegraphics[width=240pt,height=200pt]{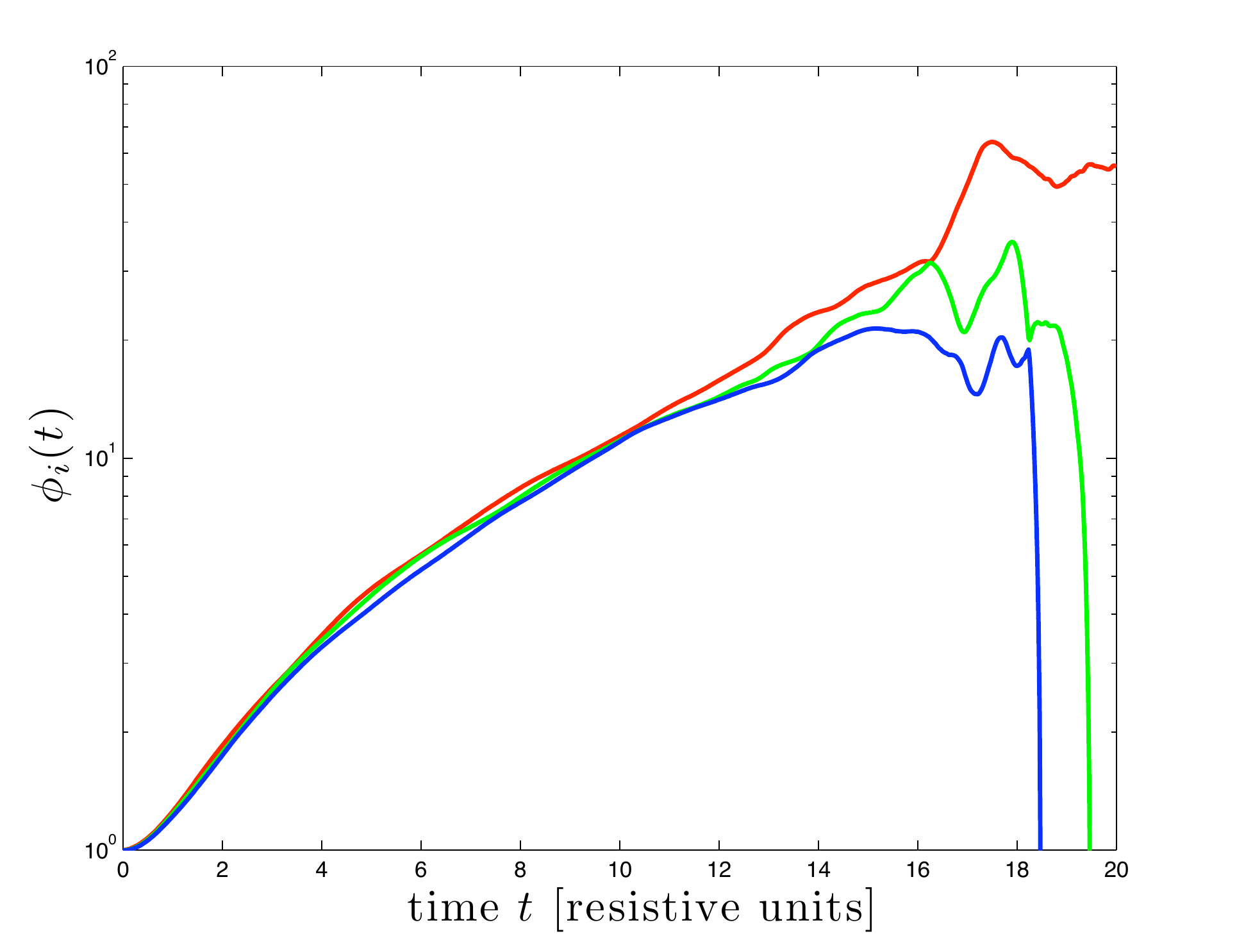}\\
\end{center}
\caption{Error estimation. (a) The ratio $\rho(t)$ of the norms of the
anti-symmetric
and symmetric parts of the approximate Cauchy-Green tensor. (b) The eigenvalues
$\phi_i(t)$,  $i=1,2,3$ of the approximate Cauchy-Green tensor, with
\textcolor{red}{red}
for largest, \textcolor{green}{green} for middle, and \textcolor{blue}{blue}
for smallest.}
\end{figure}

We now present our results for $Pr_m=1.$ We first demonstrate convergence of
our
algorithm in $S$ and $N.$ There are several ways to estimate the errors
associated
with the averaging over space and random samples. One approach is to consider
the
approximation to the Cauchy-Green matrix ${\bf F}$ obtained by omitting the
$\frac{1}{3}$
factor and the trace in eq.(\ref{Bsq-tau-approx}). As we have discussed above,
the exact
Cauchy-Green matrix should be a positive-definite, symmetric matrix. Thus, if
we form the symmetric and anti-symmetric parts
$$ {\bf F}_S=\frac{1}{2}({\bf F}+{\bf F}^\top),\,\,\,\, {\bf
F}_A=\frac{1}{2}({\bf F}-{\bf F}^\top), $$
one measure of the relative error in our calculation is the ratio of matrix
norms
$$ \rho(t) = \frac{\|\langle {\bf F}_A(t_f,\tau)\rangle\|}{\|\langle{\bf
F}_S(t_f,\tau)\rangle\|}. $$
Furthermore, if the small-scale turbulence is statistically isotropic (as is
known
for the database employed), then the space-average Cauchy-Green matrix should
satisfy
$$\langle F_{ij}\rangle = \frac{1}{3}{\rm tr}\,(\langle {\bf
F}\rangle)\delta_{ij}. $$
In particular, each of the three eigenvalues $\phi_i(t),$ $i=1,2,3$ of $\langle
{\bf F}(t_f,\tau)\rangle$
should be equal to $\langle B^2(t)\rangle. $

In Fig.~5 (a) and (b) we plot our results for $\rho(t)$ and $\phi_i(t),$
$i=1,2,3, $ respectively.
Note that time $t$ in these plots and in all those following have been
non-dimensionalized
by the resistive time $t_\eta=\sqrt{\lambda/\bar{\varepsilon}}=4.49\times
10^{-2}$ (which is also
the viscous time since $Pr_m=1$). The results for $\rho(t)$ show that the
relative error in our
calculation is less than a few percent up until about 15 resistive times. This
is confirmed by the
plot of the three eigenvalues $\phi_i(t),$ $i=1,2,3, $ in panel (b), which are
in quite close agreement
until that time.  The eigenvalues also remain all positive until after 18
resistive times, consistent with
positivity of $\langle B^2(t)\rangle.$ These results show, incidentally, that
there was no need for
us to take the initial magnetic field $\bB_0$ to be random and statistically
isotropic. The
same dynamo growth is observed for any deterministic uniform field pointing in
any direction.

The error in our approximation for $\langle B^2(t)\rangle$ can also be
estimated in another
way. Since the number of pairs in (\ref{Bsq-tau-approx}) is quite large, one
can guess that
the dominant error arises from the average over $S$ space points in
(\ref{Bsq-approx}).
A central limit theorem argument then suggests that the error is approximated
by
\be \delta\langle B^2(t)\rangle\simeq \sqrt{\frac{{\rm Var}\,\{\langle
B^2(t_f)\rangle_\tau\rangle\}}{S}}
\lb{Bsq-error} \ee
with the spatial variance
\be  {\rm Var}\,\{\langle B^2(t_f)\rangle_\tau\rangle\}\doteq
     \frac{1}{S-1}\sum_{s=1}^S \Big| \langle B^2(\bx_s,t_f)\rangle_\tau-\langle
B^2(t)\rangle
     \Big|^2. \lb{Bsq-error-tau} \ee
This latter quantity has some independent physical interest, because it
quantifies the
spatial intermittency of the dynamo effect. In Fig.~6 we plot our approximation
(\ref{Bsq-approx})
for $\langle B^2(t)\rangle$ along with plus-or-minus the error estimate $\delta
\langle B^2(t)\rangle$
above. These are consistent with the estimates in the previous Fig.~5.

\begin{figure}
\begin{center}
\includegraphics[width=240pt,height=200pt]{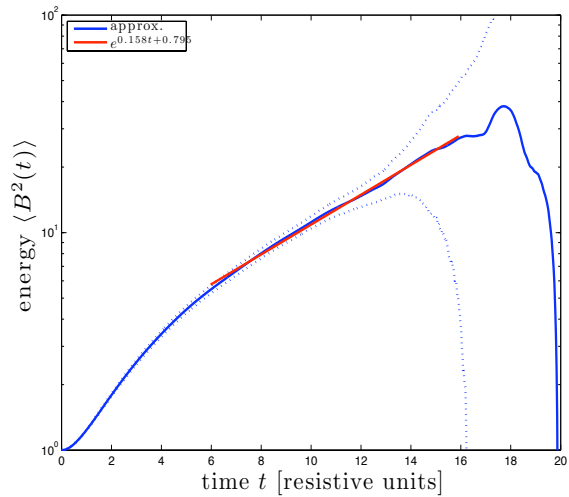}\\
\end{center}
\caption{Mean magnetic energy. Plotted in \textcolor{blue}{blue} is the mean
magnetic energy
(solid line) calculated from (\ref{Bsq-tau-approx}),(\ref{Bsq-approx}), along
with plus and minus
the error (dotted line) estimated from (\ref{Bsq-error}),(\ref{Bsq-error-tau}).
The straight line in
\textcolor{red}{red} shows the least-square linear fit over the time interval
$t=6$ to $t=16.$}
\end{figure}\lb{energy}

\begin{figure}
\begin{center}
\includegraphics[width=240pt,height=200pt]{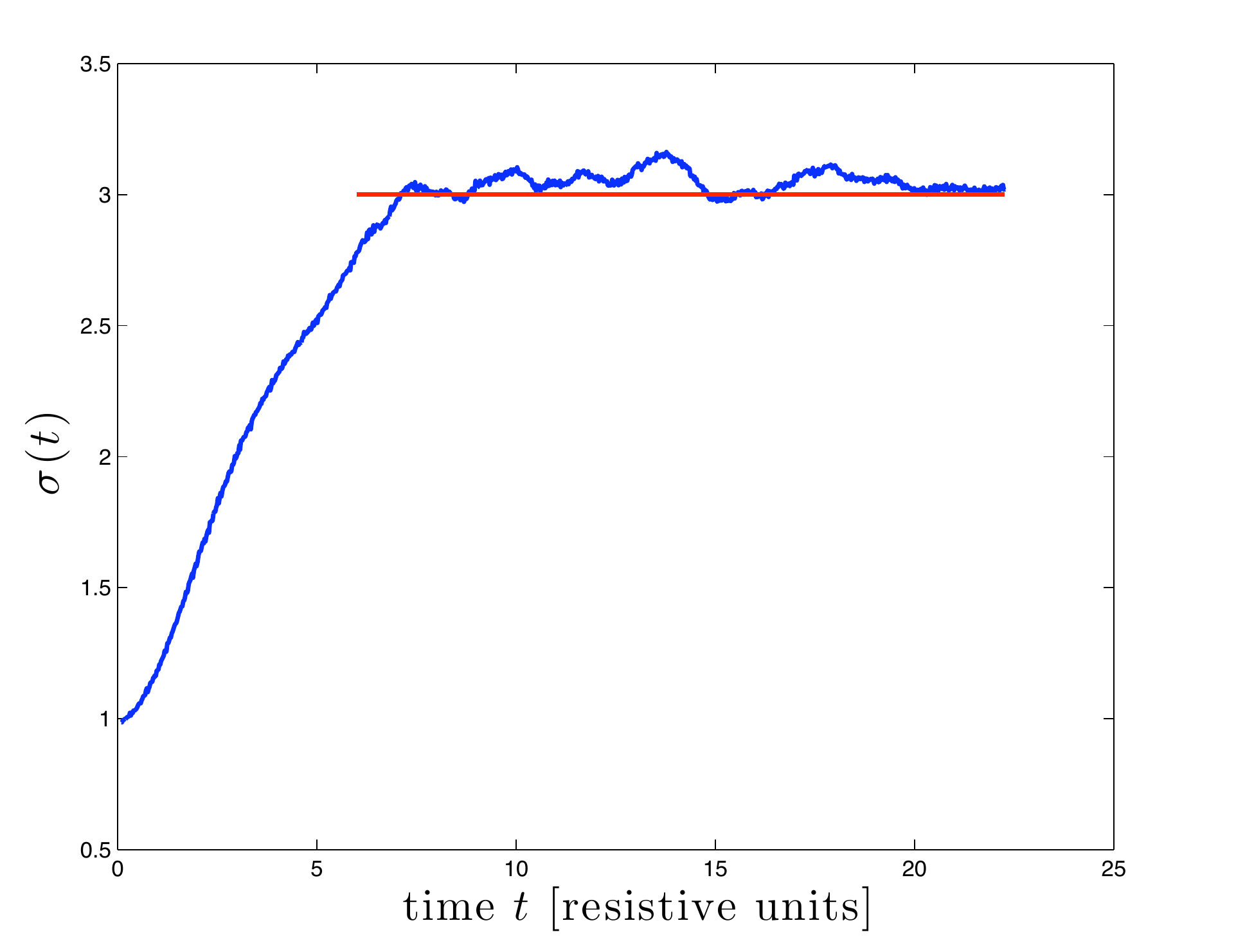}\\
\end{center}
\caption{Local slope of backward dispersion. Plotted in \textcolor{blue}{blue}
is
the local slope of the backward dispersion $\langle r^2(t)\rangle,$ as defined
in (\ref{sigma-def}).
Plotted in \textcolor{red}{red} is the horizontal line-segment from $t=6$ to
$t=25$
for the Richardson value $\sigma=3.$}
\end{figure}\lb{localslope}

The  plot of magnetic energy from kinematic dynamo effect in Fig.~6 is our
central result.
The expected exponential growth $\sim e^{\gamma t}$  of magnetic energy is
clearly observed
after about 6 resistive times.  A linear fit over the range of times 6 to 16 is
also plotted
in Fig.~6, yielding an estimated growth rate of $\gamma t_\eta\doteq 0.158$.
At earlier times the growth rate is significantly larger. For example,  a
linear fit over
the range of 1 to 4 resistive times yields an estimate $\gamma t_\eta\doteq
0.344.$ This is closer
to the magnitude of the typical viscous strain rate eigenvalue $\sqrt{\langle
S^2\rangle/3}$, which in
units of  the viscous/resistive  rate $\sqrt{\bar{\varepsilon}/\nu}$ is equal
to $1/\sqrt{6}\doteq 0.408.$
The physical interpretation of these results is clear. According to the
stochastic Lundquist
formula (\ref{stoch-Lundquist}), field lines that are carried into a point
$\bx$ along individual
Lagrangian trajectories are stretched at the viscous strain rate. However, the
resistive average
over the ensemble of stochastic trajectories leads to cancellation and
suppression of the growth
rate. As time advances, the spatial region sampled by the wandering
trajectories increases
in size and the suppression effect increases. Indeed, the asymptotic
exponential
growth range begins precisely at the onset of turbulent Richardson diffusion of
the trajectories.
To demonstrate this, we consider the (backward) dispersion $\langle
r^2(t)\rangle$
of the stochastic Lagrangian trajectories which determines the typical linear
dimension
$L(t)=\sqrt{\langle r^2(t)\rangle}$ of the region from which field-lines
arrive. The
local-in-time scaling exponent
\be \sigma(t)\equiv \frac{d \log(\langle r^2(t)\rangle)}{d \log(t)}.
\lb{sigma-def} \ee
will equal 1 at early times and transition to 3 when Richardson diffusion sets
in.
The numerical results for this quantity are plotted in Fig.~7, along with a
horizontal
line (red) at level $\sigma=3$ beginning at $t=6. $ It is clear that the
asymptotic
exponential growth range of magnetic energy and the Richardson diffusion
of trajectories start at the same time. The inertial-range properties of
turbulent
2-particle diffusion are thus critical in determining the ultimate
growth rate of the $Pr_m=1$ turbulent kinematic dynamo.

Although initial field lines arrive to a point at time $t$ from a large region
of size $L(t),$
not all of these field lines contribute equally to the growth of magnetic
energy.  To quantify
the contribution of line-vectors at initial separation $r,$ we use the
line-vector correlation
function which was proposed as a  ``dynamo order-parameter''
\cite{EyinkNeto10}:
$$\mathcal{R}_{k\ell}(\br,t)=F^{ii}_{k\ell}(\br,t_0|\bzed,t), $$
where $F$ is the homogeneous 2-particle Green's function from  (\ref{2-GF-hom})
\footnote{This Green's function was denoted in our previous work
\cite{EyinkNeto10}
as $\bar{F}$ and $F$ was used for the corresponding adjoint Green's function.}.
$\mathcal{R}_{k\ell}(\br,t)$ represents the scalar correlation at time $t$
between material
line-vectors $\bell(t),\,\bell'(t)$ which started as unit vectors
$\hat{\bE}_k,\hat{\bE}_\ell$
at positions displaced by $\br$ at the initial time $0$ and which arrive at the
same final
point. Setting $r=0$ in eq.(\ref{Bcorr-hom}) gives, in general,
$$ \langle B^2(t)\rangle = \int d^3r\, \langle B_0^k(\brho)
B_0^\ell(\bzed)\rangle
     \mathcal{R}_{k\ell}(\brho,t). $$
This formula separates the effect of the initial correlations of the magnetic
field and
the effect of the turbulent advection and stretching.
For the case of isotropic and non-helical velocity statistics, we may decompose
the tensor $\mathcal{R}$ into contributions from line-vectors initially
longitudinal and
transverse to the separation vector $\br:$
$$ \mathcal{R}_{k\ell}(\br,t)= R_L(r,t)\hat{r}_k\hat{r}_\ell
   +R_N(r,t)(\delta_{k\ell}-\hat{r}_k\hat{r}_\ell), $$
where $\hat{\br}$ is the unit vector in the direction of $\br$. For our
particular
choice of initial magnetic field,
\be \langle B^2(t)\rangle =\frac{1}{3}\int_0^\infty 4\pi
r^2dr\,[R_L(r,t)+2R_N(r,t)].
\lb{Bsq-Rcorr-uni} \ee

\begin{figure}
\begin{center}
(a)
\includegraphics[width=240pt,height=200pt]{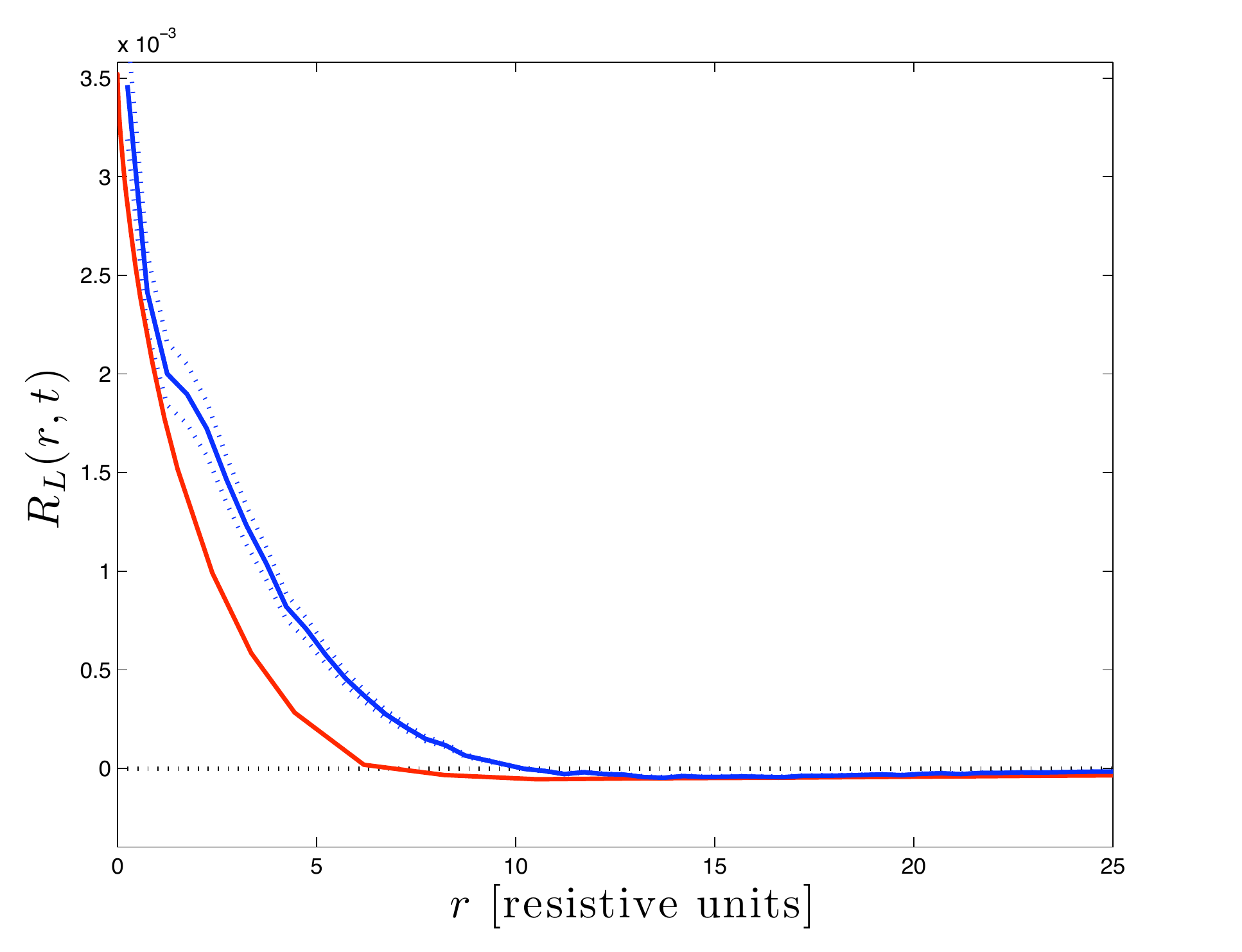}\\
(b)
\includegraphics[width=240pt,height=200pt]{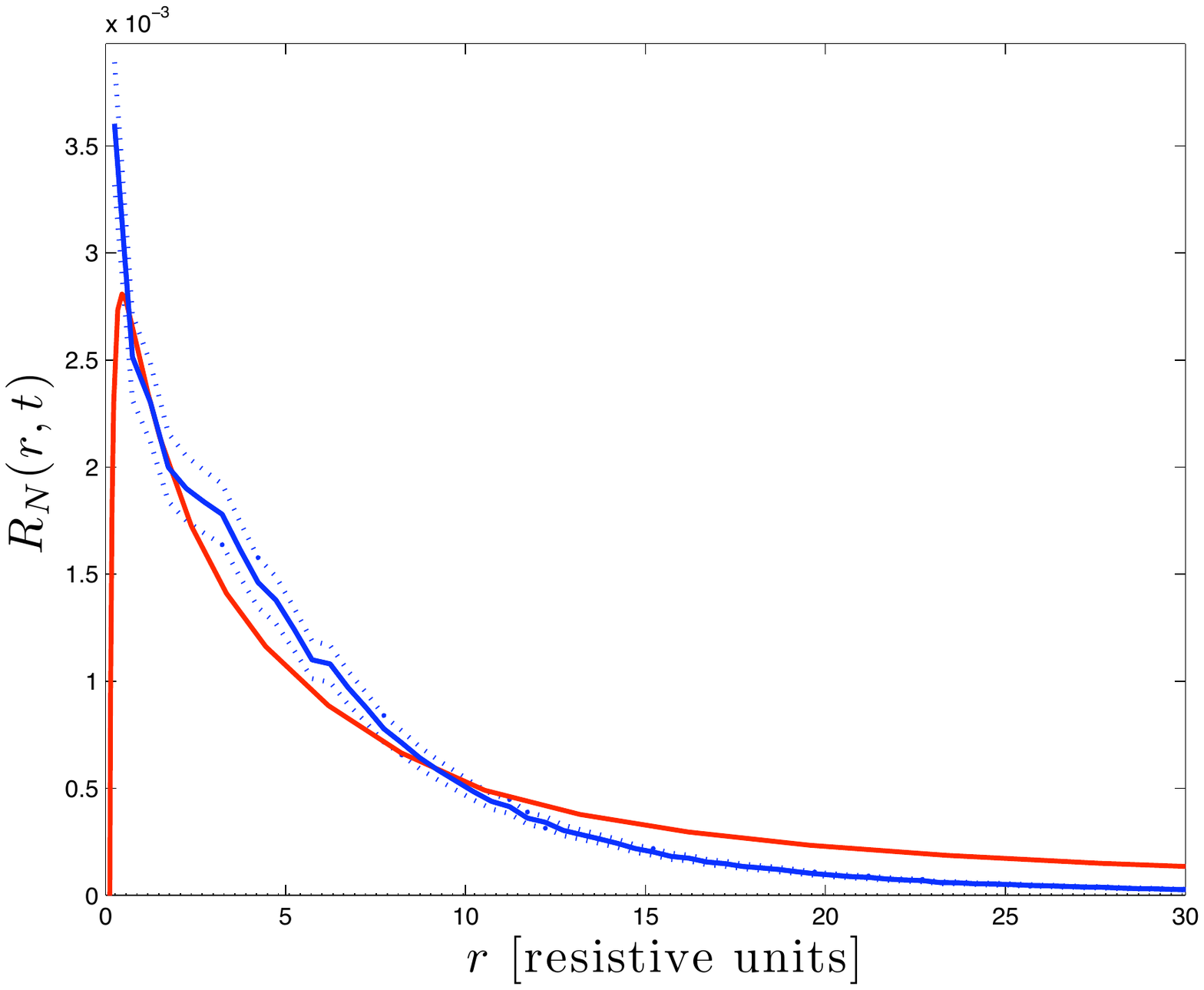}\\
\end{center}
\caption{Line-vector correlations: (a) longitudinal, (b) transverse.
The line-vector correlations $R_L(r,t)$ and $R_N(r,t),$ respectively, as
calculated
numerically from formula (\ref{GL}) at time $t=11.12,$ are plotted in
\textcolor{blue}{blue}
with solid lines and the correlations plus and minus their estimated errors
with dotted
lines. Also plotted with \textcolor{red}{red} lines are analytical results
\cite{Eyink10}
for these correlations in the Kazantsev-Kraichnan model at $Pr_m=0$, normalized
as described in the text.}
\end{figure}

This line-correlation can be calculated numerically by the same procedure that
we have used to obtain the magnetic energy itself in eqs.(\ref{Bsq-tau-approx})
and (\ref{Bsq-approx}). For example, the longitudinal line-correlation can be
approximated by
\begin{eqnarray}
&& R_L(r,t)\doteq \cr
&& \frac{1}{S}\sum_{s=1}^S\frac{2}{N(N-1)}\sum_{n<m}
\hat{\br}\boJ(\wt{\ba}^n_s,t_f,\tau)
\boJ^\top(\wt{\ba}^m_s,t_f,\tau)\hat{\br}^\top \cr
&&   \,\,\,\,\,\,\,\,\,\,\,\,\,\,\,\,\,\,\,\,\,
\,\,\,\,\,\,\,\,\,\,\,\,\,\,\,\,\,\,\,\,\,
\,\,\,\,\,\,\,\,\,\,\,\,\,\,\,\,\,\,\,\,\,
\times\frac{\delta(|\wt{\ba}^n_s(\tau)-\wt{\ba}^m_s(\tau)|-r)}{4\pi r^2}, \cr
&&
\lb{GL} \end{eqnarray}
taking $\hat{\br}$ as a row vector. The corresponding transverse correlation
$R_N(r,t)$ is obtained by replacing $\hat{\br}$ in (\ref{GL}) with two
orthogonal
unit vectors $\hat{\bE}_i,$ $i=1,2$ which span the subspace orthogonal to
$\hat{\br}$
and by then summing over these two contributions. In practice these continuous
distributions must be sampled in discrete bins. We took 200 bins of size
$\Delta r=
\ell_\eta/2,$ or one-half of the resistive length-scale. To capture the
contributions
from $r>100\ell_\eta$ but to avoid large fluctuations in the results, we added
three
extra large bins of size $100,200,$ and $400$ $\ell_\eta,$ centered at $150,
300$
and $600$ $\ell_\eta,$ thus covering the whole range of $r$ in the database.

\begin{figure}
\begin{center}
\includegraphics[width=240pt,height=200pt]{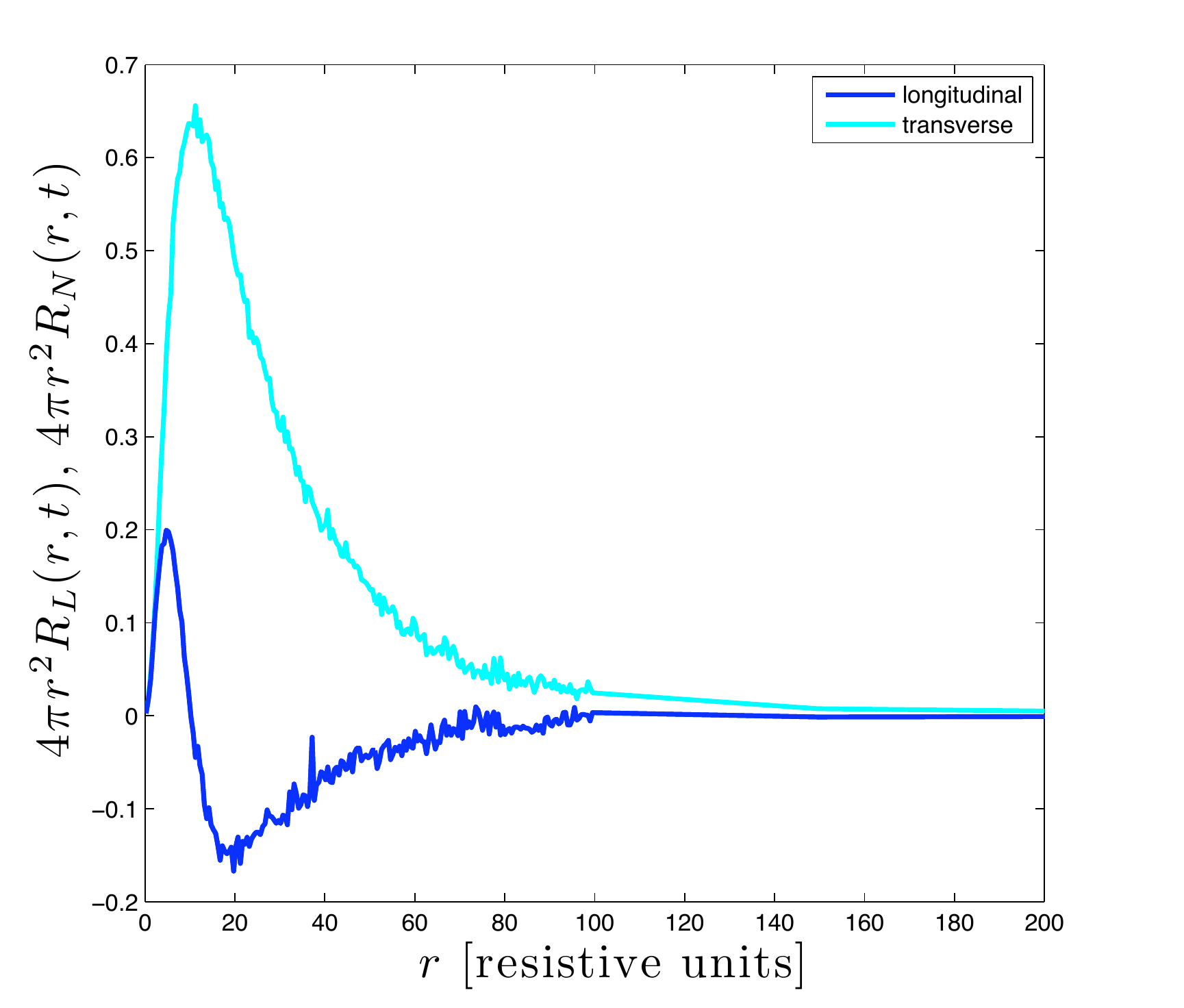}\\
\end{center}
\caption{Contributions to magnetic energy. Plotted in \textcolor{blue}{blue} is
the function $4\pi r^2 R_L(r,t)$ and in \textcolor{cyan}{cyan} the function
$4\pi r^2 R_N(r,t),$
both at time $t=11.12.$ These represent the contributions to magnetic energy in
(\ref{Bsq-Rcorr-uni}) from pairs of line-vectors at initial separations $r$ and
initially parallel
and perpendicular, respectively, to the separation vector $\br.$}
\end{figure}

We present in Fig.~8 our numerical results for $R_L(r,t)$ and $R_N(r,t)$ at a
time
$t=11.12$ in resistive time-units. As should be clear from Figs.~6 and 7, this
time
lies well within the range both of exponential growth of energy and of
Richardson diffusion
of particle pairs. The most interesting feature of these correlations is their
considerable
diffuseness in $r.$ This is shown even more clearly in Fig.~9, which plots the
integrands
$4\pi r^2 R_L(r,t)$ and $4\pi r^2 R_N(r,t)$ in eq.(\ref{Bsq-Rcorr-uni}) for
$\langle B^2(t)\rangle.$
In order to get 50\% of the magnetic energy one must integrate in that formula
out to
$r=15.71\ell_\eta.$ Likewise, to get 75\% of the energy one must integrate out
to $r=39.15\ell_\eta$
and to get 90\% one must integrate all the way out to $r=66.59\ell_\eta.$ Thus,
line-vectors separated
initially by many resistive lengths are brought together by turbulent advection
to produce the dynamo
growth. Another very interesting feature is the large {\it negative}
contribution of the initially longitudinal
line-vectors, seen in Fig.~8(a) and even more clearly in Fig.~9. A very similar
negative
contribution was found \cite{Eyink10} for the Kazantsev-Kraichnan dynamo model
at $Pr_m=0$ and $Re_m=\infty.$ It was suggested there that negative values of
$R_L(r,t)$
for sufficiently large $r$ are due to an effect of bending and looping of field
lines.

It is interesting to make a more detailed comparison of our numerical results
for
hydrodynamic turbulence at $Pr_m=1$ and of the analytical results in the
Kazantsev-Kraichnan
(KK) dynamo model for $Pr_m=0.$  The comparison is not completely
straightforward, so a few
words of explanation are required. We consider the KK model with spatial
H\"older exponent
$h=2/3$ in the velocity correlation (\ref{KK-space-corr}). With this choice of
$h$,
Richardson's $t^3$-law holds for particle dispersion and it is generally the
appropriate
choice for comparisons with physical fluid turbulence. In this case, the
coefficient $D_1$
in (\ref{KK-space-corr}) has the same dimensions as $\varepsilon^{1/3},$ with
$\varepsilon$
the mean energy dissipation per unit mass. Our previous results for the KK
model
\cite{EyinkNeto10,Eyink10} were obtained with lengths non-dimensionalized by
$(\lambda/D_1)^{3/4}$ and times by $(\lambda/D_1^3)^{1/2}.$ However, $D_1$ is
not
numerically equal to $\varepsilon^{1/3},$ but only dimensionally the same. In
order
to relate them quantitatively, we note that the KK particle diffusion equation
(\ref{Kr-Rich-eq})
reduces in the isotropic sector to Richardson's original equation
(\ref{Rich-eq}) with coefficient
of eddy-diffusivity $K_0=2D_1.$ In that case, one recovers the $t^3$-law
(\ref{toy-law}) with a
Richardson-Obukhov constant $g_0$ if one makes the definition for the KK model
\be  \varepsilon \equiv  143\frac{64}{81g_0} D_1^3. \lb{eps-KK} \ee
We thus obtain
$ \ell_\eta\equiv \left(\lambda^3/\varepsilon\right)^{1/4}=\beta
           \left(\lambda/D_1\right)^{3/4}$
with $\beta=\frac{3}{2\sqrt{2}}\left(\frac{g_0}{143}\right)^{1/4}.$  For the
backward 2-particle
diffusion in our kinematic dynamo study we have found that $g_0\doteq 1.57$
(somewhat
larger than the value 1.35 reported in section II.C for an independent
experiment over a
different time range) and thus $\beta\doteq 0.343.$ For comparison with the
present results,
therefore, the results of Eyink \cite{Eyink10}, Fig.~10 must have $x$-axis
scaled by $1/\beta$
and $y$-axis scaled by $\beta^3$. Furthermore, the quantities
$\widetilde{G}_L(r),$
$\widetilde{G}_N(r)$ previously calculated \cite{Eyink10} are dynamo growth
eigenmodes
which dominate the behavior of $R_L(r,t),$ $R_N(r,t)$ at long times. Since
there
is an arbitrariness in the normalization of the eigenmodes, we have additional
freedom in
the vertical scale. This is fixed by imposing on the KK eigenmodes the
normalization
condition (\ref{Bsq-Rcorr-uni}) at $t=11.12$ resistive time units.

We have plotted the results for the line-correlations from the KK model in
Fig.~8 (red lines),
together with our numerical results for hydrodynamic turbulence (blue lines).
Clearly,
the two sets of results are qualitatively very similar. In both cases,  the
transverse correlation
$R_N(r,t)$ is everywhere positive, sharply peaked at small $r,$ but with a slow
decay at
large $r.$ The longitudinal correlations $R_L(r,t)$ of the two sets share all
these same features
except for their sign, with both exhibiting a long negative tail at large $r$.
The major difference
between the results for hydrodynamic turbulence at $Pr_m=1$ and the KK model at
$Pr_m=0$ is the
distinctly slower rate of decay of correlations at large $r$ for the latter.

\begin{figure}
\begin{center}
(a)
\includegraphics[width=240pt,height=200pt]{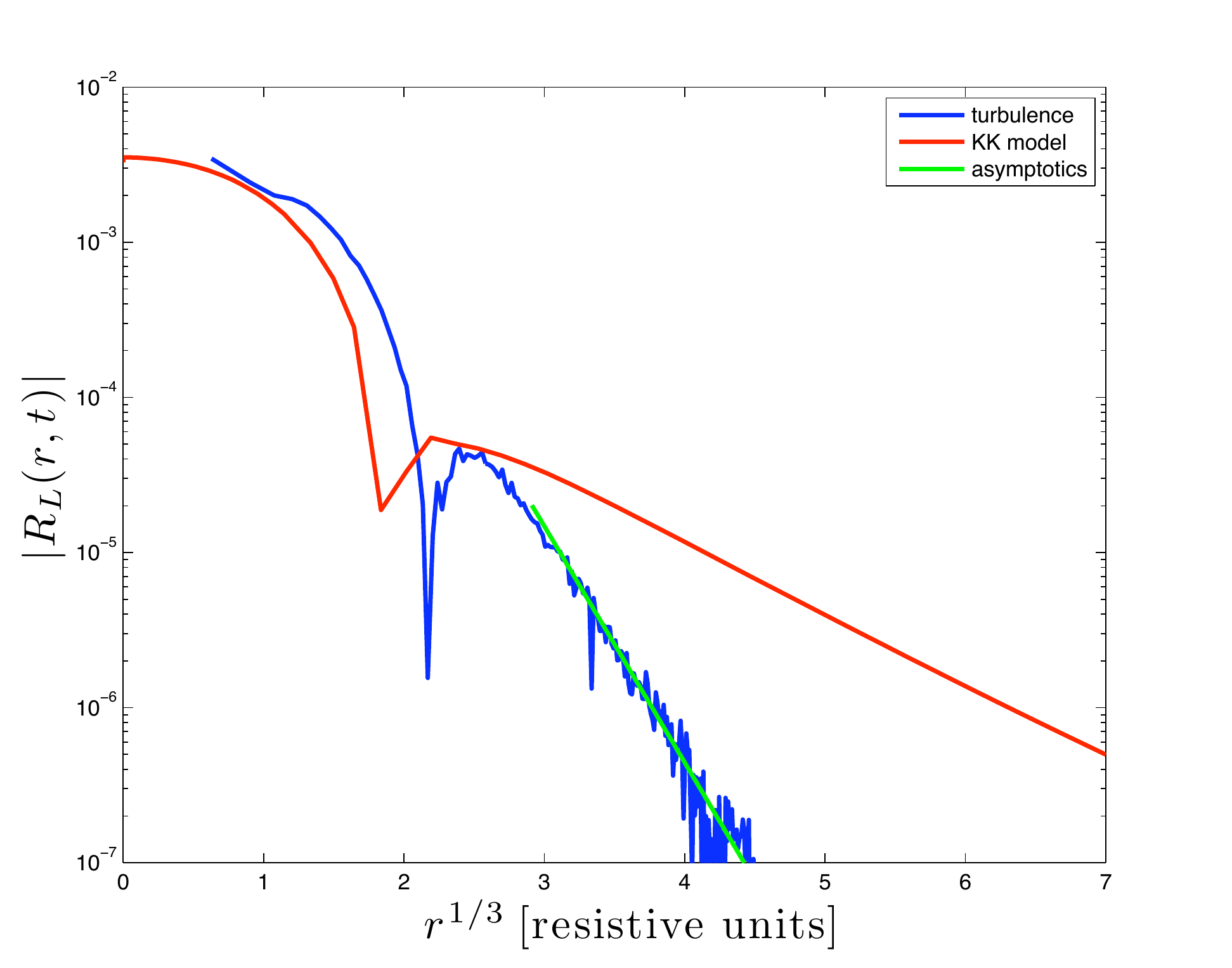}\\
(b)
\includegraphics[width=240pt,height=200pt]{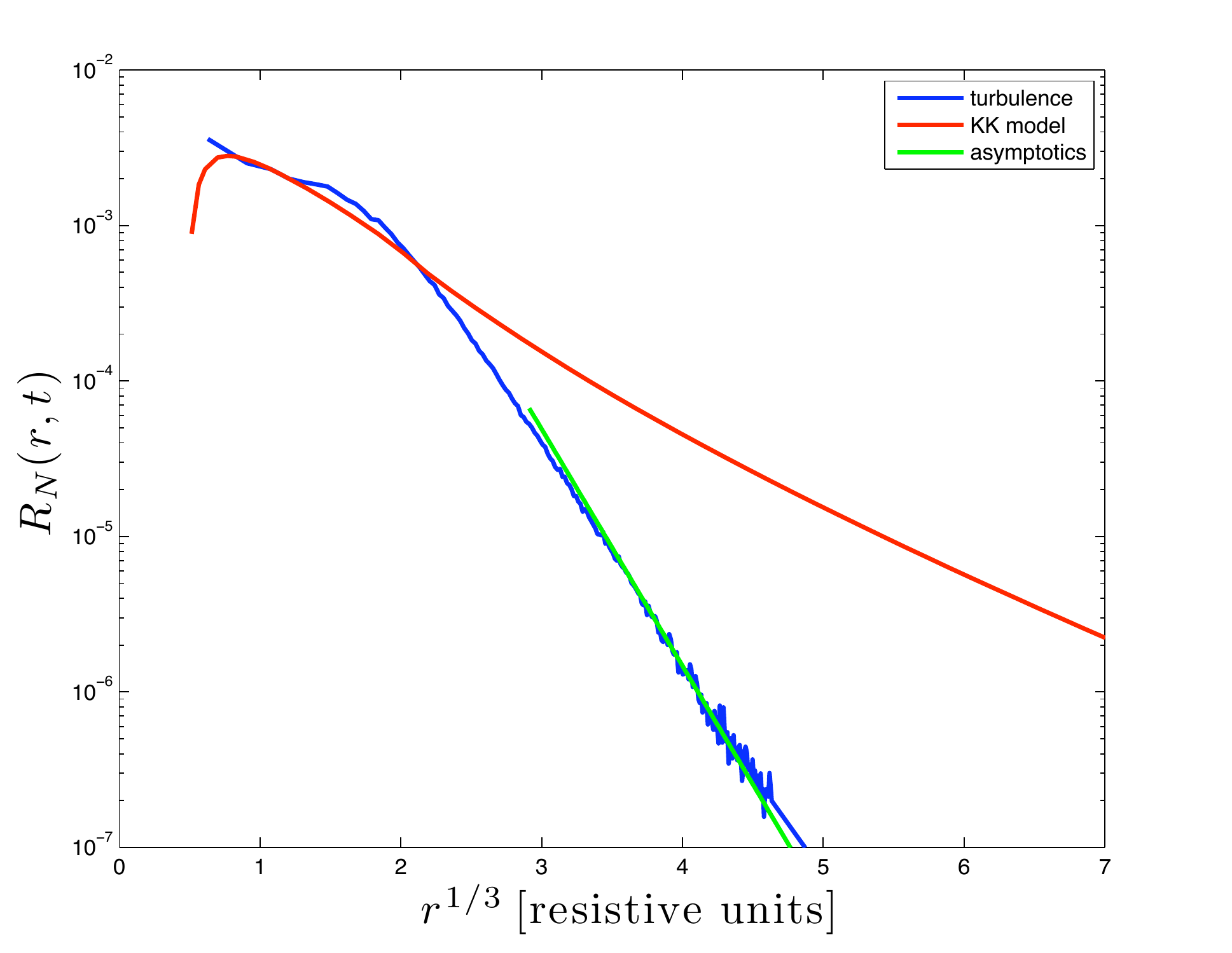}\\
\end{center}
\caption{Line-correlations for large separations: (a) longitudinal, (b)
transverse.
Plotted in \textcolor{blue}{blue} are the numerical results from (\ref{GL})
at time $t=11.12$ and in \textcolor{red}{red} the analytical results
\cite{Eyink10}
from the Kazantsev-Kraichnan model at $Pr_m=0$. Straight lines in the
log-linear
plot versus $r^{1/3}$ correspond to the large-$r$ asymptotics (\ref{large-r}).
Shown in \textcolor{green}{green} is the prediction of that formula with the
growth
rate $\gamma=0.158$ determined in Fig.~6.}
\end{figure}

The precise decay rate at large $r$ is known for the line-correlations in the
KK model.
It was shown \cite{Eyink10} that, up to power-law prefactors,  these exhibit a
very slow
stretched-exponential decay
\be \widetilde{G}_L(r),\,\widetilde{G}_N(r) \sim
\exp\left(-\frac{3\sqrt{2\gamma}}{2D_1}r^{1/3}\right),
    \,\,\,\,\, r\gg \ell_\eta, \lb{large-r} \ee
where $\gamma$ is the dynamo growth rate. A remarkable feature of this
asymptotic formula
is that it depends upon resistivity $\eta$ (or magnetic diffusivity
$\lambda=\eta c/4\pi$) only through
the growth rate $\gamma.$ To check for a similar decay in hydrodynamic
turbulence we present
in Fig.~10 (a),(b) a log-linear plot of the line-correlations $|R_L(r,t)|,$
$R_N(r,t)$ versus $r^{1/3},$
both for our numerical calculation with the hydro turbulence database (blue)
and for the KK model
(red). The straight lines in these plots verify the stretched exponential decay
with power $r^{1/3}$.
In fact, the decay law (\ref{large-r}) which was derived in the KK model for
$Pr_m=0$ holds very
well in hydrodynamic turbulence at $Pr_m=1,$ including the coefficient in the
stretched exponential.
If we use the relation (\ref{eps-KK}) to replace $D_1$ with $\varepsilon$ and
then substitute
the dynamo growth rate $\gamma t_\eta\doteq 0.158,$ we predict from
(\ref{large-r}) a line
in a log-linear plot with slope  $-3.507$ (in resistive units). A line with
this slope is plotted
(in green) in Fig.~10 (a),(b) and can be seen to match our numerical results
quite well.
The reason for the slower decay of the stretched exponential in the KK model at
$Pr_m=0$ is thus
entirely attributable to a smaller dynamo growth rate than what we find in
hydro
turbulence for $Pr_m=1.$ In particular, using the result
\cite{Vincenzi02,Eyink10} that
$\gamma(\lambda/D_1^3)^{1/2}\doteq 0.193$ for KK at $Pr_m=0$ and using again
the
relation (\ref{eps-KK}), we get that $\gamma t_\eta\doteq 0.0217. $ This is
almost an order
of magnitude smaller than the result $\gamma t_\eta\doteq 0.158$ that we found
for hydro
turbulence at $Pr_m=1$ and motivates some further discussion below.

\subsubsection{Discussion}

Let us make a quantitative comparison of our growth rate with those found in
other
studies \cite{Schekochihinetal04,Haugenetal04,Vincenzi02,Choetal09,ChoRyu09}
of small-scale dynamo in incompressible, non-helical turbulence at $Pr_m=1$. We
present
both the Reynolds numbers and the dynamo growth rates found in these studies.
We
give the ``box-size'' Reynolds number $Re=u'/\nu k_0,$ where
$u'=u_{rms}/\sqrt{3}$ is
the rms value of a single velocity component and $k_0$ is the smallest
wavenumber
in the simulation study. This is not as dynamically significant a Reynolds
number as
is $Re_f=u_{rms}/\nu k_f,$ based on the forcing wavenumber $k_f,$  but it is
the
easiest to calculate from the published data. We give the
growth rate in the dimensionless form $\gamma t_\nu$ where, as above,
$t_\nu=(\nu/\varepsilon)^{1/2}$ is the viscous time (and also the resistive
time
for $Pr_m=1$).  The data from the paper of Schekochihin et al. (2004)
\cite{Schekochihinetal04}
is taken from their Table 2 (where $1/t_\nu=\sqrt{15}\Gamma_{{\rm rms}}$
there).  The
data of Haugen et al. (2004) \cite{Haugenetal04} is taken from their Fig.3,
which
plots their results for $\alpha\equiv (\gamma/2)/(u_{rms}k_f)$ versus $Re_f.$
Note,
however, that
$$ \gamma t_\nu =2\alpha\sqrt{\frac{3}{5}} \frac{Re_T}{Re_f}, $$
where $Re_T=u'\ell_T/\nu$ is the Taylor-scale Reynolds number defined there
\cite{Haugenetal04}, with $\ell_T=\sqrt{5}u_{rms}/\omega_{rms}.$ We can also
infer
from their Table II \cite{Haugenetal04} that $Re_T=\sqrt{60 Re_f}$ so
that, putting all these relations together,
$$ \gamma t_\nu = \frac{12\alpha}{\sqrt{Re_f}}. $$
Finally, $Re=(1.5)Re_f$ since their forcing wavenumber \cite{Haugenetal04}
is $k_f=1.5.$ The paper of Vincenzi (2002) \cite{Vincenzi02} presents a
numerical
study of the KK model. Fig.~6 in that work plots the results for
$\gamma(\lambda/D_1^3)^{1/2}$
at $Re=\infty$ and a broad range of Prandtl numbers. Note that the
``viscosity'' for the KK model
is defined \cite{Vincenzi02} by $\nu_*=D_1\ell_\nu^{4/3}$ where $\ell_\nu$ is
the
short-distance cut-off length of the inertial scaling range and the
corresponding
``Prandtl number'' is defined by $Pr_{m\,*}=\nu_*/\lambda.$ It is not obvious
how to
best compare this ``Prandtl number'' with that for viscous hydrodynamic
turbulence.
In any case, the result for $Pr_{m\,*}=1$ is $\gamma(\lambda/D_1^3)^{1/2}
\doteq 0.350,$
which we transform to $\gamma t_\nu\doteq 0.0383$ as described earlier.
Finally,
the papers of Cho et al. \cite{Choetal09,ChoRyu09} present results for a large
number of runs
with both normal and hyperviscosity.
The authors have kindly provided me
with data from their highest Reynolds-number normal-viscosity simulation,
denoted
RUN $512P-B_010^{-3}$ (for $512^3$ resolution with physical viscosity and
initial
uniform seed field of strength $10^{-3}.$) This data is reported below (J. Cho,
private
communication). In one important respect this simulation is quite distinct from
the others reported
here, because the initial velocity field is not a fully developed turbulent
field but is instead
supported in the very low-wavenumber interval $2\leq k\leq 4.$  Turbulence
quickly
develops with the volume-averaged kinetic energy dissipation $\varepsilon(t)$
increasing
by a factor of 42 over their time interval of kinematic dynamo. Although this
dynamo
simulation is not in a statistically steady turbulent regime, it does give some
useful perspective
and we thus include it here. All of these data are gathered into Table I below.

\begin{table}[h]
\caption{\label{tab:table1} Reynolds numbers and dynamo growth rates
from several numerical studies with $Pr_m=1$. Shown are the box-size Reynolds
number $Re=u'/\nu k_0,$ where $k_0$ is the minimum wavenumber, and the
growth rate $\gamma$ non-dimensionalized by the viscous time $t_\nu=
(\nu/\varepsilon)^{1/2}$.}
\begin{tabular}{| l | c | c |}
\hline\hline
 Reference                            &                     $Re$ &
      $\gamma t_\nu$ \\
\hline\hline
Schekochihin et al. (2004)&                        110  &
     0.0138 \\
                                                &                        210  &
                          0.0265 \\
                                                 &                       450
&                          0.0329 \\
\hline
Haugen et al. (2004)&                                   227  &
        0.0338  \\
                                      &                                  450
&                         0.0379  \\
                                       &                                1050  &
                         0.0385 \\
\hline
Vincenzi (2002) [KK model]&                    $\infty$  &
  0.0383  \\
\hline
Cho et al. (2009)&                                          294  &
             0.550  \\
\hline
present study&                                               6380   &
             0.158  \\
\hline
\end{tabular}
\end{table}

It is widely expected that $\gamma t_\nu$
approaches a universal value as $Re\rightarrow \infty,$ but the data presented
seem not to confirm this. The study of Haugen et al. \cite{Haugenetal04} indeed
reported
seeing such a limiting behavior, as can be observed from their data in Table I.
Their results for the growth rate are also consistent with those of
Schekochihin et al.
\cite{Schekochihinetal04} and remarkably close to those of Vincenzi
\cite{Vincenzi02}
for the KK model at $Re=\infty.$ The latter must be regarded as a coincidence,
however,
due to the ambiguity in the definition of the Prandtl number for that model.
For example,
a better definition might be $Pr_m\equiv \epsilon^{1/3}\ell_\nu^{4/3}/\lambda,$
which,
using (\ref{eps-KK}), gives $Pr_m\doteq 4.16 Pr_{m\,*}.$ Thus, $Pr_m=1$
corresponds to
$Pr_{m\,*}\doteq 0.24.$ Whatever the ``best'' definition of the Prandtl number
for the KK
model might be, the value of $\gamma t_\nu$ for $Pr_m=1$ will be somewhat
smaller
than that reported above for $Pr_{m\,*}=1,$ because the latter is near the
maximum
of  $\gamma (\lambda/D_1^3)^{1/2}$ (Vincenzi \cite{Vincenzi02}, Fig.~6). Our
own result
for $\gamma t_\nu$ is about 4 times larger than the value of Haugen et al. for
their
highest Reynolds number simulation. The dimensionless growth rate increases
with $Re$ and our Reynolds number is about 6 times larger than theirs, so this
might account for some of the discrepancy. On the other hand,  our resolution
of
the viscous range is also relatively poor and this may degrade the numerical
accuracy of our result for the growth rate. To further complicate the picture,
the
value of $\gamma t_\nu$ found by  Cho et al. is about 3.5 times larger than
ours.
Note that the result reported in Table I for Cho's data used the value
$\bar{\varepsilon}$
from a time-average of $\varepsilon(t)$ over the kinematic interval of
exponential growth
in order to define $t_\nu.$ If we instead used the value $\varepsilon(t_*)$ at
the end
of the kinematic interval to define $t_\nu$, then we would obtain a somewhat
smaller
value $\gamma t_\nu\doteq 0.312$ for Cho et al. but still twice ours.

The results in the Table I appear contradictory on the face of it, but we
believe there
is a simple resolution. The key is the time range over which the exponential
growth is observed.  Our growth rate was calculated for the time interval
$6$--$16 t_\nu$ with the latter time only about 1/3 $T_u,$ where $T_u=L_u/u'$
is the
large-scale eddy-turnover time.  Haugen et al. do not publish growth curves in
their
paper, but  Schekochihin et al. present in their
Fig.~21\cite{Schekochihinetal04}
a log-linear plot of the magnetic energy for their kinematic runs, with fitting
ranges
for exponential growth indicated. The cited growth rate for their 450
Reynolds-number
simulation (Run A) was obtained over a  time-interval of $66.6$--$505.9 t_\nu.$
This corresponds in their simulation to about $5.86$--$44.5 T_u$. (Here we have
used the well-known result that $T_u\doteq 0.4 (u')^2/\varepsilon$ and the data
in
their Table 1 \cite{Schekochihinetal04} to estimate $T_u\doteq 0.267$ for Run
A.)
At earlier times in their simulation, the growth of magnetic energy is much
faster. For
example, if we use the data in their Fig.~21 for times $<5.86 T_u$ to estimate
the
growth rate, we obtain a value $\gamma t_\nu=0.0691,$ about half of ours.

We therefore conjecture that there are two distinct kinematic regimes with
exponential growth of magnetic energy at different rates,  one for times $t\ll
T_u$
and another for times $t\gg T_u.$ This makes perfect sense from the Lagrangian
point of view developed in this paper. For times $t\ll T_u,$ the (backward)
particle
dispersion $\langle r^2(t)\rangle$ is growing as $\sim \varepsilon t^3$ (see
Fig.~7).
This means that the magnetic field strengths are obtained by averaging initial
field
vectors arriving from inertial-range separations growing as a power-law.
However, for
times $t\gg T_u,$ the mean-square separation $\langle r^2(t)\rangle$ saturates
to a
value of order $\sim L_B^2,$ where $L_B$ is the size of the periodic box. In
this time
regime,  magnetic field strengths are obtained by averaging seed field vectors
arriving
to each point from the entire flow domain. It stands to reason that the growth
rate would
be reduced in this strongly mixed regime. Which of the two kinematic regimes is
most
relevant in practice depends upon the strength of the seed magnetic field
$B_0.$
It is observed \cite{Schekochihinetal04,Haugenetal04,Choetal09} that the
kinematic
interval of exponential growth ends once the magnetic field energy $\langle
B^2(t)\rangle$
grows to equipartition with the viscous-range kinetic energy
$u_\nu^2=(\varepsilon \nu)^{1/2}$.
If $\ln(u_\nu/B_0)\ll \gamma T_u=O((Re)^{1/2}),$ then this occurs at times much
less than $T_u,$ the regime that we have studied. If the seed field is
extremely small
or the Reynolds number not so large, however, then viscous equipartition will
only occur
for times $\gg T_u.$ This is the regime studied in Shekochihin et al.
\cite{Schekochihinetal04}
and Haugen et al. \cite{Haugenetal04}. In support of this conclusion, we note
that the
RUN $512P-B_010^{-3}$ of Cho et al., with a not too small seed field, reached
saturation
with viscous kinetic energy at a time around $12t_\nu$ or $1.2T_u$. The growth
rate
cited in Table I was obtained from a fit  over the range $1.72$--$11.7t_\nu$ or
$0.173$--$1.18T_u,$ comparable to the time range that we study, and
with a similarly large growth rate.

Clearly more study of this matter would be desirable. Unfortunately, the
turbulence
database that we employ only stores the velocity field for about one $T_u,$
or $44.8t_\nu,$ so we cannot study times $\gg T_u.$ We do believe, however,
that our
results have attained the asymptotic regime $t_\nu\ll t\ll T_u.$ In support of
this claim,
we note that we are observing already at time $11.12t_\nu$ the expected
inertial-range
behaviors, such as Richardson diffusion and the stretched-exponential
correlation
decay in eq.(\ref{large-r}) (which was derived in the KK model for
$L_u=\infty$).


The most important general conclusion from our results presented above is that
the inertial-range phenomenon of Richardson diffusion plays a critical role in
the
small-scale fluctuation dynamo of hydrodynamic turbulence at $Pr_m=1.$ Because
field lines are only ``frozen-in'' stochastically, an infinite number of lines
enters
each point from a very large region of size $L(t)\sim (\varepsilon t^3)^{1/2}$
at
time $t.$ This mixing of field lines from far away opposes the growth by
stretching
of the individual lines and suppresses the dynamo growth rate. Because of
nearly
complete cancellation, the lines arriving from separations of order $\sim L(t)$
contribute a vanishingly small amount to the magnetic energy. Nevertheless,
the asymptotic formula
(\ref{large-r}) shows that field-lines separated at inertial-range distances
$r\gg \ell_\eta$
give a non-negligible contribution. That formula implies also---very
remarkably---
that the dynamo growth rate can be inferred directly from the
stretched-exponential
decay of the magnetic line-vector correlations in the inertial range.

A second important conclusion is that there are very fundamental similarities
between
the mechanisms of the small-scale fluctuation dynamo for Prandtl numbers
$Pr_m=1$ and $Pr_m=0$ in high Reynolds-number turbulence. The physics that we
have discussed above for hydrodynamic turbulence at $Pr_m=1$ and large $Re_m$
agrees very closely with what was established  \cite{EyinkNeto10,Eyink10}
for the Kazantsev-Kraichnan model at $Pr_m=0$ and $Re_m=\infty$. Richardson
diffusion
and stochasticity of flux-freezing play a critical role in both.  We thus
disagree with
Shekochihin et al. \cite{Schekochihinetal04} who drew the conclusion that
``The small-scale
dynamo at $Pr_m\sim 1$ belongs to the same class as the large-$Pr_m$ dynamo.''
There is a flaw in the argument made by those authors that inertial-range
time-scales
are too long to affect the operation of the dynamo, which proceeds at a faster
viscous rate
(see quote in the opening paragraph of Section IV.B). Because the exponential
dynamo
growth is a {\it long-time} phenomenon, there is sufficient time for advection
by
inertial-range eddies to affect and modify its rate. Indeed, we have seen in
the
present study at $Pr_m=1$ that the inertial-range phenomenon of Richardson
diffusion
commences at times $t\gtrsim 6 t_\nu$ and not at times orders of magnitude
greater
than the viscous time.

In our view, the $Pr_m=1$ small-scale dynamo---and, more generally, the finite
$Pr_m$ dynamo---
is a transitional case which shares some of the attributes of both of the
extreme limits
$Pr_m=0$ and $Pr_m=\infty.$  A corollary of this view is that there should be
an important
influence of the inertial range even for small-scale dynamos with $Pr_m\gg 1,$
if also
$Re\gg 1.$  This is the situation for many low-density, high-temperature
astrophysical
plasmas with large magnetic Prandtl numbers but with, also, a substantial
inertial range.
As we have discussed in section II.B, the exponential separation of particles
typical of the large-$Pr_m$ ``Batchelor regime'' is an intermediate asymptotics
for
a range of times $t_\nu\ll t\ll  \ln(Pr_m) t_\nu$ in high Reynolds-number
turbulence.
After a time $t\gtrsim t_\nu \ln(Pr_m),$ magnetic line-elements in such a
turbulent flow
will begin to experience Richardson diffusion and a concomitant decrease of the
dynamo growth rate. Because the dependence upon $Pr_m$ is logarithmic, the time
to enter this inertial-range influenced regime is only a relatively small
multiple of
$t_\nu$ even at very large $Pr_m.$ (For example, with a value $Pr_m=10^{14}$
typical
of the warm interstellar medium, $\ln(Pr_m)\doteq 32$.) If the initial magnetic
seed
field is weak enough, then its backreaction on the turbulence up to that time
may
be neglected and the turbulent kinematic dynamo process essentially as we have
discussed in this section of our paper will proceed, even at very large Prandtl
numbers.

\section{Nonlinear MHD Turbulence, Dynamo and Reconnection}

Although we have focused in our discussion of dynamo effect on the kinematic
stage, almost all of the results of this paper extend---with appropriate
modifications---
to fully nonlinear MHD turbulence with backreaction on the flow from the
Lorentz force.
A full treatment is not possible here, but we shall try to stress what is
general in our
previous  presentation,  sketch any necessary modifications, discuss relevant
references, and point out some important directions for further work.

The phenomenon of 2-particle Richardson diffusion and ``spontaneous
stochasticity''
will doubtless exist in MHD turbulence. Their cause is the roughness of the
advecting
velocity field and all theories, simulations and observations of MHD turbulence
(whatever
their other differences) agree that both the velocity and magnetic fields are
indeed
rough in the inertial-range of  such flows. The quantitative growth law of
particle
dispersion will depend upon the precise characteristics of MHD turbulence, such
as
the spectral slopes, degree of anisotropy, etc. These depend upon the ultimate
theory of MHD turbulence, which is an open problem. We base our discussion
below
on the Goldreich-Sridhar (GS) theory of strong MHD turbulence
\cite{GoldreichSridhar95,
GoldreichSridhar97}, but alternative theories
\cite{Iroshnikov64,Kraichnan65,Boldyrev05,
Boldyrev06} would lead to similar results.

For simplicity, we assume incompressible and sub- or trans-Alfv\'enic MHD
turbulence, with rms velocity fluctuations $u'\leq v_A,$ where
$v_A=B_0/\sqrt{4\pi\rho}$
is the Alfv\'en velocity based on the external magnetic field strength $B_0$.
We also assume that large-scale anisotropy is such that ``critical balance''
holds throughout the inertial range and the turbulence is strong. Finally,
we assume that mean cross-helicity is zero and there is an equal flux of
upward and downward propagating Alfv\'en waves.  Under these conditions,
the GS theory predicts that velocity increments for separations $\ell$ scale as
$$\delta u(\ell)\sim (\varepsilon\ell_\perp)^{1/3}\sim (\varepsilon
\ell_\|/v_A)^{1/2}, $$
where $\varepsilon$ is energy dissipation per mass and $\ell_\perp,\ell_\|$ are
the separations perpendicular and parallel to the magnetic field, respectively.
We shall also assume that perpendicular (shear-Alfv\'en) and parallel
(pseudo-Alfv\'en)
components of the velocity increments scale in the same way, or  $\delta
u_\perp(\ell)
\sim \delta u_\|(\ell).$ In that case, we can give a simple theory of
particle-pair
dispersion like that  presented for hydro turbulence in section II.A. If we let
$r_\perp(t)$
and $r_\|(t)$ be the particle separations perpendicular and parallel to the
field, respectively,
then
$$ \frac{d}{dt}r_\perp \sim \delta u_\perp(r)\sim (\varepsilon r_\perp)^{1/3}
$$
(assuming that $r_\|\equiv 0$) implies that
\be r_\perp^2(t) \sim \varepsilon t^3, \lb{Rich-perp} \ee
and in the same manner
$$ \frac{d}{dt}r_\|\sim \delta u_\|(r)\sim (\varepsilon r_\|/v_A)^{1/2} $$
(assuming that $r_\perp\equiv 0$) implies that
\be r_\|^2(t) \sim (\varepsilon/v_A)^2 t^4.  \lb{Rich-para} \ee
The above scaling laws will hold for intermediate times where $t\gg
(r_\perp^2(0)/\varepsilon)^{1/3},\,
(v_A r_\|(0)/\varepsilon)^{1/2}$ but $t\ll T_u=L_u/u',$ the large-eddy turnover
time. The most
important feature observed in both cases is that initial separations are
``forgotten'', the
physical basis of the phenomenon of spontaneous stochasticity. A full theory of
2-particle
dispersion in MHD turbulence will obviously be quite intricate and will depend
upon the
regime of turbulence considered and the phenomenological assumptions employed.
Within theories of weak MHD turbulence
\cite{Galtieretal00,Galtieretal02,Galtier09}, it is
should be possible to give an analytical treatment using well-established
methods
\cite{Balk02}. However, whatever final form the theory of MHD turbulence may
take,
spontaneous stochasticity seems to be a likely consequence.

There have been a few numerical studies of 2-particle dispersion in MHD
turbulence,
both with \cite{BusseMueller08} and without \cite{MuellerBusse07,Busseetal07}
external magnetic field. In the presence of an external magnetic field, Busse
and
M\"uller find that $r_\perp(t)$ grows faster than $r_\|(t)$ (see their Fig.~4
\cite{BusseMueller08}),
in qualitative agreement with our formulas (\ref{Rich-perp}) and
(\ref{Rich-para})
\footnote{Recall that in GS theory, the mean energy dissipation for
sub-Alfv\'enic turbulence
scales as $\varepsilon\sim (u')^4/v_AL_u.$ Using this relation,
(\ref{Rich-perp}) becomes
$r_\perp^2(t)\sim L^2_i(u'/v_A)(t/T_u)^3$ and (\ref{Rich-para}) becomes
$r_{\|}^2(t)
\sim L^2_i (u'/v_A)^4 (t/T_u)^4$. Since $u'\leq v_A,$ these formulas imply that
$r_{\|}(t)\ll r_\perp(t)$ for $t\ll T_u.$}.
The situation is not entirely clear, however.
Following Cho and Vishniac \cite{ChoVishniac00}, it is generally believed that
similar alignments and anisotropies will hold at small scales in MHD turbulence
without an external magnetic field, just as for the external field case, if the
alignments
are taken with respect to a local magnetic field. Paradoxically, however, Busse
et al. find in MHD turbulence without external field that $r_\perp(t)\ll
r_\|(t)$ when
these quantities are defined with respect to the local field. The total
displacement
vector ${\bf r}(t)$ thus becomes preferentially aligned with the local magnetic
field.
It is not obvious how to explain this observation, but it possibly has
something
to do with the dynamical alignment of velocity and magnetic-field increments
at small scales \cite{Boldyrev06,BersnyakLazarian06,Masonetal06}. Clearly,
more study of these issues is required.

Our results in section III.A on stochastic flux freezing in resistive
magnetohydrodynamics
were derived in a fully nonlinear setting. No kinematic assumption was made
there.
It has elsewhere been shown \cite{Eyink09} that all effects of the Lorentz
force on
fluid motion are described by a second stochastic conservation law which
generalizes
the Kelvin circulation theorem, at least for incompressible fluids with
$Pr_m=1.$
This result extends to resistive MHD the ``generalized Kelvin theorem'' derived
for ideal MHD \cite{BekensteinOron00,KuznetsovRuban00}.  As we shall show in
a future paper, this additional stochastic Kelvin theorem also holds in
compressible plasma
fluids with $Pr_m=1$ if they are barotropic (pressure depending only on density
and not
on temperature) and in non-isothermal fluids if the thermal Prandtl number is
also unity.
The existence of  two stochastically ``frozen-in'' fields provides strong
constraints
which deserve to be further explored. We remark finally, and most importantly,
that our discussion in section III.B on the high-Reynolds limit of plasma
turbulence made
no kinematic assumption. The prediction that stochastic flux-conservation holds
in that
limit depends only upon the phenomenon of spontaneous stochasticity, which we
have argued should inevitably occur in high Reynolds-number plasma turbulence.

The Lagrangian formalism of magnetic dynamo presented in section IV.A is valid
in
nonlinear MHD turbulence, if averages over velocity ensembles are conditional
on a given initial magnetic field. An interesting question to pursue, even in
the kinematic
stage, is the role of compressibility on the small-scale, turbulent dynamo. The
behavior of
Lagrangian particles in strongly compressible flows are dramatically different
from those
in incompressible flows and this should have a signficant impact on dynamo
action, e.g. possible
Prandtl-number effects
\cite{GawedzkiVergassola00,EvandenEijnden01,LeJanRaimond04}.
Our Lagrangian numerical methods in section IV.B are specific to the kinematic
dynamo.
However, it would be possible, and interesting, to employ them together with
simple
phenomenological models of nonlinear saturation that have been proposed
 \cite{Subramanian99,Schekochihinetal02}. This would obviate the need for
additional
 closure approximations and would help to understand some of the physics of the
saturation process.

Perhaps the most important implications of the present work are for the problem
of turbulent
magnetic reconnection. Our results and arguments show compellingly, we believe,
that the
constraint of flux-freezing in a turbulent plasma at high-conductivity must be
quite different
than is generally understood. The naive estimate of flux-line slippage due to
resistivity,
$\langle r^2(t)\rangle\sim \lambda t,$ is incorrect, by many orders of
magnitude. The correct
estimate will depend upon the ultimate theory of MHD turbulence, but it must
have the general
form of (\ref{Rich-perp}) and (\ref{Rich-para}) above. The quantity
$\sqrt{\langle r^2_\perp(t)\rangle}$
can be interpreted as the lateral distance that magnetic field-lines diffuse
through a turbulent plasma
in time $t,$ and (\ref{Rich-perp}) should be particularly useful in estimating
reconnection rates
for astrophysical phenomena. Note that both (\ref{Rich-perp}) and
(\ref{Rich-para}) are completely
independent of the resistivity (and of any other microscopic plasma mechanism
of
line-slippage). These results give support, therefore, to all theories
\cite{MatthaeusLamkin86,LazarianVishniac99,LazarianVishniac00,ShibataTanuma01}
and observations
\cite{Lapenta08,Servidioetal09,Loureiroetal09,Kowaletal09,Bhattacharjeeetal09}
of fast magnetic reconnection in MHD turbulence.
In fact, our results show that fast reconnection is necessary for (implied by)
all standard theories of MHD turbulence
\cite{GoldreichSridhar95,GoldreichSridhar97,
Iroshnikov64,Kraichnan65,Boldyrev05,Boldyrev06}.

There are particularly close connections with the Lazarian-Vishniac
theory \cite{LazarianVishniac99,LazarianVishniac00} based on stochastic
wandering of field-lines.
We have stressed that Lagrangian particle trajectories become intrinsically
stochastic in a
rough velocity field, due to the phenomenon of Richardson 2-particle
dispersion. The ``stochastic
wandering'' invoked by Lazarian-Vishniac is an analogue of Richardson diffusion
for
the field lines themselves in a rough magnetic field. Because of this effect,
there is not just
one field-line passing through each point in the limit of zero resistivity, but
instead an
infinite ensemble of random field-lines. The connection between our ideas and
those of
Lazarian-Vishniac \cite{LazarianVishniac99,LazarianVishniac00} deserve to be
further examined.

\section{Final Discussion}

Two ideas are commonplace in the literature on plasma magnetohydrodynamics. One
is
that flux-freezing must hold approximately for $Re_m\gg 1,$ an assumption
widely employed
in treatments of magnetic dynamo effect. Another idea frequently advanced in
discussion of
magnetic reconnection is that the flux-freezing constraint is broken by rapid
diffusion of field
lines across thin current sheets. There has long been a tension between these
two ideas,
never fully reconciled. Both ideas are sometimes invoked in the same setting.
For example,
it is recognized that  while stretching of  ``nearly'' frozen-in lines drives
magnetic dynamo action,
nevertheless fast magnetic reconnection is necessary to relieve tangled
field-line structure
\cite{Parker92,LazarianVishniac00,Galloway03}. We have argued that {\it both}
ideas are
correct, if suitably understood. Magnetic flux through individual material
loops, advected by
the plasma in the usual sense, will {\it not} be conserved for $Re_m\rightarrow
\infty,$ because
of the development of singular current sheets and vortex sheets. However,
magnetic flux
will nevertheless be conserved on average for a random ensemble of loops, in a
novel
statistical sense associated to the spontaneous stochasticity of Lagrangian
flows for
high-Reynolds-number turbulence.

Spontaneous stochasticity is a fluid-dynamical phenomenon due to the explosive
separation of particles produced by turbulent Richardson diffusion. Because of
this effect,
fluid particles that start infinitesimally close together will separate to a
finite distance in a
fixed amount of time, independent of the Reynolds number. In the limit
$Re\rightarrow \infty$
there is an infinite ensemble of Lagrangian trajectories for each initial
particle position.
There is already good evidence for this effect from turbulence simulations and
laboratory
experiments, much of which is reviewed in Section II.C together with own
simulation results. We
expect confirmation from future studies at higher Reynolds numbers. The
G\"ottingen Turbulence
Tunnel now in operation should reach Taylor-scale Reynolds numbers $Re_T\sim
10^4$ and turbulent Richardson diffusion shall be one of the main subjects of
investigation
(E. Bodenschatz, private communication). Richardson diffusion and spontaneous
stochasticity
should also occur in high-Reynolds number MHD turbulence and can be studied by
both
simulation and experiment. Recent laboratory studies of magnetic dynamo and
turbulent
induction with low Prandtl-number liquid metals
\cite{Peffleyetal00,Bourgoinetal02,
Nornbergetal06,Monchauxetal09} have Reynolds numbers high enough and
inertial-ranges sufficiently extensive to support the phenomena. Although
extremely
challenging, it would be quite informative to develop particle-tracking
techniques
for such flows which could investigate Lagrangian mechanisms.

In an earlier  work \cite{EyinkAluie06} we have considered Alfv\'en's theorem
for
turbulent plasma flows from a complementary perspective. The approach used
there was
a spatial coarse-graining of the MHD equations over a continuous range of
scales $\ell,$
similar in spirit to a renormalization group analysis. The effective induction
equation
at length-scale $\ell$ contains a ``turbulent EMF''  $\boepsilon_\ell$ induced
by subscale
plasma motions. As long as energy remains finite for high kinetic and magnetic
Reynolds
numbers, the turbulent EMF can be shown to be much larger at inertial-range
length-scales
$\ell$ than viscous and resistive terms, or than other possible microscopic
dissipation terms.
In that case, the effective flux-conservation equation at inertial-range
length-scales $\ell$
takes the form
\be  \frac{d}{dt}\oint_{\overline{\bx}_\ell(C,t)}
\overline{\bA}_\ell(\bx,t)\bdot d\bx
    = \oint_{\overline{\bx}_\ell(C,t)} \boepsilon_\ell(\bx,t)\bdot d\bx,
\lb{alfven-ell} \ee
where $\overline{\bA}_\ell$ is the coarse-grained magnetic vector-potential and
$\overline{\bx}_\ell(\ba,t)$ is the Lagrangian flow map generated by the
coarse-grained
velocity field $\overline{\bu}_\ell.$ For a smooth, laminar solution of the
ideal MHD
equations, $\boepsilon_\ell\rightarrow \bzed$ everywhere in space as
$\ell\rightarrow 0$
and flux-conservation in the standard sense is recovered. However, it was shown
\cite{EyinkAluie06} that the righthand side of (\ref{alfven-ell}) need not
vanish
for very singular velocity and magnetic fields, in particular, those with
(coincident) current
sheets and vortex sheets.  If such severe singularities appear in MHD
turbulence at
infinite Reynolds numbers, then Alfv\'en's theorem in its conventional sense
can
break down. Magnetic flux
$\overline{\Phi}_\ell(C,t)=\oint_{\overline{\bx}_\ell(C,t)}
\overline{\bA}_\ell(\bx,t)\bdot d\bx$ need no longer be conserved for
individual loops
$C$ as $\ell\rightarrow 0.$ Here we extend that conclusion by arguing that
flux-conservation
will hold {\it on average} for a suitable random ensemble of loops. Similarly
as in section III.B,
we may consider the set of loops $\overline{\ba}_\ell(C+\epsilon
\widetilde{C},t)$ at time $t_0$
obtained by adding random perturbations of size $\epsilon$ to $C$ and then
advecting
backward in time with
$\overline{\ba}_\ell(\cdot,t)=\overline{\bx}_\ell^{-1}(\cdot,t).$ We
expect that this ensemble of loops in the limits first of vanishing magnetic
diffusivity
and viscosity $\lambda,\nu\rightarrow 0$, then $\ell\rightarrow 0$ and finally
$\epsilon\rightarrow 0$ will exactly coincide (at least for incompressible
flow) with
the ensemble $\widetilde{\ba}^\lambda(C,t)$ for $\lambda,\nu\rightarrow 0$
considered
in section III.A.
If so, then flux-conservation will hold in an average sense similar to
(\ref{alfven-resist}).
The advantage of the present argument is that it makes no explicit reference to
Spitzer resistivity or to any other microscopic plasma mechanism. Stochastic
flux-freezing is fundamentally a phenomenon of the nonlinear MHD dynamics.

Stochastic flux-conservation is expected to be a property of singular solutions
of the ideal MHD equations that describe turbulent plasmas asymptotically
at high kinetic and magnetic Reynolds numbers. The existence of such singular
solutions is an old conjecture of Lars Onsager
\cite{Onsager49,EyinkSreenivasan06,
EyinkPhysD08}. Such solutions must be quite different from smooth laminar
solutions of the ideal MHD equations that are familiar from current analysis,
however, and
must possess many ``strange'' properties. We are still learning how to deal
with such solutions.
We have recently shown how to derive the stochastic Kelvin theorem for the
incompressible
Navier-Stokes equation using a stochastic least-action principle
\cite{EyinkPhysD09}.
It is expected that there will be a similar stochastic least-action principle
for Onsager's
singular Euler and ideal MHD solutions. This is a fundamental motivation to
expect stochastic flux-conservation at infinite Reynolds numbers.

There must be many important applications of stochastic flux-freezing in plasma
physics,
astrophysics and geodynamo studies. We have here presented one concrete
application
to the finite-$Pr_m,$ kinematic, fluctuation dynamo in non-helical,
incompressible fluid
turbulence. Stochastic flux-freezing is critically important to the mechanism
of small-scale
turbulent dynamo, because distinct field-lines that are initially separated by
inertial-range
distances arrive to the same point and resistively merge to produce the net
magnetic field.
See Eq.(\ref{large-r}). Our results and analyses point to essential
similarities between
the finite-$Pr_m$ and $Pr_m=0$ dynamo in their Lagrangian mechanisms.
Understanding
the saturation effect of the Lorentz force from a Lagrangian perspective is an
obvious next
step. In addition, there will be interesting applications of stochastic
flux-freezing to many
other problems, e.g. to the theory of fast turbulent reconnection. This will be
 the subject
of future work.


%
%

%

\begin{acknowledgments}
This work was partially supported by NSF Grant No. AST-0428325 at Johns Hopkins
University.
All of our numerical results for hydrodynamic turbulence were obtained using
public data
available online in the JHU Turbulence Database Cluster. I wish to acknowledge
useful
conversations with H. Aluie, A. Balk, A. Beresnyak, E. Bodenschatz,  S.
Boldyrev, A. Busse,
S. Chen, J. Cho, G. Hornig, A. Lazarian, C. Meneveau, A. Neto,  A. Newell, E.
Vishniac,  M.-P. Wan,
and Z. Xiao. Most of the writing was completed during the author's spring 2010
sabbatical
at the Center for Magnetic Self-Organization in Laboratory and Astrophysical
Plasmas
at the University of  Wisconsin - Madison. We acknowledge the warm hospitality
of the
center and of Ellen Zweibel, Center Director.
 \end{acknowledgments}

\appendix*

\section{Path-Integral Formulas}\lb{PathInt}

\noindent We here sketch the derivation of the path-integral formulas
(\ref{P-pathint}) and
(\ref{theta-FK}) in the text. There are discussions already available in the
literature
\cite{Drummond82,ShraimanSiggia94}, but we stress here some connections not
found
in those works, with rigorous stochastic analysis, on the one hand, and with
Feynman
path-integral methods, on the other.

Our starting point is the SDE (\ref{SDE}), which we
discretize using the Euler-Maruyama scheme:
\be \bX_n =\bX_{n-1}+\bu(\bX_{n-1},t_{n-1})\delta t
+\sqrt{2\kappa}(\bW_n-\bW_{n-1}).
 \lb{SDE-EM} \ee
Note that the Brownian-motion variables at the discrete set of times have the
Gaussian density (with $\bW_0\equiv\bzed$)
\begin{eqnarray}
&& \mathcal{P}(\bW_1,...,\bW_N)=\cr
 && (const.) \exp\left(-\frac{1}{2\delta t}\sum_{n=1}^N
          \left| \bW_n -\bW_{n-1}\right|^2\right). \cr
 &&
\end{eqnarray}
with respect to $D\bW= \prod_n d\bW_n.$
We can obtain the density $\mathcal{P}(\bX)$ from the change of variables
formula
$\mathcal{P}(\bX)=\mathcal{P}(\bW)/\left|{\rm \det}\,\left(\frac{\partial
\bX}{\partial \bW}\right)
\right|.$ It is easy to see from (\ref{SDE-EM}) that the $(3N)\times(3N)$
Jacobian matrix $\frac{\partial\bX}{\partial\bW}$
is block lower-triangular, with diagonal blocks
$\frac{\partial\bX_n}{\partial\bW_n}=
\sqrt{2\kappa}\bI,$ for the $3\times 3$ identity matrix $\bI.$  Thus, ${\rm
\det}\,
\left(\frac{\partial\bX}{\partial\bW}\right)=const.$ and
\begin{eqnarray}
&& \mathcal{P}(\bX_1,...,\bX_N)\propto   \cr
&&\exp\left(-\frac{1}{4\kappa}\sum_{n=1}^N \delta t
          \left|  \frac{\bX_n -\bX_{n-1}}{\delta
t}-\bu(\bX_{n-1},t_{n-1})\right|^2\right) \cr
&&  \,\,\,\,\,\,\,\,\,\,\,\,\,\,
\lb{P-X} \end{eqnarray}
Integrating this density with respect to $D\bX=\prod_n d\bX_n$ and taking the
continuous-time limit $\delta t\rightarrow 0,N\rightarrow \infty,$ one formally
obtains
formulas like (\ref{P-pathint}) and (\ref{theta-FK}).

The mathematically rigorous versions of these path-integral formulas is the
classical {\it Girsanov transformation} \cite{Girsanov60}, or see Chung and
Williams
\cite{ChungWilliams90} for a modern proof. We shall just remind the reader of
this result,
in its simplest terms. Suppose that $\mathcal{W}_\kappa$
is a Wiener measure over a rescaled Brownian
motion $\sqrt{2\kappa}\bW(t)$ and $\mathcal{W}_{1/2}=\mathcal{W}$ is the
standard
Wiener measure. Suppose that $\bu(\bx,t)$ is any smooth vector field and
$\bW(t)$
is defined in terms of $\bX(t)$ by integrating (\ref{SDE}):
$$ \bW(t)=\frac{1}{\sqrt{2\kappa}}\left[ \bX(t)-\bx_0-\int_0^t
ds\,\bu(\bX(s),s)\right]. $$
Then Girsanov's theorem states that
\begin{eqnarray}
&& \mathcal{DW}_\kappa(\bX)\exp\left[\frac{1}{2\kappa}\left(
     \int_0^T \bu(\bX,t)\bdot d\bX -\frac{1}{2}\int_0^T
u^2(\bX,t)\,dt\right)\right] \cr
&&    \,\,\,\,\,\,\, \,\,\,\,\,\,\, \,\,\,\,\,\,\, \,\,\,\,\,\,\,
\,\,\,\,\,\,\, \,\,\,\,\,\,\, \,\,\,\,\,\,\, \,\,\,\,\,\,\, \,\,\,\,\,\,\,
        \,\,\,\,\,\,\, \,\,\,\,\,\,\, =\mathcal{DW}(\bW), \lb{girsanov}
\end{eqnarray}
where
$$  \int_0^T \bu(\bX,t)\bdot d\bX =\lim_{N\rightarrow\infty}
      \sum_{n=1}^N \bu(\bX_{n-1},t_{n-1})\bdot(\bX_n-\bX_{n-1}) $$
is the usual Ito stochastic integral \cite{Girsanov60,ChungWilliams90}. This is
easily
seen to be equivalent to (\ref{P-X}) by expanding the square in the exponent of
the latter
and noting that $D\bX \exp\left(-\frac{1}{4\kappa\delta t}\sum_{n=1}^N
\left|\bX_n -\bX_{n-1}\right|^2\right)$ converges to the Wiener measure
$\mathcal{DW}_\kappa(\bX)$ in the continuum limit.

Although the Euler-Maruyama scheme provides the simplest derivation of such
path-integral formulas, other discretizations are possible and yield the same
results.
E.g., suppose that the trapezoidal rule is employed:
\begin{eqnarray}
 \bX_n &= & \bX_{n-1}+\frac{\bu(\bX_n,t_n)+\bu(\bX_{n-1},t_{n-1})}{2}\delta t
\cr
 && \,\,\,\,\,\,\,\,\,\,\,\, \,\,\,\,\,\,\,\,\,\,\,\, \,\,\,\,\,\,\,\,\,\,\,\,
      +\sqrt{2\kappa}(\bW_n-\bW_{n-1}). \lb{SDE-trap} \end{eqnarray}
The sum in the exponent of (\ref{P-X}) is replaced by
$$ \sum_{n=1}^N \delta t \left|  \frac{\bX_n -\bX_{n-1}}{\delta t}
-\frac{\bu(\bX_n,t_n)+\bu(\bX_{n-1},t_{n-1})}{2}\right|^2. $$
Expanding and taking the continuum limit, this converges formally to
$$ \int_0^T |\dot{\bX}|^2\,dt-2\int_0^T \bu(\bX,t)\bcirc \,d\bX
+\int_0^T u^2(\bX,t)\,dt, $$
where
\begin{eqnarray}
 && \int_0^T \bu(\bX,t)\bcirc \,d\bX= \cr
 && \lim_{N\rightarrow\infty}
      \sum_{n=1}^N
\frac{\bu(\bX_n,t_n)+\bu(\bX_{n-1},t_{n-1})}{2}\bdot(\bX_n-\bX_{n-1}) \cr
 &&
\end{eqnarray}
is the Stratonovich stochastic integral. Since formally
$\mathcal{D}Q_\kappa(\bX)=
\mathcal{D}\bX\exp\left(-\frac{1}{4\kappa} \int_0^T |\dot{\bX}|^2\,dt\right),$
it appears
that we recover the same result as before but with the Stratonovich integral
substituted
for the Ito integral in equation (\ref{girsanov}). However, we have not yet
computed
the contribution of the Jacobian determinant to the change of variables. With
the discretization (\ref{SDE-trap}) the Jacobian matrix
$\frac{\partial\bX}{\partial\bW}$
is block lower-triangular, with diagonal blocks
\begin{eqnarray*}
&&
\frac{\partial\bX_n}{\partial\bW_n}=\sqrt{2\kappa}\left(\bI-\frac{1}{2}\delta
t\,
       \frac{\partial \bu}{\partial \bx}(\bX_n,t_n)\right)^{-1} \cr
&& = \sqrt{2\kappa}\exp\left(\frac{1}{2}\delta t\,
       \frac{\partial \bu}{\partial \bx}(\bX_n,t_n) + O(\delta t^2)\right).
\end{eqnarray*}
Thus, using ${\rm det}\,(\exp\bA)=\exp({\rm tr}\,\bA),$
\begin{eqnarray*}
{\rm \det}\,\left(\frac{\partial\bX}{\partial\bW}\right) &\propto&
\exp\left(\frac{1}{2}\sum_{n=1}^N \delta t\,
(\grad\bdot\bu)(\bX_n,t_n)+O(\delta t)\right) \cr
 &\longrightarrow&  \exp\left(\frac{1}{2}\int_0^T
(\grad\bdot\bu)(\bX,t)\,dt\right)
\end{eqnarray*}
in the continuum limit $N\rightarrow \infty.$ The integral in the latter
exponent can be
expressed in terms of the quadratic variation process \cite{ChungWilliams90}
$$ \langle X,Y\rangle_T=\lim_{N\rightarrow \infty}
    \sum_{n=1}^N (X_n-X_{n-1})(Y_n-Y_{n-1}). $$
Using $d\langle X_i,X_j\rangle_t=2\kappa \delta_{ij}dt,$ one obtains
$$ \frac{1}{2}\int_0^T (\grad\bdot\bu)(\bX,t)\,dt=
     \frac{1}{4\kappa}\int_0^T d\langle \bu(\bX,t);\bX\rangle. $$
But the standard relation between Ito and Stratonovich integrals,
$$\int_0^T \bu(\bX,t)\bcirc \,d\bX - \frac{1}{2}\int_0^T
d\langle\bu(\bX,t);\bX\rangle
    = \int_0^T \bu(\bX,t)\bdot \,d\bX, $$
then recovers (\ref{girsanov}) exactly as before.

For physicists, an illuminating derivation of the path-integral formulas can be
based on
Feynman's famous formula for the transition amplitude of a quantum,
nonrelativistic,
charged particle moving in a scalar potential $V$ and in a magnetic field with
vector
potential $\bA.$ Feynman's result was
$$ \langle \bx,t|\bx_0,0\rangle =
\int^{\bx(t)=\bx}_{\bx(0)=\bx_0}\mathcal{D}\bx
    \exp\left(\frac{i}{\hbar}\int_0^t ds\,\mathcal{L}(s)\right),$$
where $\mathcal{L}(t)$ is the classical Lagrangian
$$ \mathcal{L}(t)=\frac{1}{2}m|\dot{\bx}|^2
+\frac{e}{c}\bA(\bx,t)\bdot\dot{\bx}-V(\bx,t)$$
and the amplitude satisfies the Schr\"odinger equation
$$ i\hbar\partial_t\Psi
=\frac{1}{2m}\left(-i\hbar\grad-\frac{e}{c}\bA(\bx,t)\right)^2\Psi
      +V(\bx,t)\Psi.$$
See Feynman \cite{Feynman48}, p.376  or later treatments
\cite{GaveauSchulman89,Gaveauetal04}.
Taking imaginary time $T=it,$ $\bu=i\frac{e\bA}{mc},$ $\kappa=\frac{\hbar}{2m}$
and
$$ V=-\frac{e^2}{2mc^2}A^2-i\frac{\hbar e}{2mc}\grad\bdot\bA, $$
yields the path-integral formula (\ref{P-pathint}) and
converts the Schr\"odinger equation into the diffusion equation
$\partial_T\Psi=\kappa\triangle\Psi-(\bu\bdot\grad)\Psi.$ This is very
straightforward
to check in the Coulomb gauge $\grad\bdot\bA=0.$ For a general choice of gauge,
note that the vector potential term in the classical action $\mathcal{L}(s)$
yields a term in the
exponent of Feynman's path-integral formula which must be interpreted as a
Stratonovich  integral:
$$ \int_0^t \bA(\bx,s) \bdot\dot{\bx}(s)\,ds \equiv \int_0^t \bA(\bx,s) \bcirc
\,d\bx(s). $$
This is implicit in Feynman's original derivation, who used the midpoint
discretization
to define the above integral \cite{Feynman48}. This point has been carefully
discussed
elsewhere \cite{GaveauSchulman89,Gaveauetal04}.
If this Stratonovich integral is combined with the
$\grad\bdot\bA$ term from the potential, one gets a net contribution to the
action proportional to
$$ \int_0^T \bu(\bX,t)\bcirc \,d\bX + \frac{1}{2}\int_0^T
d\langle\bu(\bX,t);\bX\rangle
    = \int_0^T \bu(\bX,t)\bdot \,\widehat{d}\bX,$$
\vspace{.1in} where now $t=is$ and
$$  \int_0^T \bu(\bX,t)\bdot \,\widehat{d}\bX =
     \lim_{N\rightarrow\infty}
      \sum_{n=1}^N \bu(\bX_n,t_n)\bdot(\bX_n-\bX_{n-1}) $$
is the {\it backward Ito integral}. One obtains a result just like the Girsanov
formula (\ref{girsanov}) but with the usual (forward) Ito integral replaced by
a backward Ito integral.

Feynman's result is correct. As we stressed in the text, our path-integral
formulas (\ref{P-pathint})
and (\ref{theta-FK}) correspond to solving the SDE (\ref{SDE}) backward in
time, e.g. with the
backward Euler-Maruyama scheme
\be \bX_{n-1} =\bX_n-\bu(\bX_n,t_n)\delta t +\sqrt{2\kappa}(\bW_{n-1}-\bW_n)
 \lb{SDE-BEM} \ee
for $t_{n-1}=t_n-\delta t.$ If we repeat the steps that led us to (\ref{P-X}),
we now obtain
\begin{eqnarray}
&& \mathcal{P}(\bX_1,...,\bX_N)\propto   \cr
&&\exp\left(-\frac{1}{4\kappa}\sum_{n=1}^N \delta t
          \left|  \frac{\bX_n -\bX_{n-1}}{\delta
t}-\bu(\bX_{n},t_{n})\right|^2\right). \cr
&&  \,\,\,\,\,\,\,\,\,\,\,\,\,\,
\lb{P-X-b} \end{eqnarray}
Integrating this density with respect to $D\bX=\prod_n d\bX_n$ and taking the
continuous-time limit $\delta t\rightarrow 0,N\rightarrow \infty,$ one formally
obtains
a Girsanov-type formula with the forward Ito integral replaced by a backward
Ito
integral. This is the rigorous version of our formulas (\ref{P-pathint}) and
(\ref{theta-FK}),
which correspond exactly to Feynman's old result.

\bibliography{StochDyn_PoP}

\end{document}